\documentclass[english,thm-restate,nolineno]{socg-lipics-v2021}

\usepackage[LGR, T1]{fontenc}
\usepackage[utf8]{inputenc}
\usepackage{lmodern, amsmath}

\usepackage{hyperref}
\usepackage{xcolor}
\definecolor{darkgreen}{RGB}{0,100,0}
\definecolor{darkred}{RGB}{139,0,0}

\usepackage{graphicx}
\usepackage{tikz}
\usetikzlibrary{arrows.meta}
\usetikzlibrary{calc}
\usetikzlibrary{decorations.pathmorphing}
\usetikzlibrary{positioning}
\usepackage{mathtools, amsmath, amsthm, amssymb}

\theoremstyle{plain}
\newtheorem{convention}[theorem]{\textbf{Convention}}

\newcommand\R{\mathbb{R}}

\newcommand{\M}{\mathcal{M}}

\newcommand{\symm}{\operatorname{sym}}
\newcommand{\Gsymm}{\operatorname{gsym}}
\newcommand{\foc}{\operatorname{foc}}

\newcommand{\Skip}[1]      {}

\def\barroman#1{\sbox0{#1}\dimen0=\dimexpr\wd0+1pt\relax
  \makebox[\dimen0]{\rlap{\vrule width\dimen0 height 0.06ex depth 0.06ex}%
    \rlap{\vrule width\dimen0 height\dimexpr\ht0+0.03ex\relax 
            depth\dimexpr-\ht0+0.09ex\relax}%
    \kern.5pt#1\kern.5pt}}

\title{A Reidemeister-type theorem for vineyards
}
\titlerunning{A Reidemeister-type theorem for vineyards
}

\author{Erin Chambers}{University of Notre Dame\\{[Notre Dame, IN, USA]}}{echambe2@nd.edu}{https://orcid.org/0000-0001-8333-3676}{Supported by NSF award 2444309.}

\author{Christopher Fillmore}{Independent Researcher\\{[Vienna, Austria]}}{cdfillmore@gmail.com}{https://orcid.org/0000-0001-7631-2885}{}

\author{Shankha Shubhra Mukherjee}{University of Notre Dame\\{[Notre Dame, IN, USA]}}{mail@mail.com}{https://orcid.org/3.14}{}

\author{
Rohit Roy}{Inria Centre d'Universit{\'e} C{\^o}te d'Azur\\{[Sophia Antipolis, France]}  }{rohit.a.roy@inria.fr}{https://orcid.org/0009-0002-1927-2172}{Supported by the French National Science Agency (ANR) under the StratMesh grant (ANR-24-CE48-1899).}

\author{Elizabeth Stephenson}{Orteliu\\{[Heggedal, Norway]}}{elizasteprene@gmail.com}{https://orcid.org/0000-0002-6862-208X}{}

\author{
Mathijs Wintraecken}{Inria Centre d'Universit{\'e} C{\^o}te d'Azur\\{[Sophia Antipolis, France]}  }{mathijs.wintraecken@inria.fr}{https://orcid.org/0000-0002-7472-2220}{Supported by the French National Science Agency (ANR) grant StratMesh (ANR-24-CE48-1899) and the welcome package from IDEX of the Universit{\'e} C{\^o}te d'Azur (ANR-15-IDEX-01).
}

\funding{} 

\authorrunning{E. Chambers, C. Fillmore, S. Mukherjee, R. Roy, E. Stephenson, and M. Wintraecken} 

\Copyright{ 
Erin Chambers, Christopher Fillmore, Shankha Shubhra Mukherjee, Elizabeth Stephenson, and Mathijs Wintraecken} 

\acknowledgements{We thank Francesca Bertoglio, Ellen Gasparovich and Andr{\'e} Lieutier for discussion. 
Mathijs Wintraecken expresses his gratitude for the administrative support he received from Sophie Honnorat, Stephanie Verdonck, and Vanessa Wallet at Inria in general and the administrative staff at the University of Notre Dame  during his visit there. 
} 

\hideLIPIcs
\EventLogo{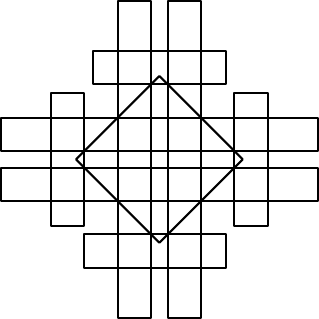}

\begin{document}
\maketitle
\begin{abstract} 

Closed vineyards record how persistent homology changes along a periodic family of filtrations. Their topology can be surprisingly complicated, with recent work showing that vines may exhibit nontrivial monodromy even in very simple settings, and more generally that every link can occur in a closed vineyard. In this paper, we study how the global topology and structure of a vineyard change when the periodic family of functions changes. We study radial persistence for a generic planar curve $\M$, where an observation point $x\in\mathbb{R}^2$ defines the squared-distance function $d_\mathbb{E}^2 (\cdot ,x)|_\M$. Moving the point $x$ along a closed simple observation loop $\gamma$ produces a closed vineyard.  Utilizing tools from singularity theory, we show that the topology of this vineyard is controlled by the interaction of $\gamma$ with the generalized symmetry set of $\M$, the geometric discriminant formed by the union of focal and symmetry sets. We establish a Reidemeister-type theorem: under a generic deformation of $\gamma$, every topological change in the vineyard is generated by a finite collection of elementary local moves. 
Thus, despite the potentially unbounded global complexity of closed vineyards, their evolution is governed by a finite collection of local geometric events.
We classify these events and determine the corresponding changes to the vineyard, giving a complete local description of how radial vineyards evolve under generic deformation.  In addition, we  derive sufficient geometric conditions under which the resulting vineyard has no monodromy and trivial topology.
The 2D case already exhibits very rich behaviour, and we conjecture that in fact it captures all possible generic changes in vineyards.
\end{abstract}

\section{Introduction}
\label{sec:Introduction}

In recent years, persistent homology~\cite{Edelsbrunner2002} has emerged as one of the primary tools in the field of topological data analysis.  Persistence diagrams are both computable and stable~\cite{CohenSteinerStability}, and have seen utility in a huge range of applications~\cite{DONUT}; while a full overview of this rich topic is beyond our scope, we refer the interested reader to recent books on this topic~\cite{Boissonnat2018,dey2022computational}.

When the input data set is time-varying, one can study the evolution of the persistence diagram, which due to known stability results will move continuously, forming an object that is generally referred to as a vineyard~\cite{CohenSteiner2006}.  Generically, each point in the persistence diagram traces a curve, and if the function is periodic, the vines join into closed curves, which we call a closed vineyard, as studied in~\cite{Burghelea2013, turner2023representing}.

Recent insights into these objects have shown they can be surprisingly complex.  In~\cite{Arya2024}, the authors proved that even curves in the plane can have periodic functions which exhibit \emph{monodromy} in the vineyard, where the points in the persistence diagram permute after a single period of the function.   Following this, other work~\cite{Chambers2026} showed that in fact the vineyard of an embedded space in $\R^d$ can contain arbitrary braiding for a loop in the radial persistence filtration~\cite{Shoving}, where one filters based on distance to a point in the ambient space and then builds a closed vineyard by following the point along a loop, that is an embedded $S^1 \subseteq \R^d$. {We call such a loop an observation loop. } Most recently, \cite{local} studied a precise characterization of when small loops generate monodromy in the vineyard, utilizing tools from singularity theory in order to better characterize the precise geometry that causes this behaviour.

\subparagraph{The persistent homology transform} 
We note that our results are situated within a larger context of studying persistence transforms, see~\cite{Turner2014, Curry2022}, which are based on \cite{Schapira1991,Schapira1995}. In the directional transform,  the time-varying persistence diagrams are extracted from some shape embedded in $\R^d$ by picking a direction $\omega \in \mathbb{S}^{d-1}$.   
Earlier work also connected persistence modules to monodromy in bifiltered complexes~\cite{Cerri2013,Cerri2019}, which are considerably more complex than our single parameter setting.
In~\cite{Shoving}, the authors introduce a transform based on distance to a point, line, or hyperplane in the ambient space, and prove that these transforms have some significant potential advantages in terms of power and computational cost.  Later work~\cite{Chambers2026,local} used the radial transform, based upon distance to a point, in order to better understand the types of monodromy which can occur and their geometric causes.

\subparagraph{Singularities in the symmetry and focal sets} 
The core tool of our analysis is the generalized symmetry set, which is the set of points $x$ where the behaviour of the squared Euclidean distance function $d_\mathbb{E}^2 (\cdot ,x)|_\M$ restricted to the set $\M$ changes. 
We focus on the case where $\M$ is a smooth {generic} manifold  in $\mathbb{R}^{d}$. In this setting the generalized symmetry set is the union of the symmetry set—the closure of the centres of spheres tangent to $\mathcal{M}$ at multiple points—and the focal set, where 
the function $d_\mathbb{E}^2 (\cdot ,x)|_\M$ has a degenerate critical point. Together, these form the bifurcation set, recording both coincidences of critical values and degeneracies of critical points. These structures arise naturally in classical singularity theory~\cite{Looijenga1974, mather1983distance, arnol2003catastrophe}. {The } generic planar configurations {of the generalized symmetry set} were studied by Bruce, Giblin, and Gibson~\cite{Bruce1985symmetry} and Giblin and Diatta~\cite{Diatta2005}. {
The different degenerate critical points have been analysed, as well as the way they can interact. Their behaviour is recorded using Arnold's notation $A_1^2$, $A_1^3$, $A_1^2/A_2$, $A_2/A_2$, $A_1A_2$, $A_1^2/A_1^2$, $A_2$, $A_3$, {indicating the local structure as depicted in Figure \ref{fig:classification2DEarly}. }

\begin{figure}[h!] 
    \centering
    \includegraphics[width=0.9\linewidth]{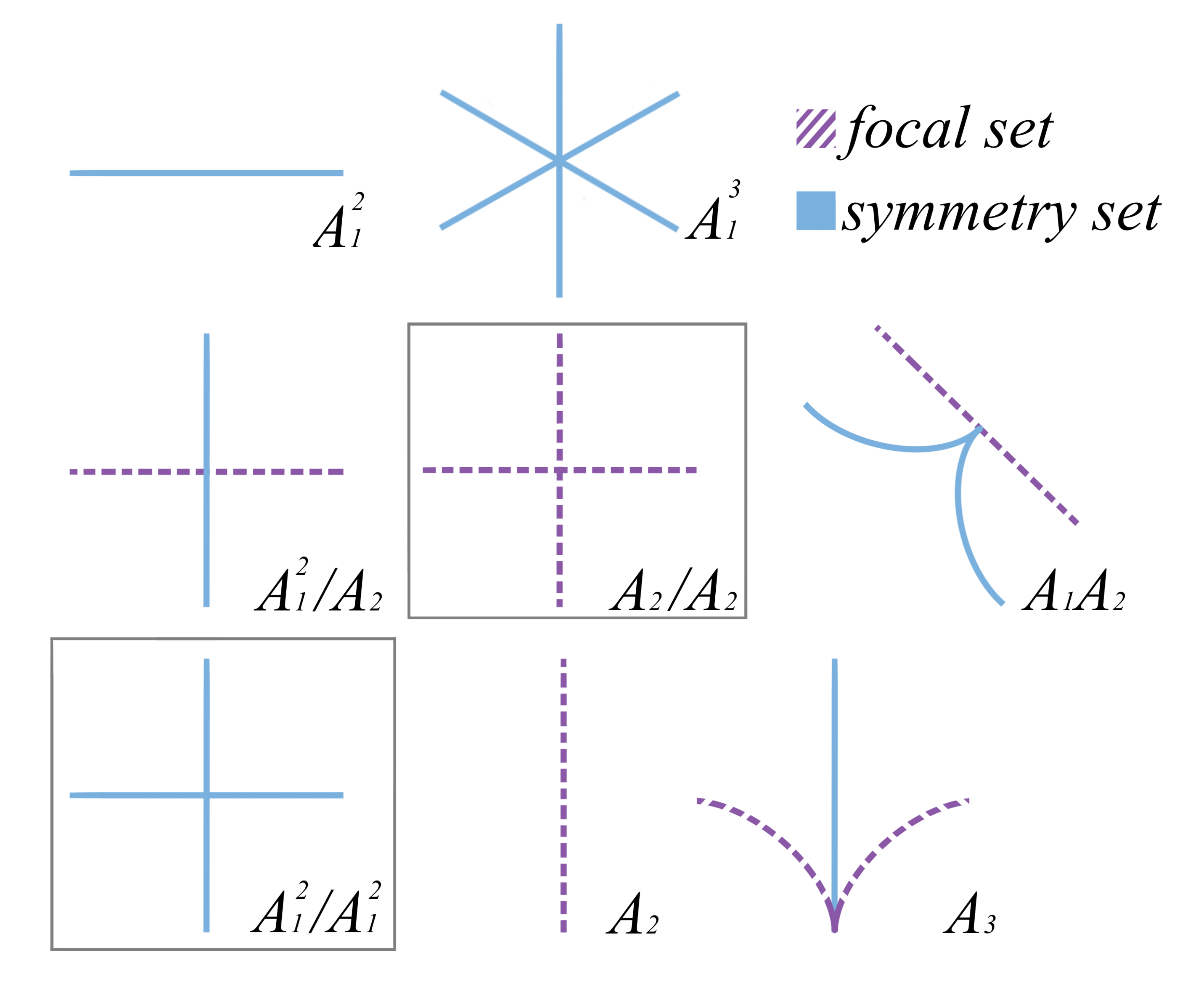}
    \caption{The singularities of the symmetry and focal sets in the plane. We use Arnold's notation for these singularities. If a loop does not contain singularities of type $A_1^2/A_1^2$ or $A_2/A_2$ {(boxed in grey)} it has neither monodromy nor non-trivial topology. (Figure adjusted from~\cite{local}.) 
    \label{fig:classification2DEarly} } 
\end{figure}

\subparagraph{Contribution and outline} 

\begin{figure}[h!] 
    \centering
    \includegraphics[width=\linewidth]{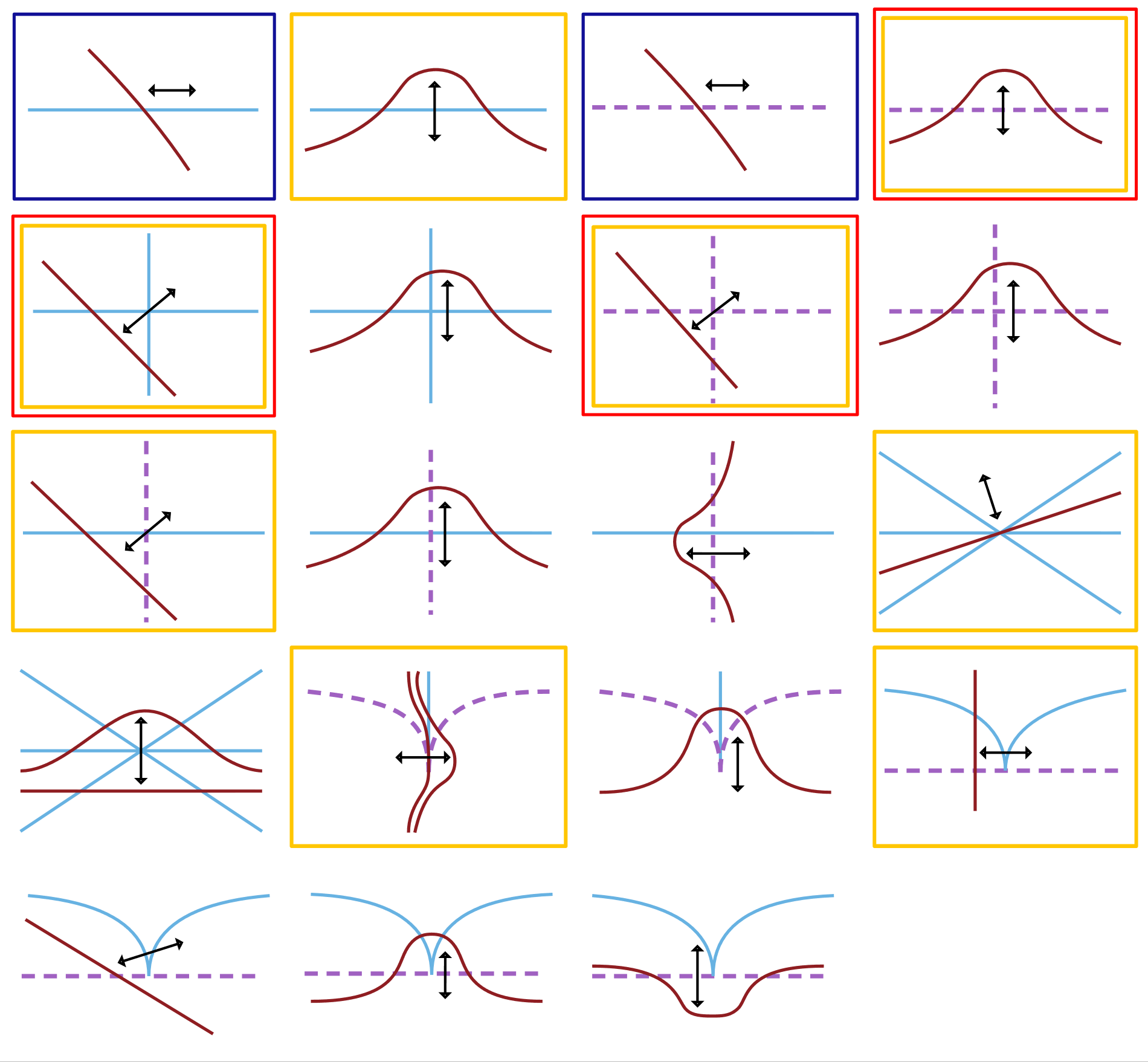}
    \caption{The complete set of elementary moves of an observation loop (shown in burgundy) in 2D with respect to the generalized symmetry set, consisting of the symmetry set (in light blue) and the focal set (in dashed purple). The trivial elementary moves 
    are indicated via a navy blue box, and the ones that are irreducible, meaning that they cannot be written as a combination of other moves, are indicated in yellow. The irreducible elementary moves of the observation loop that induce topology or monodromy in the vineyard are on top of this marked in red.  
    \label{fig:AllGenReidemeisterMoves} } 
\end{figure}

Our first main contribution is a complete list of generic elementary moves of an observation loop relative to the focal and symmetry sets in the radial persistence transform; see Figure~\ref{fig:AllGenReidemeisterMoves}.
Although inspired by the classical Reidemeister moves~\cite{Reidemeister1927}, we include two special moves to simplify the analysis. Section~\ref{sec:MainResults} states the classification and its consequences.
In particular, Theorem~\ref{prop:MovesOnVineyard} describes the correspondence between elementary moves of the observation loop and local changes in the vineyard. 
This correspondence need not be one-to-one, since the resulting change also depends on the global topology of $\mathcal{M}$.

There are only three elementary irreducible moves that induce a real topological change or induce monodromy in the vineyard (other moves may change how some coordinates of vines may concur, but not the topology). 
Here by topology of the vineyard we mean that there is an isotopy of $S^1 \times \mathbb{R}^2$ minus the diagonal which maps the two vineyards to each other, where we view them as subsets of  $S^1 \times \mathbb{R}^2$ minus the diagonal and $S^1= \mathbb{R}/ (2 \pi \mathbb{Z})$. 
Critically, we do not demand that the time slices are preserved in this isotopy.  This separates the {changes that can induce knotting or linking (in the persistence diagram seen as a subset of a solid torus in $\mathbb{R}^3$)}, from changes in the monodromy {caused by vines ending on the diagonal}, cf. Figures~\ref{fig:CorrespondenceA2SA2} and~\ref{fig:bishop-staff}.  
These irreducible moves have been indicated in red in Figure \ref{fig:AllGenReidemeisterMoves}. The correspondence between the moves of the curve and the elementary changes in the vineyard has been depicted in Figures \ref{fig:CorrespondenceA12SA12}, \ref{fig:CorrespondenceA2}, and \ref{fig:CorrespondenceA2SA2}. 
Thus, the evolution of an arbitrarily complex global persistence object can be described by a finite local calculus.

\begin{figure}[h!] 
    \centering
    \includegraphics[width=0.7\linewidth]{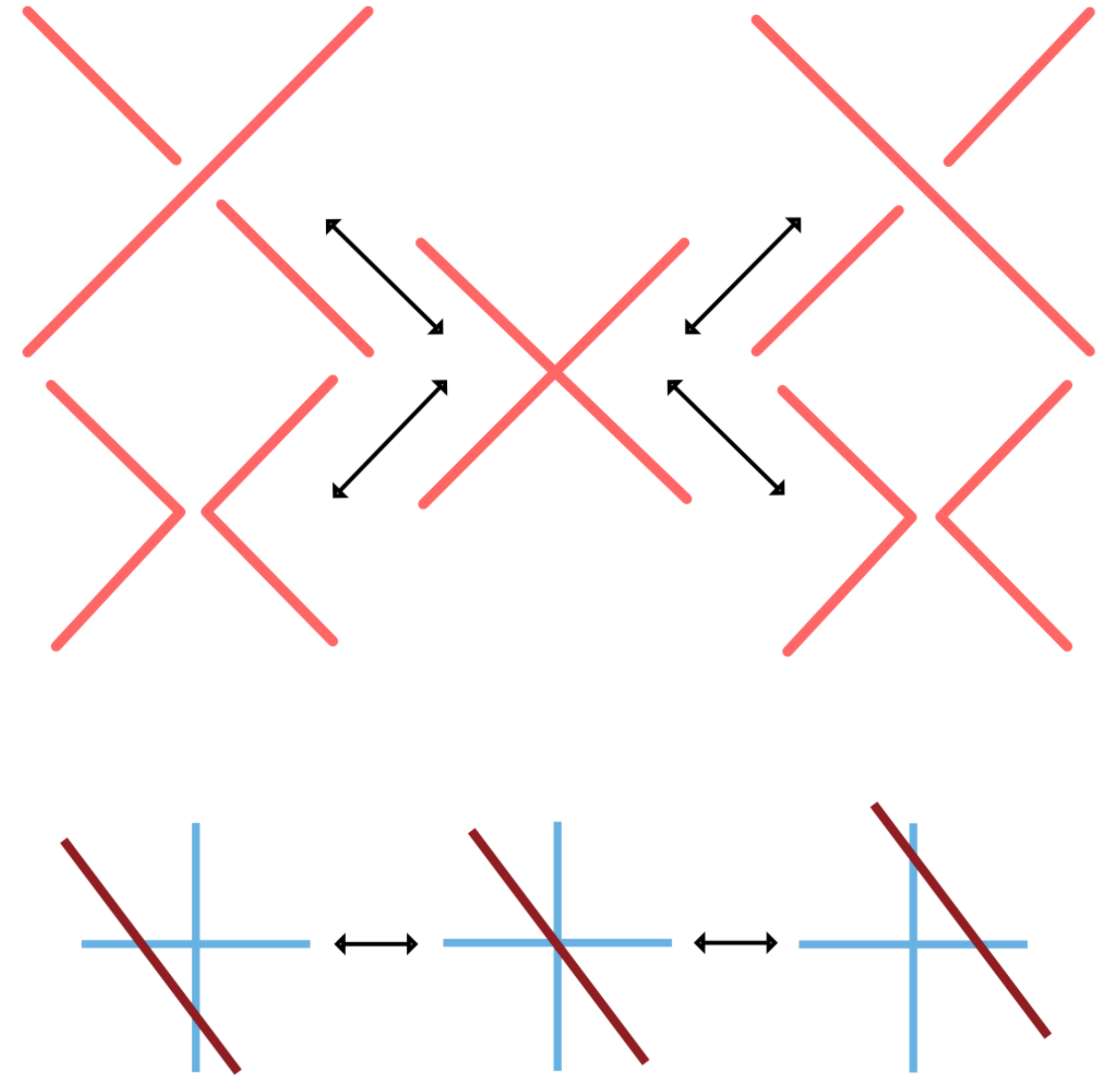}
    \caption{The topological changes in the vineyard (top) and the corresponding elementary moves of the observation curve with respect to a singularity of type $A_1^2/A_1^2$ in the generalized symmetry set (bottom). The arrows indicate the possible changes, and the different possibilities depend on the global topology.  However, {following} the two lower arrows on both sides is not generic. 
    \label{fig:CorrespondenceA12SA12} } 
\end{figure}

\begin{figure}[h!] 
    \centering
    \includegraphics[width=0.9\linewidth]{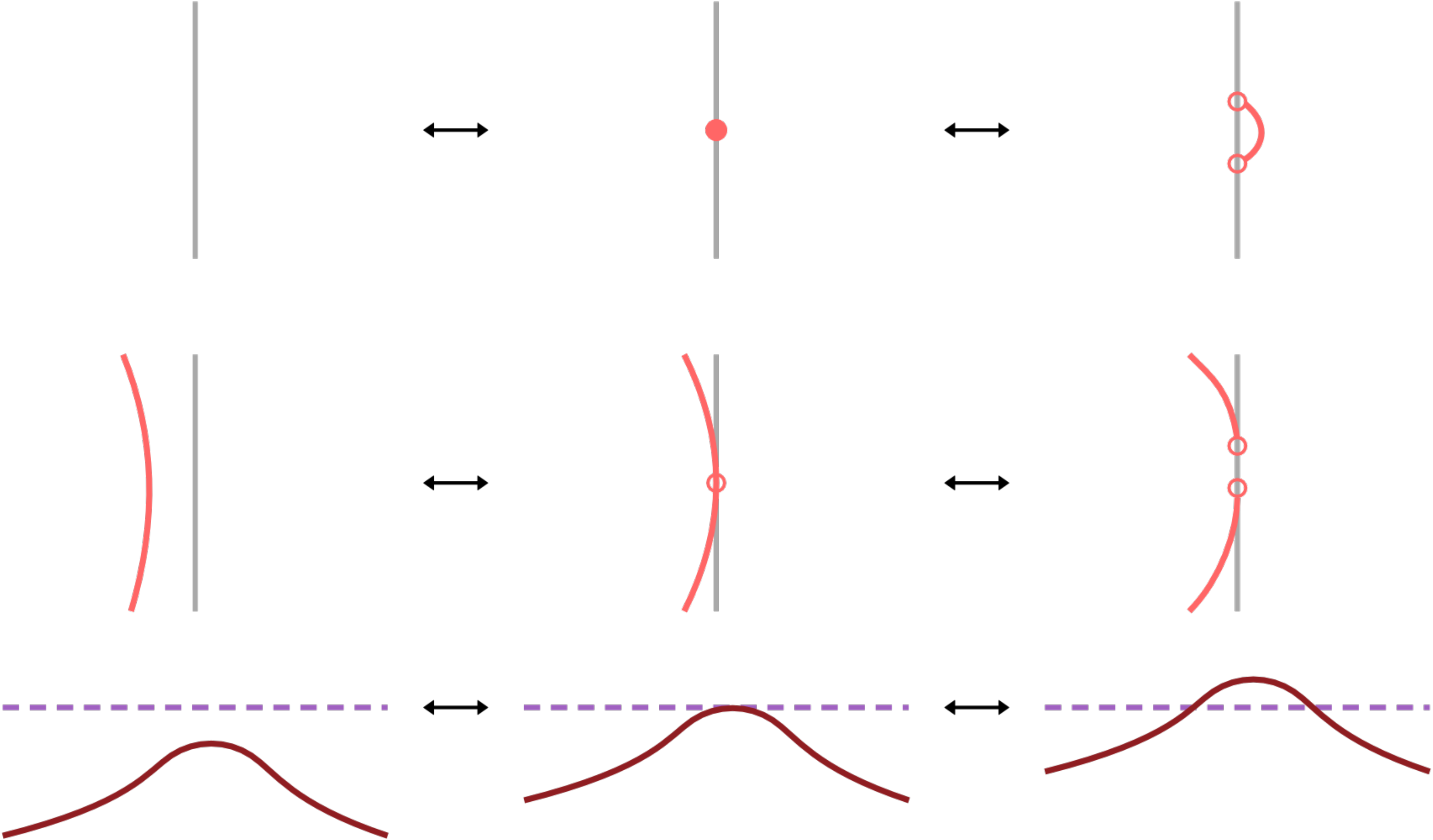}
    \caption{The topological changes in the vineyard (the top two lines) and the corresponding elementary moves of the observation curve with respect to a singularity of type $A_2$ in the generalized symmetry set (bottom). The diagonal is indicated in grey and the corresponding view vector is along the diagonal plane in the vineyard. 
    \label{fig:CorrespondenceA2} } 
\end{figure}

\begin{figure}[h!] 
    \centering
    \includegraphics[width=0.9\linewidth]{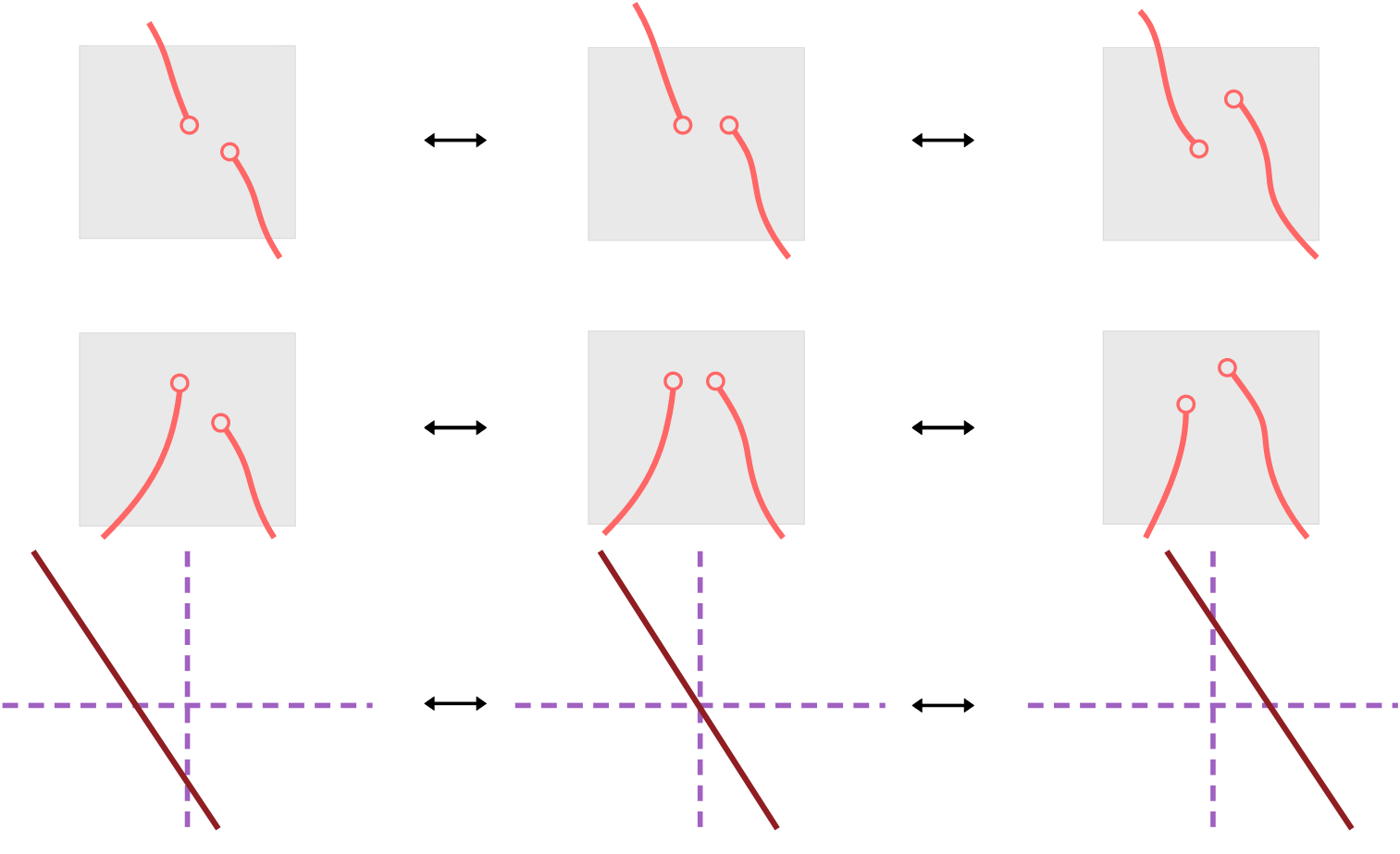}
    \caption{The change in the vineyard that can induce monodromy (the top line), a change in the vineyard that does not induce monodromy (note that the symmetrical case where the vines extend upwards can also occur) and the corresponding elementary moves of the observation curve with respect to a singularity of type $A_2/A_2$ in the generalized symmetry set (bottom). The diagonal is indicated in grey, that is although schematic, the corresponding view vector is normal to the diagonal plane in the vineyard.
    \label{fig:CorrespondenceA2SA2} } 
\end{figure}

The second main result of the paper is a set of criteria which, if satisfied by a curve, imply that the vineyard of the radial transform of that curve does not contain monodromy. Indeed, 
\begin{restatable}{theorem}{MainTHM}\label{thm:main}  
Let $\gamma_\mu$ be a generic ``isotopy'' from a point to a curve $\gamma$, where $\gamma_\mu$ is an isotopy except for the limit of $\mu \to 0$.  
No monodromy or non-trivial knotting or linking can occur in the radial transform vineyard of a generic compact manifold $\M$ unless one of the following events occurred in the isotopy:
\begin{itemize} 
\item $\gamma_\mu$ has passed over a singularity of type $A_1^2/A_1^2$ 
\item $\gamma_\mu$ has passed over a singularity of type $A_2/A_2$ 
\end{itemize}
In particular, if a simple curve $\gamma$ in the plane encloses no singularities of type $A_1^2/A_1^2$ or of type $A_2/A_2$ then the vineyard of the distance transform does not contain monodromy and the vines are all unlinked.
\end{restatable}

The remainder of the paper is organized as follows.
Section~\ref{sec:VineyardTransitions} introduces the local changes in a vineyard and specifies our conventions for connecting vines along the diagonal.
Section~\ref{sec:MainResults} shows how the changes in the vineyard arise from elementary moves of the observation loop with respect to singularities of the generalized symmetry set.
Full proofs, constructions, and examples appear in the Appendices.
Section~\ref{sec:future_work} concludes with open problems and directions for future research.


\section{Preliminaries}
\label{sec:preliminaries}

In this section, we review some of the essential tools underlying our work, with particular emphasis on those with which the reader may be less familiar.
Due to space constraints, we defer the definitions of monodromy and vineyards to Appendix \ref{App:VineyardsMonodromyBackground}.  

In this paper, we will restrict our definitions to the two dimensional setting, but we note that all the definitions extend to arbitrary dimensions \cite{porteous2001geometric}. In light of the generalization we will however refer to the 1-dimensional manifold (generic)  $\M$ as a manifold, not as a curve. 

\begin{definition}[Restricted distance function]
For a $p \in \mathbb{R}^2$ we can consider the squared (Euclidean) distance function to this point $d_\mathbb{E}^{2} (\cdot,p): \mathbb{R}^2 \to \mathbb{R}: x \mapsto d_\mathbb{E}^2 (x, p)$. Let $\M \subset \mathbb{R}^2$ be a smooth {(generic)} manifold. Then we call the {squared} distance function restricted to $\M$, that is $d_\mathbb{E}^{2}   (\cdot ,p) |_\M  : \M \to \mathbb{R}: x \mapsto d_\mathbb{E}^{2}  (x, p)$, the restricted distance function.   
\end{definition}

\subsection{Focal and Symmetry Sets}

We begin by recalling the geometric objects that describe the exceptional behaviour of the restricted distance function.
We restrict ourselves to two dimensions, so that a (non-trivial) manifold $\M \subset \mathbb{R}^2$ is a curve, which we will parametrize according to arclength with the parameter $s$. 
For a planar curve $\M$, the \emph{focal set} $\foc(\M)$ is the \emph{evolute}, which is the set of centres of curvature of $\M$, see \cite{porteous2001geometric}. For a general pedagogical introduction to the geometry of curves we refer the reader to e.g. \cite{DoCarmoCurves, oNeill}. 

We write $\mathbf{t}(s)$ for the unit tangent vector of $\M$ at the point with parameter $s$ and let $\mathbf{n}(s)$ denote the unit normal obtained by rotating $\mathbf{t}(s)$ ninety degrees counter-clockwise.

\begin{definition}[Curvature and Radius of Curvature]
The \textbf{curvature} $\kappa(s)$ is the rate of change of the tangent vector in the direction of the normal:
\begin{equation}
    \mathbf{t}'(s) = \kappa(s) \mathbf{n}(s)
\end{equation}
The \textbf{radius of curvature} $\rho(s)$ is the reciprocal of the curvature:
\begin{equation}
    \rho(s) = \frac{1}{\kappa(s)}, \quad \text{for } \kappa(s) \neq 0
\end{equation}
\end{definition}

In the following, we denote the parametrization of $\M$ by $\hat{\gamma}$. 
\begin{definition}[Evolute]
The \textbf{evolute} of $\mathcal{M}$ is the curve $E(s)$ defined by the centres of osculating circles {of the plane curve} $\mathcal{M}$,
$    E(s) = \hat{\gamma}(s) + \rho(s) \mathbf{n}(s)$,
where $\mathbf{n}(s)$ is the unit normal vector to the curve at $\hat{\gamma}(s)$.
\end{definition}

The evolute may be described as the envelope of the lines normal to the curve. 
The \emph{focal set} is also the locus of centers of circles that have second-order contact with $\mathcal{M}$.
Another characterization is as the set of points where the natural map from the normal bundle into Euclidean space fails to be an immersion; cf. \cite[Section 10.4]{DoCarmoRiemannian}.  
We recall that, given a closed {sub}set of Euclidean space, the medial axis \cite{blum1967transformation, erdos1945some,erdos1946,Federer1959} is the set of all points in the ambient space such that there is no unique closest point on the set.
The symmetry set will also be central to our analysis. This superset of the medial axis (assuming the closed set in fact has a manifold without boundary as its boundary) is defined as 
\begin{definition}[Symmetry set, \cite{Bruce1985symmetry}] \label{def:symmetry_set}
Let $\M \subset \mathbb{R}^d$ be a smooth {(generic)} manifold. The symmetry set $\symm (\M)$ of $\M$ is (the closure\footnote{Sources differ on whether or not to take the closure, this is not essential to our analysis.} of) the set of points $p$ in $\mathbb{R}^d$, such that there is a radius $r$ so that $\partial B(p,r)$ is tangent to $\M$ in at least two places. 
\end{definition}

\begin{figure}[htbp]
    \centering
    \includegraphics[width=0.65\textwidth]{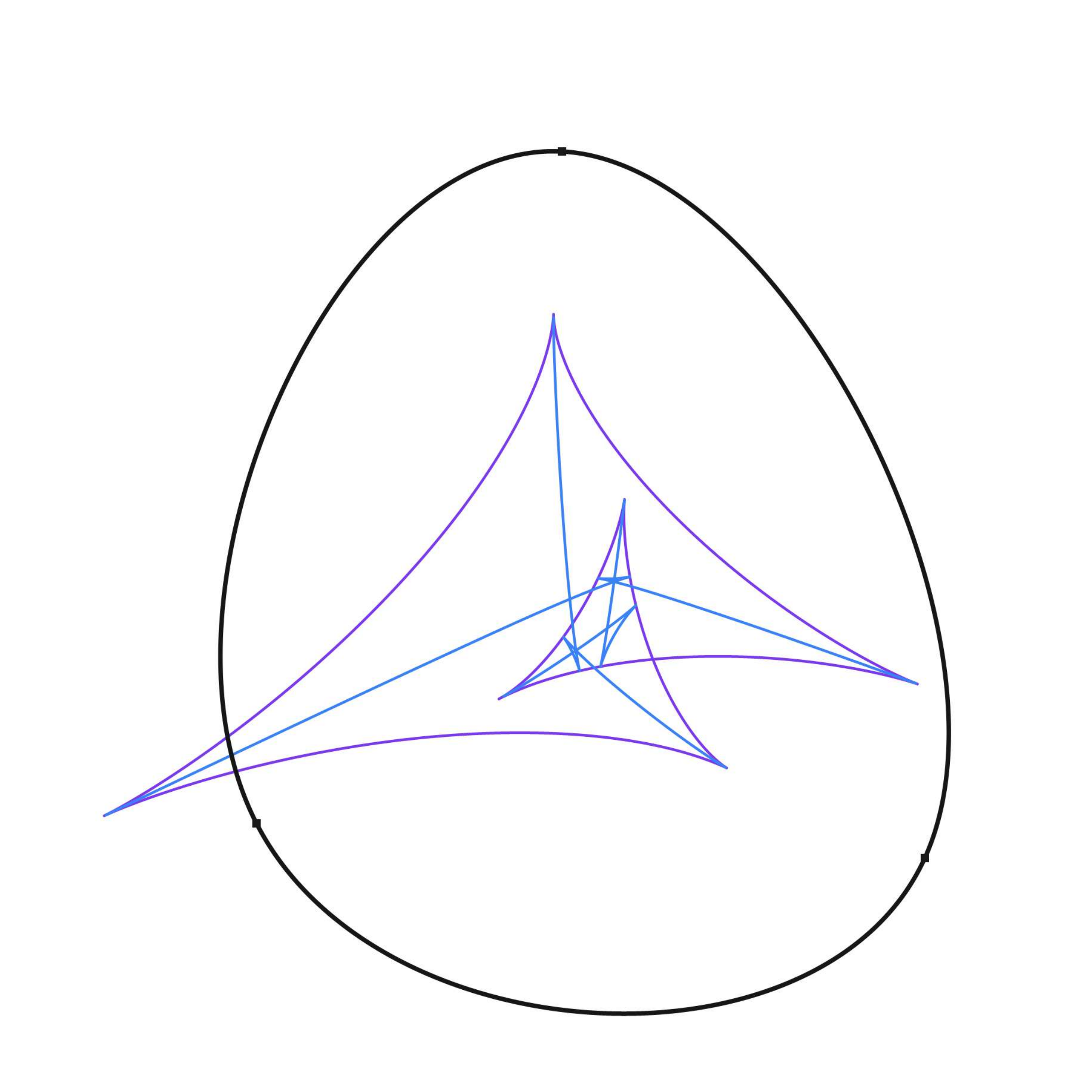}
    \caption{Visualizing the focal set (evolute) and symmetry set of a spline-defined curve $\mathcal{M}$ (in black). The purple locus represents the evolute, and the blue locus represents the symmetry set. }
    \label{fig:focal_set_visualizer}
\end{figure}

The focal and symmetry sets therefore encode two complementary ways in which the restricted distance function ceases to behave generically.
The former records degenerate critical points, whereas the latter records distinct critical points occurring at a common critical value. 
Since both phenomena will play a role in the changes of vineyards studied below, it is convenient to treat their union as a single geometric object, {see Figure \ref{fig:focal_set_visualizer}.}

\begin{definition}[Generalized symmetry set]
Let $\M \subset \mathbb{R}^d$ be a smooth manifold. The generalized symmetry set $\Gsymm (\M)$ of $\M$ is the union of $\symm (\M)$ and  $\foc(\M)$. 
\end{definition}

\subsection{The classification of singularities of the generalized symmetry set}
\label{sec:Singularities_Plane}
\textbf{From this point onwards we assume that $\M$ is generic in the sense of~e.g.~\cite{Bruce1985symmetry}.}
The local structure of the generalized symmetry set, or rather the topology of its singularities, is critical to the two main results of this paper: The description of the elementary moves which change the vineyard topology, and the (sufficient) condition which implies that a vineyard does not exhibit monodromy and that the vines are unlinked. We therefore recall the classification of singularities of the generalized symmetry set as given in \cite{Bruce1985symmetry}. Because we also rely on \cite{local}, we will try to stay as close as possible to that paper.

The union of the symmetry and focal sets, which we call the generalized symmetry set, is referred to as the full bifurcation set by Bruce, Giblin, and Gibson \cite{Bruce1985symmetry} and by Giblin and Diatta~\cite{giblin2003symmetry}.
The notation ($A_1^2$, $A_1^2/A_1^2$, $A_1^3$, $A_2$, $A_1A_2$, $A_2/A_2$, $A_1^2/A_2$, $A_3$) that we use to indicate the classes of singularities goes back to Arnold, Thom, and collaborators \cite{arnol2003catastrophe,arnol1993singularity,Arnold2012,arnold1998singularity}.
We point out the relation between the singularity theory and Cerf theory~\cite{Cerf_1970}, which is discussed in more detail in Appendix \ref{sec:Cerf}.  

The notation $A_k$ in 1D refers, roughly speaking, to the leading order in the Taylor approximation of a function $f(\tau)$. More formally, it means that the function is $\mathcal{R}$-equivalent (that is, equivalent up to a coordinate diffeomorphism, see \cite{bruce1992curves} for a full definition) to 
\[\pm \tau^{k+1}+c,\] 
with $c$ some constant. For us, the function {$f=f_p$} is  the Euclidean squared distance function $d_\mathbb{E}^2  (\cdot,p)|_\M$ restricted to the 1-dimensional manifold $\M$ and $\tau$ parametrizes $\M$. This also means that $c$ can be interpreted as the radius squared.

Geometrically, an $A_1$ contact is an ordinary tangency between $\M$ and a circle centred at $p$.
An $A_2$ contact occurs when $p$ is a centre of curvature of $\M$, or, equivalently, when the corresponding circle has second-order contact with the curve.
An $A_3$ contact is the next degeneracy. 
The osculating circle has third-order contact with $\M$, meaning that $p$ is a centre of curvature, but also that curvature attains a local extremum in the corresponding point on $\M$.
Contacts of higher order do not occur generically in the planar setting considered here~\cite{Bruce1985symmetry}.

In the case where a circle is tangent to $\M$ in multiple places denote the type of contact for each contact consecutively, that is $A_i A_j \dots A_l$, where we generally assume $i \leq j \leq \dots$. Moreover, $k$ repeated entries are abbreviated, that is $A_i A_i \dots A_i= A_i^k$.    
A slash separates contacts occurring at different radii but with the same centre.
For instance, $A_1^2/A_1^2$ describes two concentric circles of different radii, each tangent to $\M$ at two points.

\begin{figure}[h!] 
    \centering
    \includegraphics[width=1\linewidth]{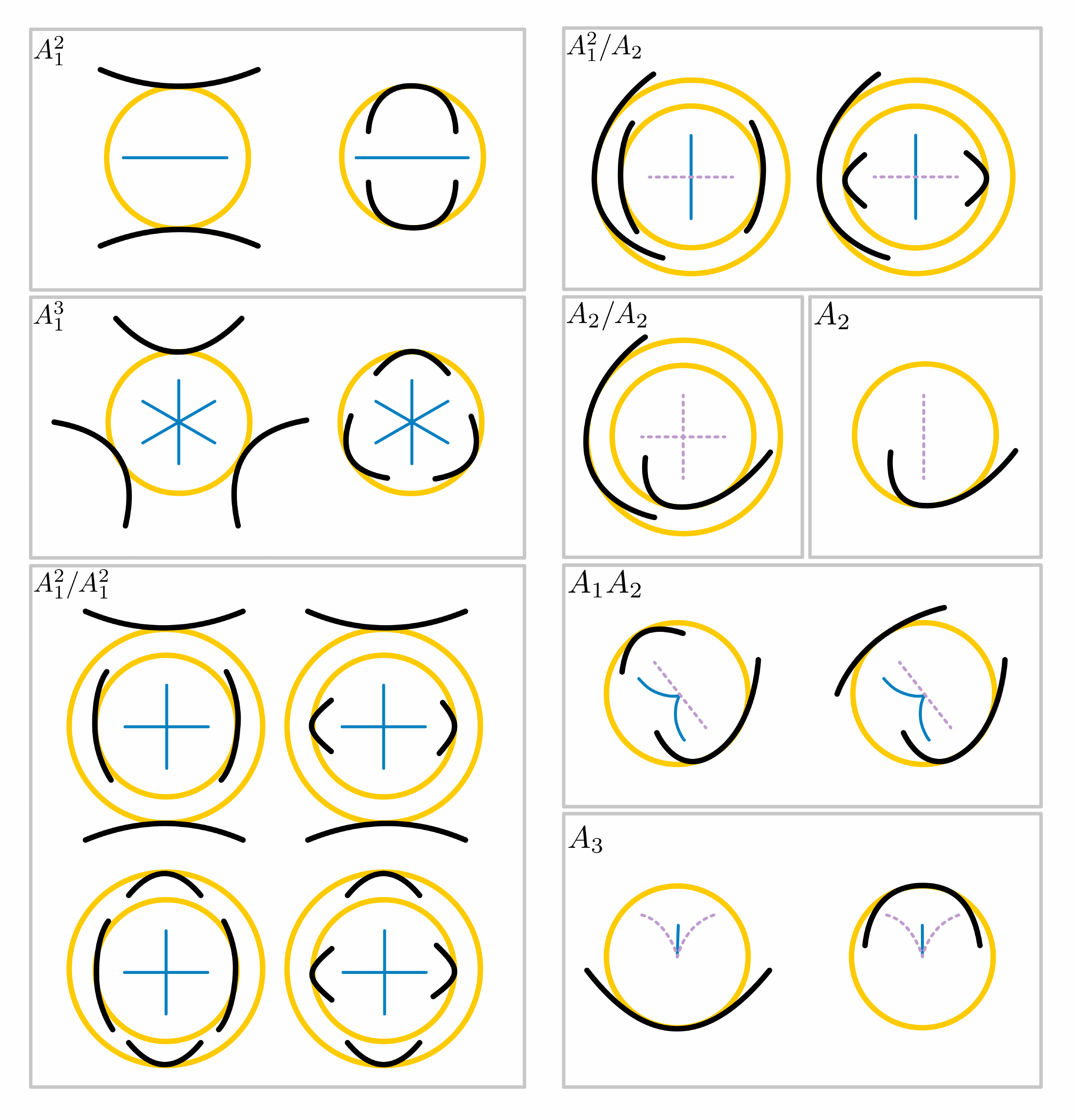}
    \caption{
    Replicated from \cite{local}, including caption: 
    The singularities of the symmetry and focal set in the plane. We adopt Arnold's notation for these singularities. {We denote the symmetry set in blue, the focal set in dashed purple, the manifold in black, and the relevant circles in yellow. We simplify the drawing using straight lines when possible to represent the singularities clearly rather than forcing completely correct geometry.} {We have indicated a number of different cases for $A_1^2$, $A_3$, $A_1^2/A_2$, $A_1^2/A_1^2$, but not all (given the number of cases involved). Note that every contact point of type $A_1$ can be either a local minimum or maximum. }
    \label{fig:classification2D} } 
\end{figure}

The local geometry represented by each singularity is depicted in  Figure~\ref{fig:classification2D}. We refer to Appendix~\ref{sec:DetailsSingularities} for a full description.

\subsection{Topological stability outside the generalized symmetry set}

The generalized symmetry set identifies precisely the locations in the ambient space at which the combinatorics of the restricted distance function may change.
Away from this set, the squared distance function is Morse and its critical values remain distinct.
Consequently, moving the observation point within a single connected component of $\mathbb{R}^2\setminus\operatorname{gsym}(\M)$ cannot change the relative ordering of the critical values, and hence cannot change the corresponding combinatorial structure of the persistence diagram.

We recall the following result from the work of Chambers et al.~\cite{local}, which makes this observation precise.

\begin{proposition}[\cite{local}]\label{prop:ConstOrd}
Let $\M$ be a manifold in $\mathbb{R}^2$ and $U$ be a connected component of the complement of $\operatorname{gsym}(\mathcal{M})$. Suppose that $y, y' \in U$ and let $(b_i,d_i)$, $(b_i',d_i')$ be the persistence points in $\operatorname{Dgm}(d_{\mathbb{E}}^2(x,y)\big|_{\mathcal{M}})$, $\operatorname{Dgm}(d_{\mathbb{E}}^2(x,y')\big|_{\mathcal{M}})$ respectively, where we order them according to birth time. This induces an ordering or permutation $\pi$ ($\pi'$, respectively) such that $d_{\pi(i)} < d_{\pi(i+1)}$ ($d_{\pi'(i)} < d_{\pi'(i+1)}$, respectively) for all $i$. We have that $\pi= \pi'$, and we say that the diagrams are combinatorially equivalent.  
\end{proposition} 

This immediately constrains the topology of a vineyard whose observation loop remains entirely inside one such component.

\begin{corollary}
    The vineyard of a loop $\gamma$ that is contained in such a connected component of the complement of $\operatorname{gsym}(\mathcal{M})$ is topologically the set of $k$ unlinked circles, where $k$ is the number of points in the persistence diagram $\operatorname{Dgm}(f_{\gamma(t^*) })$ for any and thus all $t^*$. Moreover the vineyard does not exhibit monodromy. 
\end{corollary}


Proposition~\ref{prop:ConstOrd} also implies that the persistence theoretic type assigned to a smooth stratum of the symmetry set is independent of the particular point chosen on that stratum.

In short,
any change in the topology or monodromy of the vineyard must be associated with the interaction of the observation loop with the generalized symmetry set.

\section{Vineyards of generic time varying functions}
\label{sec:VineyardTransitions}

While most of the paper focuses on radial persistence, we first outline how vineyards can change for any generic time-varying function, with the intent to later connect these moves to our combinatorics of the generalized symmetry set.  We conjecture that this is the complete set of vineyard changes for a generic time-varying function, but only prove that this holds for the radial transform.

We now introduce some neccessary nomenclature. Firstly, we say that a 
manifold piece of the symmetry set is a birth-birth stratum if the two contact points on $\M$ are local minima. We define birth-death and death-death strata similarly, see Appendix~\ref{sec:DetailsSingularities}. 
\begin{remark} We note that Proposition \ref{prop:ConstOrd} implies that the distinction between birth-birth, birth-death and death-death strata is well defined, in the sense that the point where you cross the same manifold piece of the symmetry set does not matter.
\end{remark}

For a generic time-varying vineyard, we say that there is a {concurrence} at time $t^*$ if there are two vines such that one of the following three things occurs:  
    \begin{enumerate}
        \item The birth coordinates at the time $t^*$ of two vines, that is of the points in the persistence diagram concur, that is they are equal (and none of the other coordinates concur). In this case we say there is a birth-birth concurrence. 
        \item The death coordinates at the time $t^*$ of two vines, that is of the points in the persistence diagram concur (and none of the other coordinates concur). In this case we say there is a death-death concurrence. 
        \item The birth coordinate of one vine and the death coordinate of another vine concur at the time $t^*$  (and none of the other coordinates concur). In this case we say there is a birth-death concurrence. 
    \end{enumerate}

We note that for non-generic curves several of the concurrences introduced above can occur 
at the same time.

\begin{remark} We stress that concurrences may, but need not, correspond to  sharp turns in a vine, known as knees \cite{CohenSteiner2006}, cf. \cite{ElizabethThesis, edelsbrunner2025mid}.
\end{remark}

We next classify the set of concurrences based on the changes in the vineyard:

\begin{definition}[Elementary (local) changes in the Vineyard] \label{def:ChangesVineyard} 
We consider a family of vineyards depending on a parameter $\mu$. 
We introduce the following vocabulary for the changes in the vineyard (with $\mu$): 
\begin{itemize}
\item \textbf{Vine Intersection type}
Two vines for a $\mu^*$ share a common point at a time $t^*$, meaning both birth and death times for two points in the persistence diagram are equal at $f_{\mu^*}(t^*)$.
Let $\epsilon_1, \epsilon_2>0$ and consider the parametrization of the two vines meeting at the single point, where we focus on $(\mu^*-\epsilon_1, \mu^* + \epsilon_1) \times \left( (t^*-\epsilon_2,t^* ) \cup (t^*, t^* + \epsilon_2) \right) $. This yields $4$ subcurves, which are pairwise connected at $t^*$, and the connection of the subcurves forming two vines may be different before and after $\mu^*$.
An example is constructed in the proof of the first main result, see Figures \ref{fig:CorrespondenceA12SA12} and \ref{fig:A12A12transversalresultRECALL}. 
\item \textbf{Vine creation/annihilation type}
In this case, a small vine is created out of or is annihilated into the diagonal plane.
\item \textbf{Vine merge/split type} Two {ends of} vines merge because one endpoint on the diagonal of each vine first (with increasing $\mu$) tends to each other, then coincides (at a point we call the merger point), and finally the part of the merged vine close to the merger point moves away from the diagonal, see Figure \ref{fig:CorrespondenceA2}. 
A split is the inverse of a merge event. {We stress that it is possible for the two ends to belong to the same vine, in which case the vine does not strictly speaking merge or split, however we stick to the nomenclature also in this case.}
\item \textbf{Rearrangement types} Rearrangement takes place in various different ways.
\begin{itemize}
    \item \textbf{The vine diagonal-diagonal rearrangements} Here the order (in time) in which two vines enter or exit the diagonal is interchanged.
    \item \textbf{The vine diagonal-concurrence, concurrence-concurrence rearrangement or triple concurrence rearrangements} 
    We say that there is a vine diagonal concurrence rearrangement, if (with increasing $\mu$) the order (in time) in which the vine enters the diagonal and a concurrence takes place is interchanged. 
    Similarly, we say that there is a vine concurrence-concurrence rearrangement, if the order (in time) in which two concurrences take place is interchanged. 
    We say that there is a triple concurrence rearrangement if the order (in time) in which three concurrences take place is interchanged. 
\end{itemize}
\item \textbf{concurrence pair creation/annihilation} We say that there is a concurrence pair creation if (with increasing $\mu$) some coordinates of two vines start to concur. The inverse (where the coordinates cease to concur) is called a concurrence pair annihilation. 
\item \textbf{Single concurrence/knee creation} We say that there is a single concurrence creation if (with increasing $\mu$) a single coordinate of each of two vines starts to concur. This concurrence has to be a knee. The limit of these coordinates at the critical transition (that is the critical $\mu$ value) has to coincide with the coordinates of a vine ending on the diagonal.  
\end{itemize}
\end{definition}

Additionally, we introduce some nomenclature for {the} setting where there is {no} topological or combinatorial change. 
\begin{definition}
We say that there is a trivial vine diagonal-concurrence arrangement, if {a vine enters or leaves the diagonal} and a concurrence occur at the same time, for some $\mu^*$, but the order {in which the vine enters or leaves the diagonal and the concurrence} before and after {$\mu^*$ } is the same. 
\end{definition}

\paragraph*{Conventions for vine connection along the diagonal}

Because there is some freedom in how one connects the vines along the diagonal, we recall and introduce some definitions and conventions.
{
By topology of the vineyard we mean that there is an isotopy of $S^1 \times \mathbb{R}^2\setminus (S^1 \times \Delta)$, with $\Delta$ the diagonal, which maps the two vineyards to each other, where we view them as subsets of  $S^1 \times \mathbb{R}^2\setminus (S^1 \times \Delta)$ and impose that the limit points on the diagonal stay limit points on the diagonal, but can move around. Since in a generic extended persistence diagram all points will have have finite times $p_i =( b_i , d_i)$, vineyards can be viewed as subsets of solid tori in $\mathbb{R}^3$;
for standard persistence we can get a similar statement if we ignore the points at infinity.
We say that $k$ vines are unlinked there is an ambient isotopy that maps the vines $\nu_i$ to 
\[\{ (t, p_i)  | t \in   S^1\} \subset S^1 \times \mathbb{R}^2,\]
or 
\[\{ (t, (b_i,b_i +4 (t (t-1)) (d_i-b_i))  )  | t \in   (0,1)]\} \subset S^1 \times \mathbb{R}^2 \setminus (S^1 \times \Delta),\]
with $p_i$ an off-diagonal point in a persistence diagram and $i \in \{ 1,\dots,k\}$ indexes the vines, where none of the coordinates of the $p_i$ coincide.  {We stress that with these definitions a vine that leaves the diagonal, then winds around the torus several times monotonously and then enters diagonal, as depicted together with another (unlinked) vine in Figure \ref{fig:ExampleUnlinked}, is unlinked.  }

\begin{figure}[h!]
    \centering
    \includegraphics[width=0.8\linewidth]{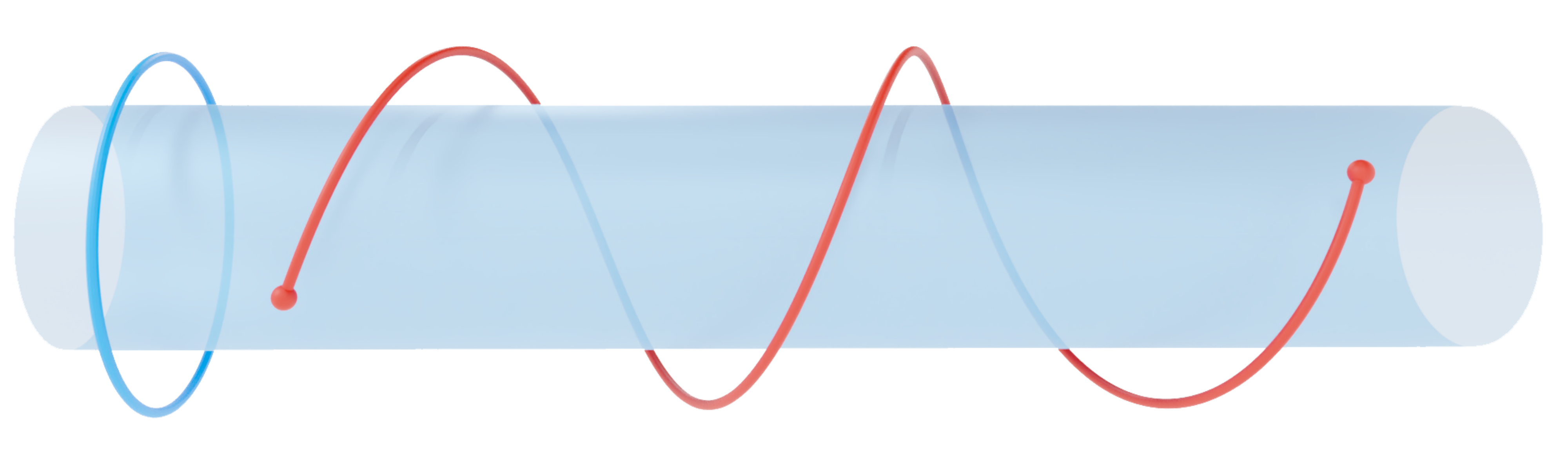}
    \caption{An unlink, with the diagonal depicted as a cylinder.  
    } 
    \label{fig:ExampleUnlinked} 
\end{figure}

\begin{convention} \label{convention:Annihilation} \label{convention:MergeSplit}\label{remark:split} 
Each vine that enters into the diagonal (forward) is connected forward in time to its other end point, where it leaves the diagonal. This may lead to intersections on the diagonal are ignored as improper. 
This has a number of consequences we spell out explicitly: 
\begin{itemize} 
\item The creation type topological change inserts an extra vine in the vineyard, or by reversing the order in $\mu$ removes a vine from the vineyard, but it does not change the topology of the other vines in the vineyard.
\item If two vines merge, then we connect the two remaining end points forward in time along the diagonal. 
    \item A merger event merges two vines into one.  If both of the two original vines that are merged are unlinked with the other vines in the vineyard before the merger, then the resulting vine is unlinked with the other vines in the vineyard after the merger. 
    \item If an unlinked vine is split, then the resulting vine(s) is (are) also unlinked. 
\end{itemize}
\end{convention}
}

\section{Main results}
\label{sec:MainResults}

Our first main 
result is that, from the perspective of the vineyard of a radial base loop, in fact we can drastically collapse the set of moves under consideration.  In essence, this is similar to Reidemeister moves, where generically only a small set of operations need to be considered.  We make this precise in our next proposition:

\begin{proposition}[Reduced from Lemma \ref{lem:MoveObservationLoopClassification} and Proposition \ref{prop:reduction} in Appendix \ref{sec:ElementaryMovesObs}]
 The list of elementary moves given in Figure \ref{fig:AllGenReidemeisterMoves} is exhaustive for a (locally) generic family $\gamma_\mu$. We have included two
 non-generic special elementary moves, one for the singularity type $A_1A_2$ and one for the singularity of type $A_3$, that are referred to as moves of special type.  
 
The elementary moves on the singularities listed in Figure \ref{fig:AllGenReidemeisterMoves} can be reduced to the elementary moves of the following types:
\begin{itemize}
    \item The tangential elementary moves on the singularities of type $A_1^2$ and of type $A_2$  (2nd, 4th panel respectively in Figure \ref{fig:AllGenReidemeisterMoves}).
    \item The transversal elementary moves on singularities of type $A_1^2/A_1^2$, $A_1^2/A_2$, and $A_2/A_2$, as well as $A_1^3$.  (5th, 9th, 7th, 12th panel respectively in Figure \ref{fig:AllGenReidemeisterMoves})
    \item The elementary move of special type on singularities of type $A_1A_2$. (16th panel in Figure~\ref{fig:AllGenReidemeisterMoves})
    \item The elementary move of special type on a singularity of type $A_3$.  (14th panel in Figure \ref{fig:AllGenReidemeisterMoves})
\end{itemize}
These irreducible elementary moves are indicated in yellow in Figure \ref{fig:AllGenReidemeisterMoves}.  
\label{prop:reductionReduced}
\end{proposition}

The proof of the first part of the statement relies on the specific form (up to ambient diffeomorphism) that the generalized symmetry set can take, see Figure \ref{fig:classification2DEarly} and the fact that locally the observation loop can be assumed to be a second order polynomial curve. 
The second part is based on explicit constructions for each case, see Figure \ref{fig:ReductionExample} for an example of such a reduction. 

\begin{figure}[h!]
    \centering
    \includegraphics[width=\linewidth]{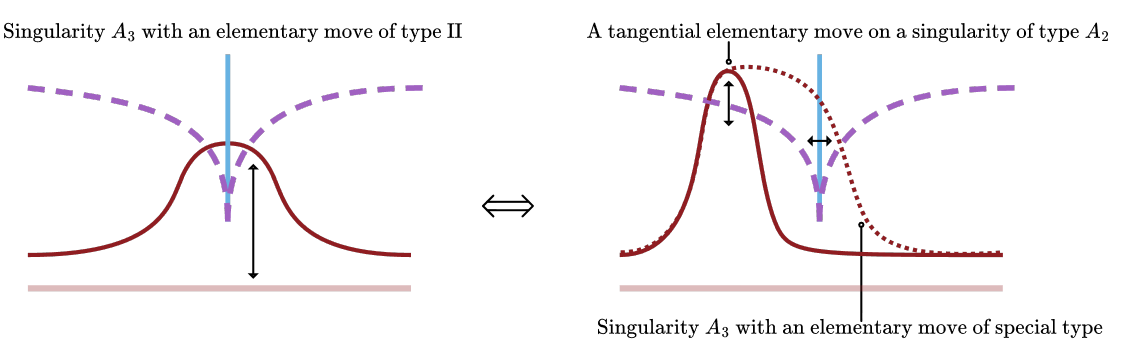}
    \caption{ The reduction of an elementary move on the singularity of type $A_3$ to two other elementary moves.  } 
    \label{fig:ReductionExample} 
\end{figure}

We can now classify the effect of {(irreducible)} moves over the singularities of the generalized symmetry set (see Figure~\ref{fig:classification2D}) from our reduced set of cases.  In considering such moves, we emphasize that when performing such a small ``edit'' in the radial base loop, the bulk of the vineyard remains unchanged, with only a local section seeing any change. More precisely: 

\begin{restatable}{theorem}{MovesOnVineyard}\label{prop:MovesOnVineyard}
The {irreducible} elementary moves of $\gamma_\mu$ with respect to the singularities induce the following changes in the vineyard:
\begin{itemize} 
\item The tangential move on the singularity of type $A_1^2$ induces a concurrence pair creation/annihilation in the vineyard. 
\item The tangential move on the singularity of type $A_2$ induces
one of the following: vine creation/annihilation or vine merge/split type changes. 
\item The transversal move on a singularity of type $A_1^2/A_1^2$ induces either a vine intersection type change, or a concurrence-concurrence rearrangement. 
\item The transversal move on a singularity of type $A_1^2/A_2$ induces a vine diagonal-concurrence rearrangement. 
\item The transversal move on a singularity of type $A_2/A_2$ induces a vine diagonal-diagonal rearrangement. 
\item The transversal move on a singularity of type $A_1^3$ induces a triple concurrence rearrangement. 
\item The elementary move of type $\textrm{I}a$ on a singularity of type $A_1A_2$ induces a trivial change on the vineyard. More specifically, for instance, it gives a vine diagonal-concurrence, but there is no reordering. 
\item The elementary move of special type on a singularity of type $A_3$ induces a single concurrence creation. 
\end{itemize}
\end{restatable}

\begin{figure}[h!]
    \centering
    \includegraphics[width=\linewidth]{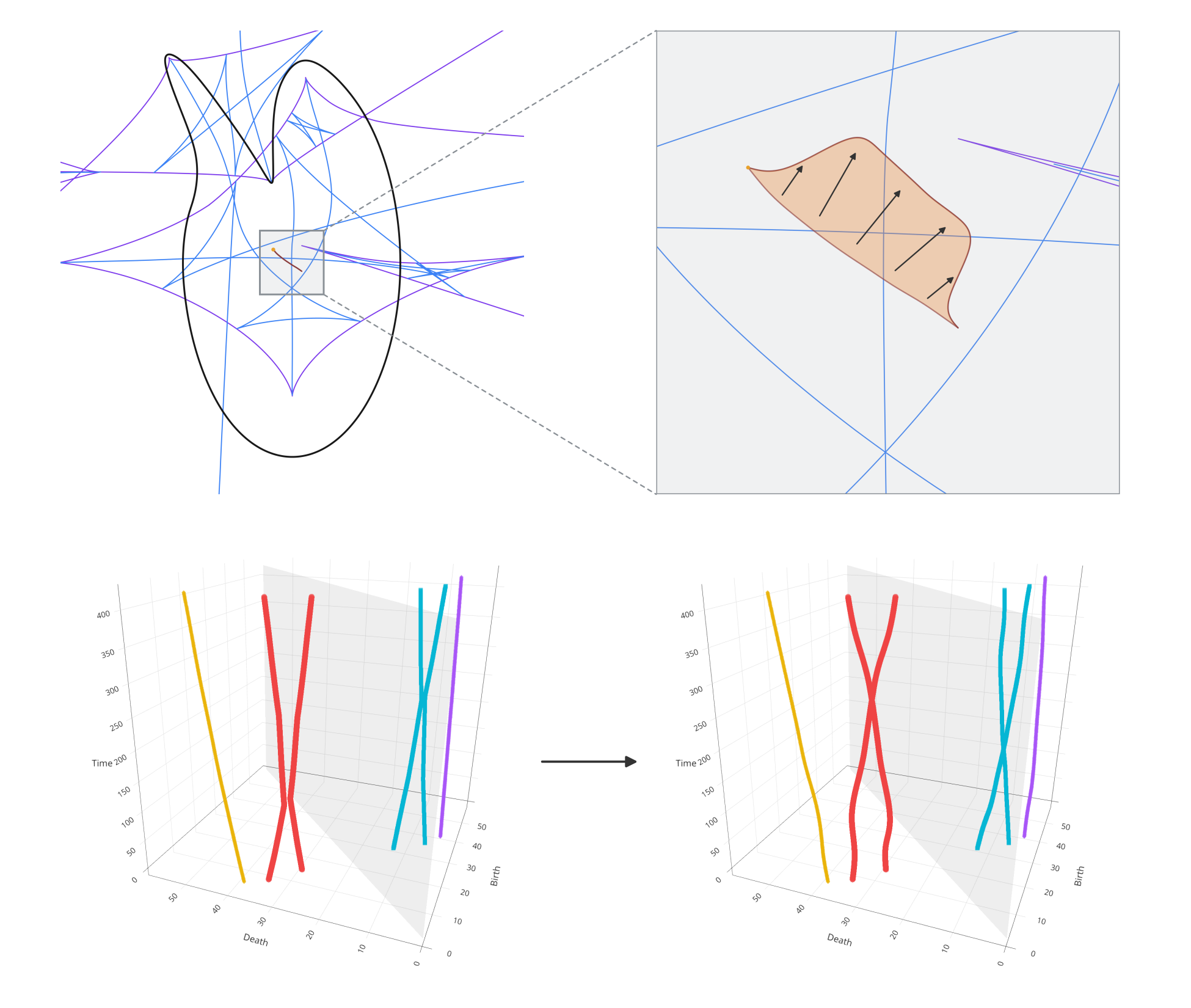}
    \caption{The transversal elementary move on a singularity of type $A_1^2/A_1^2$. In the top panels the elementary move of the observation loop is depicted, and in the bottom panels the effect of the move on the vineyard. In this case the move induces monodromy. } 
    \label{fig:A12A12transversalresultRECALL} 
\end{figure}

The proof mainly consists of very careful case analysis, which is visualised in Figures \ref{fig:A12A12transversalresultRECALL} and \ref{fig:A2A2transversalresultRECALL}. However, some general observations can be made to streamline the proof, the most important of which is: 
\begin{restatable}{observation}{VineIntersectionCriterion}
\label{ob:A12intersection} Vines only intersect if the death and birth coordinates of both vines concur. This can only happen at a singularity of type $A_1^2/A_1^2$, because this is the only singularity involving four contact points.  
Moreover, as long as the vines do not intersect (as one changes~$\mu$) and the points where vines enter or exit the diagonal change Lipschitz continuously, all of the vines move Lipschitz continuously with the parameter $\mu$, thanks to the stability of persistence diagrams \cite{CohenSteinerStability}. This in particular means that the (parts of) vines outside the diagonal are moving isotopically, although on the diagonal jumps are possible, see Figure~\ref{fig:bishop-staff}). 
\end{restatable}

The surprising result is that only a few options can change the actual topological type of the vineyard or introduce monodromy.

\begin{figure}[h!]
    \centering
    \includegraphics[width=\linewidth]{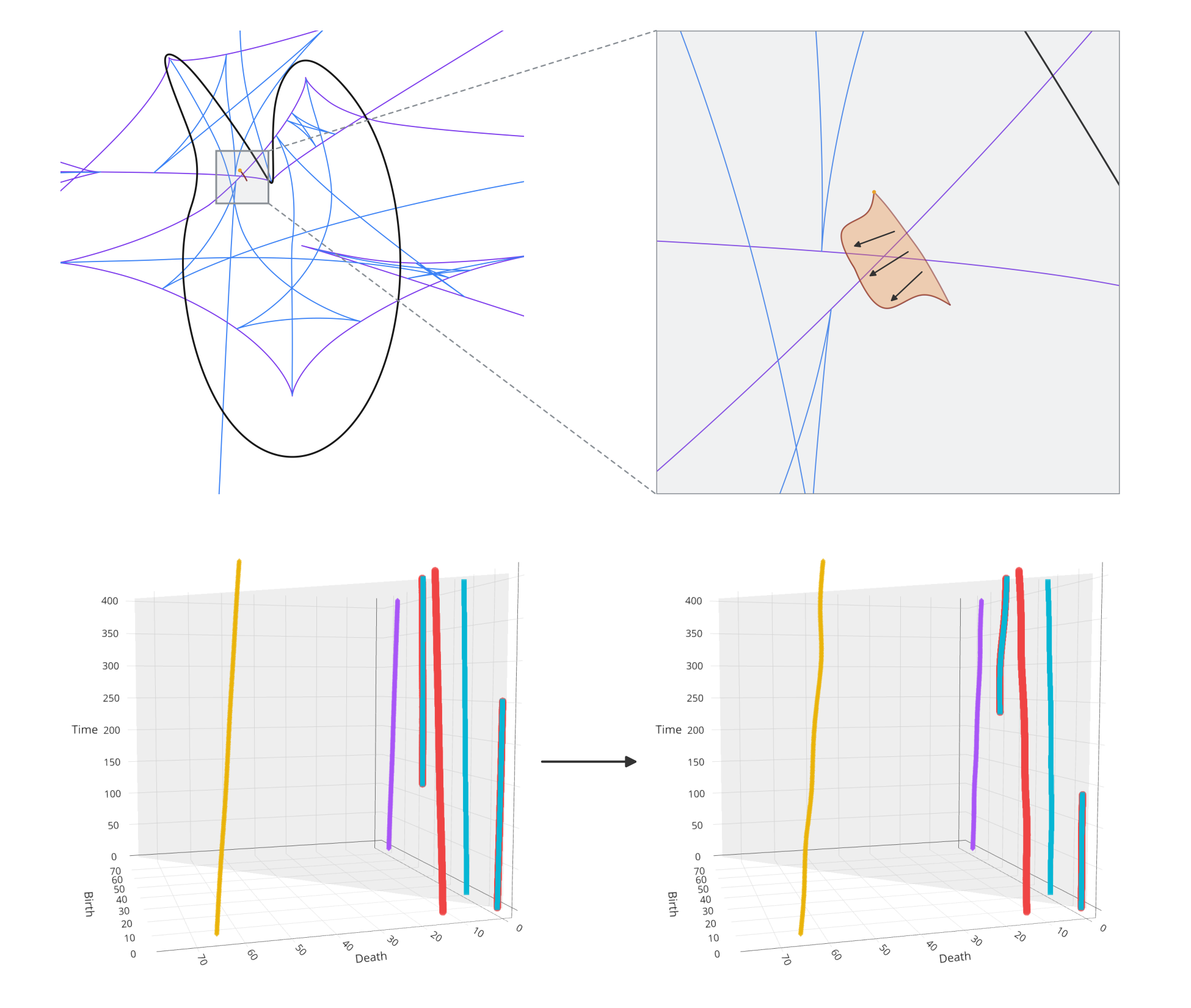}
    \caption{The transversal elementary move on a singularity of type $A_2/A_2$. In the top panels the elementary move of the observation loop is depicted, and in the bottom panels the effect of the move on the vineyard. In this case the change in the vineyard is the vine diagonal--diagonal rearrangement of Definition~\ref{def:ChangesVineyard}, which can induce monodromy.} 
    \label{fig:A2A2transversalresultRECALL} 
\end{figure}

The following proposition summarizes and describes the possible changes in topology and monodromy under a generalized generic isotopy (Definition~\ref{def:GeneralizedGenericIsotopy}), using our conventions for connecting vines along the diagonal.

\begin{restatable}{proposition}{TopologyMonodromyFromMoves}
\label{prop:TopologyMonodromFromMoves}
Let $\gamma_\mu$ be a generalized  generic isotopy relative to a generic curve $\mathcal{M}$.
The topology of the vineyard can only change with the parameter $\mu$ by two {irreducible} elementary moves:
\begin{itemize} 
\item The tangential move on the singularity of type $A_2$ which induces one of the following: vine creation/annihilation or vine merge/split type changes. Among these changes the vine creation/annihilation type changes introduce unlinked vines into, or remove them from, the vineyard. By Convention \ref{remark:split} the split does not introduce linking. 
\item The transversal move on a singularity of type $A_1^2/A_1^2$, but only if it induces a vine intersection type change.
\end{itemize}
Among these moves only the transversal move on a singularity of type $A_1^2/A_1^2$ and a tangential move on a singularity of type $A_2$ inducing a merge type change can induce monodromy (although a split can destroy it). 
Additionally, monodromy in the vineyard {(for vines ending on the diagonal)} can be created by an additional move: 
\begin{itemize}
\item The transversal move on a singularity of type $A_2/A_2$ induces a vine diagonal-diagonal rearrangement. 
\end{itemize}
\end{restatable}

Proposition~\ref{prop:TopologyMonodromFromMoves} identifies the local mechanisms relevant to Theorem~\ref{thm:main}:
\MainTHM*
To obtain that global criterion, we additionally use the geometry of the focal set to control the $A_2$ merge events. This control is based largely on the following lemma, which may be of independent interest:  
\begin{restatable}{lemma}{FocalLoops}
Suppose that a connected component of the focal set ($\mathcal{F}$) is a topological circle or loop (which necessarily has singularities) without self intersections (that is no singularities of type $A_2/A_2$) and $\M$ is compact and connected, then 
\begin{itemize}
    \item $\M$ is a convex curve (that is the boundary of a convex set). 
    \item $\mathcal{F}$ exhibits exactly $4$ singularities of type $A_3$. 
    \item The interior of $\mathcal{F}$ contains at least one singularity of type $A_1^2/A_1^2$.
\end{itemize}
\label{lem:focal-loops}
\end{restatable}


In \cite{local} it was proven that a single singularity of type $A_1^2/ A_1^2$ can suffice to generate monodromy. A similar statement is true for larger observation loops surrounding an $A_2/A_2$ singularity; in Figure \ref{fig:bishop-staff} we show an example where an observation loop encloses a single (0-dimensional) singularity, which is of type $A_2/A_2$, and the vineyard exhibits monodromy. 
We stress that the generalized symmetry set in the interior of this larger observation loop is not diffeomorphic to one of the singularities of Figure \ref{fig:classification2DEarly}, hence this does not contradict the main result of  \cite{local}. We refer to the appendix for full details.

\begin{figure}
    \centering
    \includegraphics[width=\linewidth]{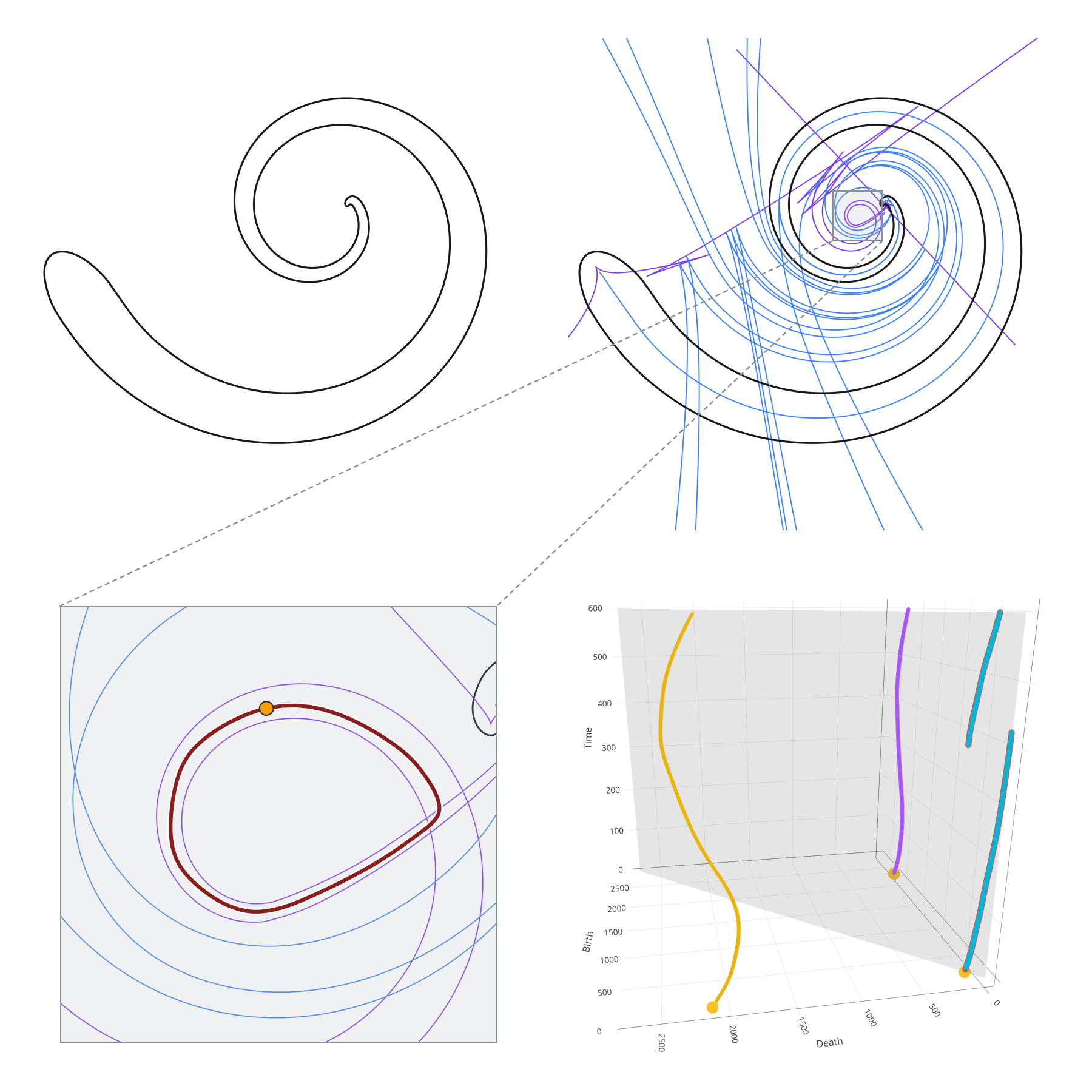}
    \caption{Monodromy arising from a single $A_2/A_2$ singularity. The curve $\mathcal{M}$ (top left) is shown alongside its focal and symmetry sets (top right). The enlarged view (bottom left) shows an observation loop enclosing only this singularity. In the corresponding vineyard (bottom right), a vine can be traced continuously for longer than one full period $2\pi$ of the observation loop, demonstrating monodromy. }
    \label{fig:bishop-staff}
\end{figure}

\section{Future work and open questions}
\label{sec:future_work}

In this paper we have described the roots of the topology and monodromy of vineyards in detail. At the same time,  the paper raises far more questions than it answers. We finish by highlighting several of the questions and directions of research that seem most promising.

In Section \ref{sec:VineyardTransitions} we provided a list of combinatorial changes in the vineyard. We conjecture that this list is complete for generic one parameter families (generalizing the parameter $\mu$ of the family of curves $\gamma_\mu$) of vineyards arising from smooth manifolds, although we defer this exploration to future work. 

Vineyards themselves have been generalized to higher dimensions \cite{Hickok2026}, which is distinct from multi-parameter persistence. 
Extending the analysis to {these generalized} 
vineyards remains an open question even for the radial persistence transform.  
The origin of monodromy in the multi-persistence setting has been studied in~\cite{Cerri2019}. 
More recently, the same question in single parameter persistence has been studied using completely different techniques in~\cite{GiuntiMunch}. 
Understanding the relation between these works and the singularity theoretical approach advocated in this paper is likely to provide deeper insight.

A further challenge is to develop a theory of braids that allows new strands to be inserted. As discussed in Remark \ref{rem:Generators}, standard braid-theoretic tools apply relatively directly when the focal set is not involved—that is, when the number of off-diagonal points in the persistence diagram remains constant. Our results show, however, that a full account of the topology of vines must also describe the insertion of new strands, corresponding in our setting to vines emerging from the diagonal.

This question is also connected to a problem with a computational-complexity flavour. We have established a correspondence between elementary moves of curves in the plane relative to the generalized symmetry set and elementary moves in the vineyard. Now suppose we are given a curve $\M$ and its generalized symmetry set, together with a closed vineyard $\mathcal{V}$. Is there an observation loop $\gamma$ in the plane, possibly with self-intersections, whose radial-transform vineyard is strongly equivalent to $\mathcal{V}$? Here, strong equivalence means that an ambient diffeomorphism of the form
$(t,x)\mapsto \bigl(\phi(t),\psi_t(x)\bigr)$
takes the induced vineyard to $\mathcal{V}$. In this definition, $\phi$ is a diffeomorphism of $S^1$, and each $\psi_t$ is a diffeomorphism that maps the part of the persistence diagram above the diagonal to itself. We conjecture that deciding whether such a loop exists is NP-hard when the focal set is present. In its absence, there is a {weak} correspondence between $A_1^2/A_1^2$ singularities and generators of the braid group.

In \cite{local}, the authors of that paper pointed out that the inverse problem for the symmetry set is a difficult one, that is given a stratified structure in the plane whose singularities are those of the classification of Bruce, Giblin, and Gibson \cite{Bruce1985symmetry}, then it is not clear there is an ambient diffeomorphism, such that after the ambient diffeomorphism, the stratified structure is the generalized symmetry set of a $1$-dimensional manifold $\M$, let alone how to construct $\M$. That is to say, finding a given knot or link in a vineyard of the radial transformation for a curve $\M$ remains an open problem, even in two dimensions.

\bibliography{bib}

\appendix

\paragraph*{AI Statement}

LLM tools were used in editing and drafting portions of the text, as well as for background literature searches.

\clearpage

\section{Full background on monodromy and vineyards}\label{App:VineyardsMonodromyBackground}

We first recall the persistence-theoretic notions needed throughout the paper; see, for example, \cite{dey2022computational, edelsbrunner2002topological} for general background. 
Let $f:\M\to\mathbb{R}$ be a continuous function. 
For $a\in\mathbb{R}$, its $a$-sublevel set is $\M_a=f^{-1}((-\infty,a])$.
Whenever $a\leq a'$, we have $\M_a\subseteq\M_{a'}$. 
The resulting nested family $\{\M_a\}_{a\in\mathbb{R}}$ is the \emph{sublevel-set filtration} induced by $f$.
Applying homology in a fixed degree $\ell$ associates to each sublevel set $\M_a$ its $\ell$-th homology group $H_\ell(\M_a)$.
For $a\leq a'$, the inclusion $\M_a\hookrightarrow \M_{a'}$ induces a linear map $H_\ell(\M_a)\longrightarrow H_\ell(\M_{a'})$, and the resulting family of homology groups and inclusion-induced maps records how $\ell$-dimensional topological features evolve across the filtration.
A homology class is said to be \emph{born} when it first appears in the filtration and to \emph{die} when it ceases to persist as an independent class.
When two classes merge, the \emph{elder rule} assigns the surviving class to the one born earlier, while the younger class is declared to die at the merge value \cite{CohenSteiner2006,edelsbrunner2002topological}.
This convention applies in every homological dimension and determines the pairing of birth and death events used to construct the persistence diagram. 
A persistent homology class with birth value $b$ and death value $d$ is represented by the point $(b,d)$, and the collection of these points is called the persistence diagram $\operatorname{Dgm}_\ell(f)$.

We write
\[
    \Delta=\{(b,d)\in\mathbb{R}^2\mid b=d\}
\]
for the diagonal of the persistence diagram.
A persistence pair with birth value $b$ and death value $d$ is represented by the point $(b,d)$, and its lifespan (or persistence) is defined as $|d-b|$.

Throughout this paper we use \emph{extended persistence} \cite{CohenSteiner2008}. 
Ordinary persistence follows the increasing sublevel-set filtration $\M_a=f^{-1}((-\infty,a])$, but some homology classes may survive for all larger values of $a$ and therefore have no finite death time. 
Extended persistence resolves this by continuing the filtration, after reaching $\M$, with a decreasing family of relative homology groups built from the superlevel-set filtration. 
Writing $\M^a=f^{-1}([a,\infty))$ for the superlevel sets of $f$, the extended filtration has the form
\[
    0 \longrightarrow H_\ell(\M_{a_1})
    \longrightarrow \cdots
    \longrightarrow H_\ell(\M)
    \longrightarrow H_\ell(\M,\M^{a_k})
    \longrightarrow \cdots
    \longrightarrow H_\ell(\M,\M)=0.
\]
Since this sequence begins and ends with the trivial group, every homology class that is born in the filtration is eventually assigned a finite death value.

The resulting persistence pairs are traditionally divided into three types: \emph{ordinary} pairs, whose birth and death both occur in the ascending sublevel-set filtration; \emph{relative} pairs, whose birth and death both occur in the descending relative filtration induced by the superlevel sets; and \emph{extended} pairs, whose birth occurs in the sublevel-set part and whose death occurs in the relative part.
In particular, extended persistence provides a complete finite pairing of the critical points of $f$.
We will usually suppress the homological degree from the notation when it is clear from context.

\subsection{Vineyards}

The notion of a vineyard arises when the filtering function varies continuously
with a one-dimensional parameter \cite{CohenSteiner2006,turner2023representing}.
Let $I\subseteq\mathbb{R}$ be an interval and let $\{f_t:\M\to\mathbb{R}\}_{t\in I}$ be a continuous one-parameter family of functions. 
For each $t\in I$, the function $f_t$ gives a persistence diagram $\operatorname{Dgm}_\ell(f_t)$ in homological degree $\ell$.
Stacking these persistence diagrams over the parameter $t$ gives the associated \emph{vineyard}.

We assume throughout that the family is generic, in the sense that each persistence diagram contains finitely many off-diagonal points and that multiplicities or other non-generic coincidences occur, if at all, only at isolated parameter values~\cite{turner2023representing}.
Away from such exceptional values, stability of persistence diagrams \cite{CohenSteinerStability} allows the persistence points to be followed continuously as the parameter varies. 
When the persistence points have multiplicity at most one, the trajectory traced by a continuously followed point is called a \emph{vine} \cite{CohenSteiner2006}.
At points of higher multiplicity, however, distinct vines need not remain individually distinguishable~\cite{GiuntiMunch}; cf.~Figure~\ref{fig:CorrespondenceA12SA12}.
Vines may enter or leave the off-diagonal part of the persistence diagram through $\Delta$.

The setting of this paper is obtained from the radial filtering function $f_p(x)=d_{\mathbb{E}}^2(x,p)\big|_{\M}$, where $\M$ is embedded in Euclidean space.
An observation path $\gamma:I\to\mathbb{R}^d$ therefore determines the one-parameter family $f_t=f_{\gamma(t)}$.
When $\gamma$ is a loop, we identify the parameter with $S^1$, so that the initial and final functions agree. 
The resulting periodic family is called a \emph{closed vineyard}; cf.~\cite{local,Chambers2026}. 
We regard such a vineyard as a subset of $S^1\times\mathbb{R}^2$, where the first coordinate records the position along the observation loop and the remaining two coordinates are the birth and death coordinates of the corresponding persistence point.

Although only one-parameter families are needed here, vineyards and related constructions have also been studied in more general parameterized settings.
These include formulations in terms of vineyard modules \cite{turner2023representing}, persistence diagram bundles over higher-dimensional parameter spaces~\cite{Hickok2026}, and more recent generalizations designed to accommodate branching and non-Hausdorff behaviour~\cite{GiuntiMunch}.

The stability theorem also controls the vineyard under perturbations of the observation loop. 
A closely related Lipschitz-continuity condition has been observed in the literature many times; see, for example, \cite{Chambers2026, CohenSteinerStability,turner2023representing}.

\subsection{Monodromy and topology of vines}
A closed vineyard need not consist of persistence points that return individually to their starting positions after one traversal of the observation loop. 
Indeed, although $\operatorname{Dgm}(f_{\gamma(0)})= \operatorname{Dgm}(f_{\gamma(2\pi)})$, continuously following a particular persistence point from $t=0$ to $t=2\pi$ may lead to a different point of the same persistence diagram.
This phenomenon is called \emph{monodromy}; see \cite{local, Chambers2026}.
Equivalently, following the persistence points through one period can induce a nontrivial permutation of
the points in a fixed time slice.

If a vine has limit points on the diagonal, we extend it along $S^1\times\Delta$ by connecting its terminal point to its initial point forward in periodic time.
With this convention, we say that a vine has \emph{no monodromy} if its extension forms a single cover of the parameter circle $S^1$. 
If instead the extended vine winds more than once around the parameter circle before returning to its initial persistence point, then it exhibits nontrivial monodromy.

Finally, the vines of a closed vineyard are curves in $S^1\times\mathbb{R}^2$, and therefore carry topology beyond monodromy: distinct vines may be linked, and individual vines may be knotted.
Indeed, \cite{Chambers2026} shows that radial vineyards can realize arbitrary knot and link types.
The purpose of the present paper is to understand which interactions of an observation loop with the generalized symmetry set can create or modify these global topological phenomena.

\section{Cerf theory} 
\label{sec:Cerf} 

Finally, we discuss in a bit more detail the relation of our work on vineyards with Cerf theory \cite{Cerf_1970, Hatcher_cerf_theory_1973}. Cerf proved that, given a one-dimensional family $g_t(\tau_1,\tau_2, \dots, \tau_d)$ of functions (hence $t$ is the parameter) connecting two Morse functions, the family of the functions can be perturbed in such a way that for all but finitely many parameter values $t^*$, $g_t$ is a Morse function.  Moreover, at those $t^*$ where the function is not Morse (with respect to all coordinates $\tau_i$), there is a third order degeneracy in $\tau_1$, and the Hessian is non-degenerate with respect to all other coordinates $\tau_i$.   
In our setting $\tau \mapsto d_\mathbb{E}^2 (\M(\tau) , \gamma(t))$, where we slightly abuse notation and write $\M(\tau)$ for a parametrization of $\M$ by $\tau$, and $t$ is the parameter of the $1$-dimensional family. Assume for the moment that $\gamma(t)=p$ is fixed. Geometrically speaking, we distinguish three cases: 
\begin{itemize} 
\item If the function $\tau \mapsto d_\mathbb{E}^2 (\M(\tau) ,p)$ does not have a critical point at $\tau'$ then the circle centred at $p$ is not tangent to $\M$ at $\M(\tau')$.
\item If the function $\tau \mapsto d_\mathbb{E}^2 (\M(\tau) ,p)$ has a Morse critical point at $\tau'$ then the circle centred at $p$ is tangent to $\M$ at $\M(\tau')$. 
\item If $\tau \mapsto d_\mathbb{E}^2 (\M(\tau) ,p)$ is not Morse at $\tau$, but is locally of third order, then the circle centred at $p$ osculates at $\M$. In other words $p$ is a centre of curvature. This corresponds therefore with the focal set. 
\end{itemize} 

Because the normal bundle (outside a region of size equal to the reach \cite{Federer1959}) is not embedded or equivalently a circle can be tangent to a curve in more than one place (which corresponds precisely to the symmetry set), one has to consider multiple parts of the manifold $\M$, whose (local) parameters will be indicated by $\tau_i$ (with distinct $i$ even if the parts belong to the same connected component). Formally, this can be treated using multi-jets, see \cite{Bruce1985symmetry} and the references mentioned there. And as was shown in \cite{Bruce1985symmetry} for one-parameter families one can get indeed two different Morse critical points for the same $p$. In other words, it follows from the work of Cerf that a generic curve only needs to intersect the focal set transversally, while to observe that a generic curve only needs to intersect the symmetry set transversally, some extra work is required. 

The general theory of singularities developed by Whitney, Thom, Arnold and others, see e.g.  \cite{Arnold1972, arnol2003catastrophe, arnol1993singularity, Looijenga1974}, in some sense generalizes the work by Cerf and shows how generic multiparameter families behave. Thus, one need not restrict attention to a curve, but may instead consider a coordinate neighbourhood of $p$. 
We stress that in our particular setting Cerf's result can be reconstructed from the classification of singularities and standard transversality arguments \cite{Hirsch1976}, but in many ways the result of Cerf is more general, because it does not require low dimensionality, which is essential for singularity theory; cf. \cite{Buchner1977, buchner1978structure, Wall1977}.

We refer to \cite{bruce1992curves} for a pedagogical introduction to singularity theory with an emphasis on $1$-parameter families. 

\section{Detailed description of the singularities of the focal and symmetry set} 
\label{sec:DetailsSingularities} 

\subparagraph{Type $A_1^2$}
There is a single circle centred at $p$ having ordinary tangency with $\M$ at two distinct points. 
Consequently, $p$ lies on a smooth stratum of the symmetry set.
Each of the two corresponding critical points of $d_{\mathbb{E}}(\cdot,p)^2|_{\M}$ may be either a local minimum or a local maximum. 
We distinguish the resulting symmetry set strata according to their persistence interpretation:
\begin{itemize}
    \item if both critical points are local minima, then $p$ lies on a \emph{birth-birth} stratum;
    \item if both critical points are local maxima, then $p$ lies on a \emph{death-death} stratum;
    \item if one critical point is a local minimum and the other is a local maximum, then $p$ lies on a \emph{birth-death} stratum.
\end{itemize}
Here, a local minimum corresponds to the birth of a $0$-dimensional class in the sublevel set filtration, while a local maximum corresponds to a death.

\subparagraph{Type \texorpdfstring{$A_1^2/A_1^2$}{A12A12}}
Two concentric circles of distinct radii are each tangent to $\M$ at two points, giving four ordinary contact points in total. 
Locally, $p$ is a transverse intersection of two smooth strata of the symmetry set.
As above, each individual contact may correspond to either a local minimum or a local maximum of the restricted squared distance function.

\subparagraph{Type \texorpdfstring{$A_1^3$}{A13}}
A single circle centred at $p$ is tangent to $\M$ at three distinct points.
Locally, three smooth branches of the symmetry set meet at $p$, one for each pair among the three tangencies.

\subparagraph{Type \texorpdfstring{$A_2$}{A2}}
The point $p$ is a centre of curvature of $\M$, the circle centred at $p$ is an osculating circle of the curve at one point.
Thus, $p$ belongs to a smooth stratum of the focal set. 

\subparagraph{Type \texorpdfstring{$A_1A_2$}{A1A2}}
A single circle centred at $p$ has an ordinary tangency with $\M$ at one point and second-order contact at another.
Locally, a cusp of the symmetry set meets a smooth branch of the focal set at $p$.

\subparagraph{Type \texorpdfstring{$A_3$}{A3}}
Here $p$ is a centre of curvature corresponding to a local extremum of curvature. 
The associated osculating circle has third-order contact with $\M$.
Geometrically, $p$ is a cusp point of the focal set and lies in the closure of the symmetry set.

\subparagraph{Type \texorpdfstring{$A_1^2/A_2$}{A12A2}}
There are two concentric circles of different radii. 
One is tangent to $\M$ at two distinct points, while the other is an osculating circle at a third point.
Accordingly, $p$ is a transverse intersection of a smooth stratum of the symmetry set with a smooth stratum of the focal set.

\subparagraph{Type \texorpdfstring{$A_2/A_2$}{A2A2}}
There are two concentric osculating circles of different radii, with their second-order contacts occurring at distinct points of $\M$.
Thus $p$ lies at a transverse intersection of two smooth branches of the focal set.

\section{The topology of the focal set and monodromy}

\begin{figure}
    \centering
    \includegraphics[width=\linewidth]{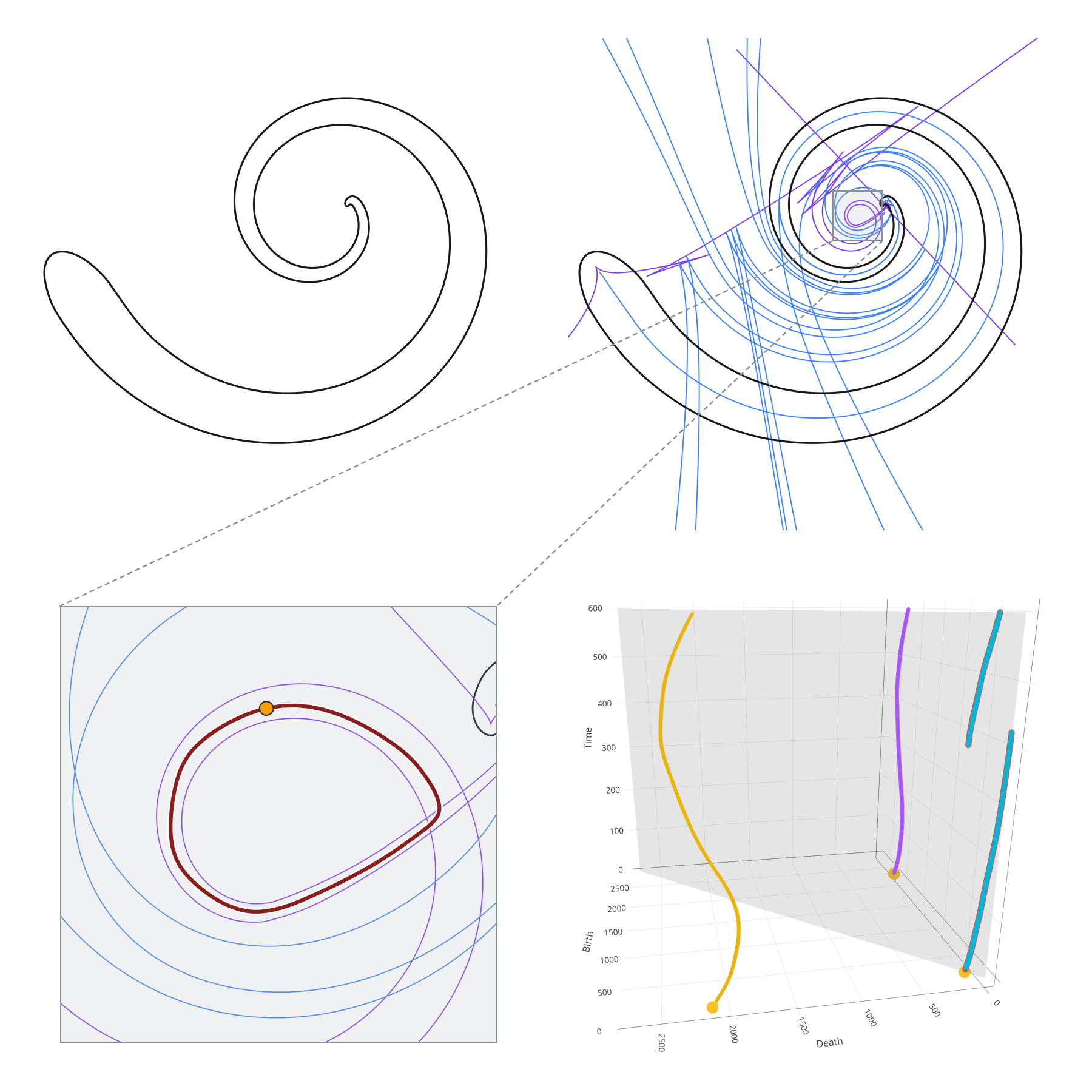}
    \caption{The shape $\M$ (top left) and the focal and symmetry to which it gives rise (top right). Bottom left: An observation loop that encloses only a single singularity of type $A_2/A_2$. Bottom right: The vineyard exhibits monodromy, because the blue vine can be traced forward in periodic time for a duration more than $2\pi$. }
    \label{fig:bishop-staff2}
\end{figure}

\subsection{Singularities of type \texorpdfstring{$A_2/A_2$}{A2SlashA2} as sources of monodromy} \label{sec:A2A2Gen}
\begin{definition}
We write $\mathcal{F}$ for a connected component of the focal set, where the connected component is determined while ignoring the $A_2/A_2$ crossings, that is, the two transversally crossing branches are not connected to each other at the crossing, although they may be connected elsewhere.
\end{definition}

We note that with this definition a connected component of the focal set {(which may contain cusp singularities)} consists of the centres of curvature of a single connected component of $\M$. Moreover the part of $\M$ to which the connected component corresponds has to have (either) positive (or negative) curvature (depending on the orientation of the curve, if we consider signed curvature). 

\begin{lemma} \label{lem:A2SA2Monodromy}
Suppose that a connected component of the focal set $\mathcal{F}$ exhibits a singularity of type $A_2/A_2$ (in other words, a self intersection) {and that 
a part of $\mathcal{F}$  forms a loop from and to this singularity. } Let $\gamma$ be an observation loop that follows this loop in $\mathcal{F}$  to and from the $A_2/A_2$ singularity, and remains on the side of the focal set where the connected component is born, but close.\footnote{Formally we need that the distance between $\gamma$ and that part of the focal set is very small compared to the difference in radii of curvature at the $A_2/A_2$ singularity.  } Then the vineyard of $\gamma$ exhibits monodromy.  
\end{lemma}

\begin{proof}
Consider the two points the persistence diagrams with low persistence (with sufficiently different birth times) for a point on $\gamma$ near the singularity of type $A_2/A_2$, as we follow these points in the vineyard while moving along $\gamma$. See Figure~\ref{fig:bishop-staff2}. The persistence of one of these points remains low because the distance to the focal set is assumed to be very small. By following $\gamma$ along the vineyard  the point, whose persistence remained low as we moved away from the $A_2/A_2$ singularity, ends in the other point of low persistence. This is because, by construction, the birth and death times of the low persistence point along $\gamma$ are almost equal to the radius of curvature and the radius of curvature changes from the one radius of curvature at $A_2/A_2$ to the other as we follow the loop in $\mathcal{F}$. {This means that there is a permutation of points in the persistence diagram induced by following the vine forward in time and the period of the vine (once the end points are connected via the diagonal forward in time) is greater than $2 \pi$. }   
\end{proof}

\begin{remark}
  Note that a single such singularity suffices to generate monodromy, as we see in Figure~\ref{fig:bishop-staff2}.  We stress that there is no linking in the resulting vineyard, as the length of a single vine exceeds the period of the vineyard.  
\end{remark}

\subsection{Closed loops of the focal set without \texorpdfstring{$A_2/A_2$}{A2SlashA2} singularities}\label{sec:ClosedLoops}

\FocalLoops*

\begin{proof}
The first point follows immediately: There is a bijection between the centres of curvature $\mathcal{F}$ and the curve $\M$ itself (which is therefore also a topological circle {with singularities}). The curvature is never zero, as the focal set associated to a point of zero curvature goes to infinity as you approach the point.  Therefore, the curve is convex. 

We now observe that, thanks to Morse theory (on the curvature, viewed as a function, on $\M$), we have that every local minimum of curvature needs to be followed by a local maximum of curvature along $\M$. We define for each cusp in the focal set (at a singularity of type $A_3$) the limit of the tangent vector of an arclength parametrization of the focal set as one tends to the $A_3$ singularity as $T_L(c)$, where $c$ indicates the centre of curvature where the $A_3$ singularity occurs. We write $x(c)$ for the corresponding point on $\M$. We let $x(c_1)$ and $x(c_2)$ be a consecutive pair of a local maximum and minimum of curvature along $\M$, corresponding to a consecutive pair of points $c_1, c_2$ on $\mathcal{F}$. We write $\mu$ for an arc length parametrization of $\M$ and if $x \in M$, with a little abuse of notation, we write $\dot{\mu} (x)$ for the tangent vector of $\mu$ at $x$. 
See Figure~\ref{fig:focal-loop-setup}.

Because $\dot{\mu} (x(c_i))$ is orthogonal to $T_L(c_i)$, and for local maxima of curvature the corresponding part of $\M$ lies in the direction in which the cusp points, while for a local minimum of curvature it lies in the opposite direction, we have that 
\begin{align}
\angle T_L(c_1), T_L(c_2) = \pi - \angle \dot{\mu} (x(c_1)), \dot{\mu} (x(c_2)),
\label{eq:angleConsec}
\end{align}

where in this case one can consider the angle as signed or an element of $S^1$. 

\begin{figure}
    \centering
    \includegraphics[width=0.7\linewidth]{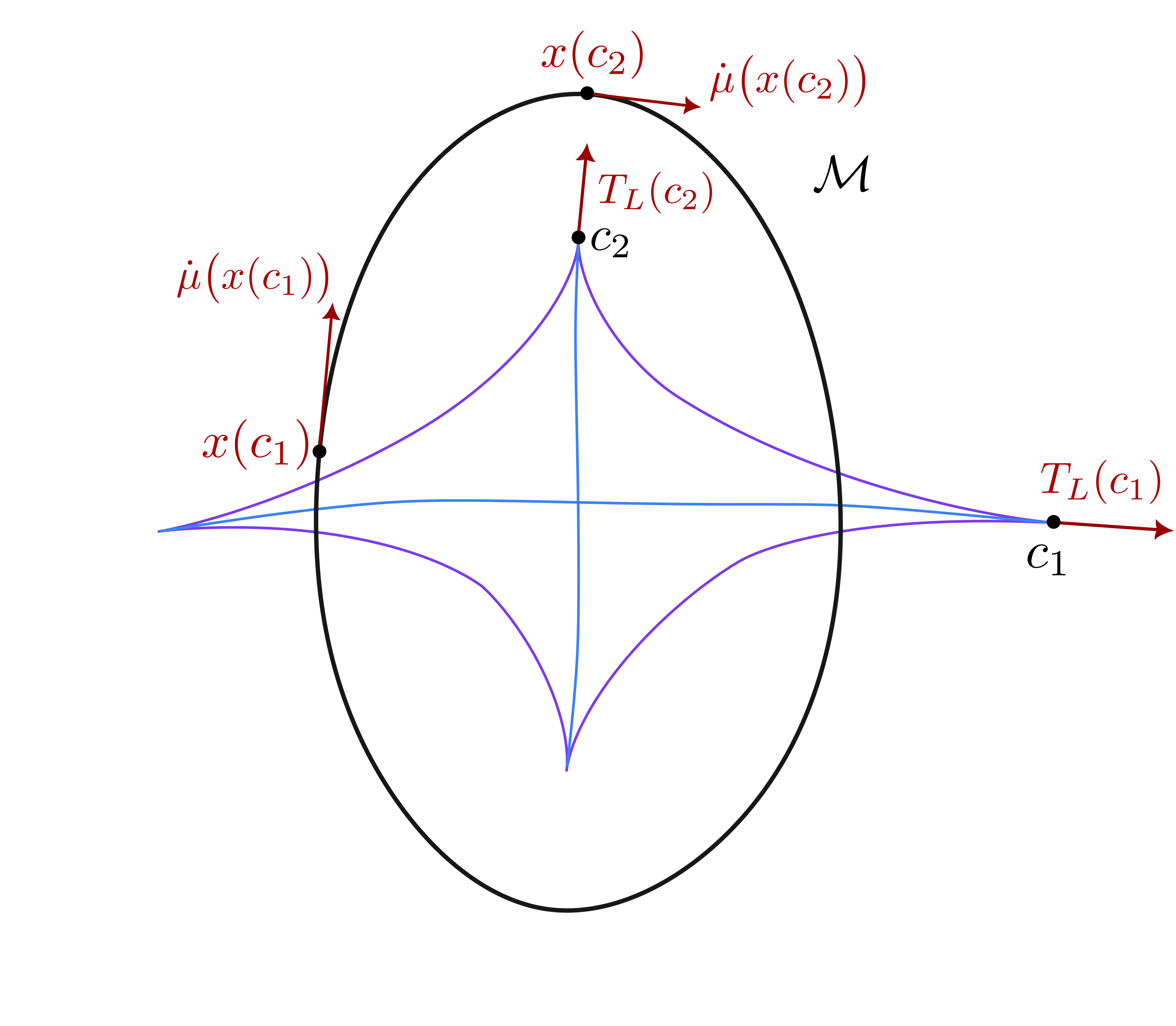}
    \includegraphics[width=0.6\linewidth]{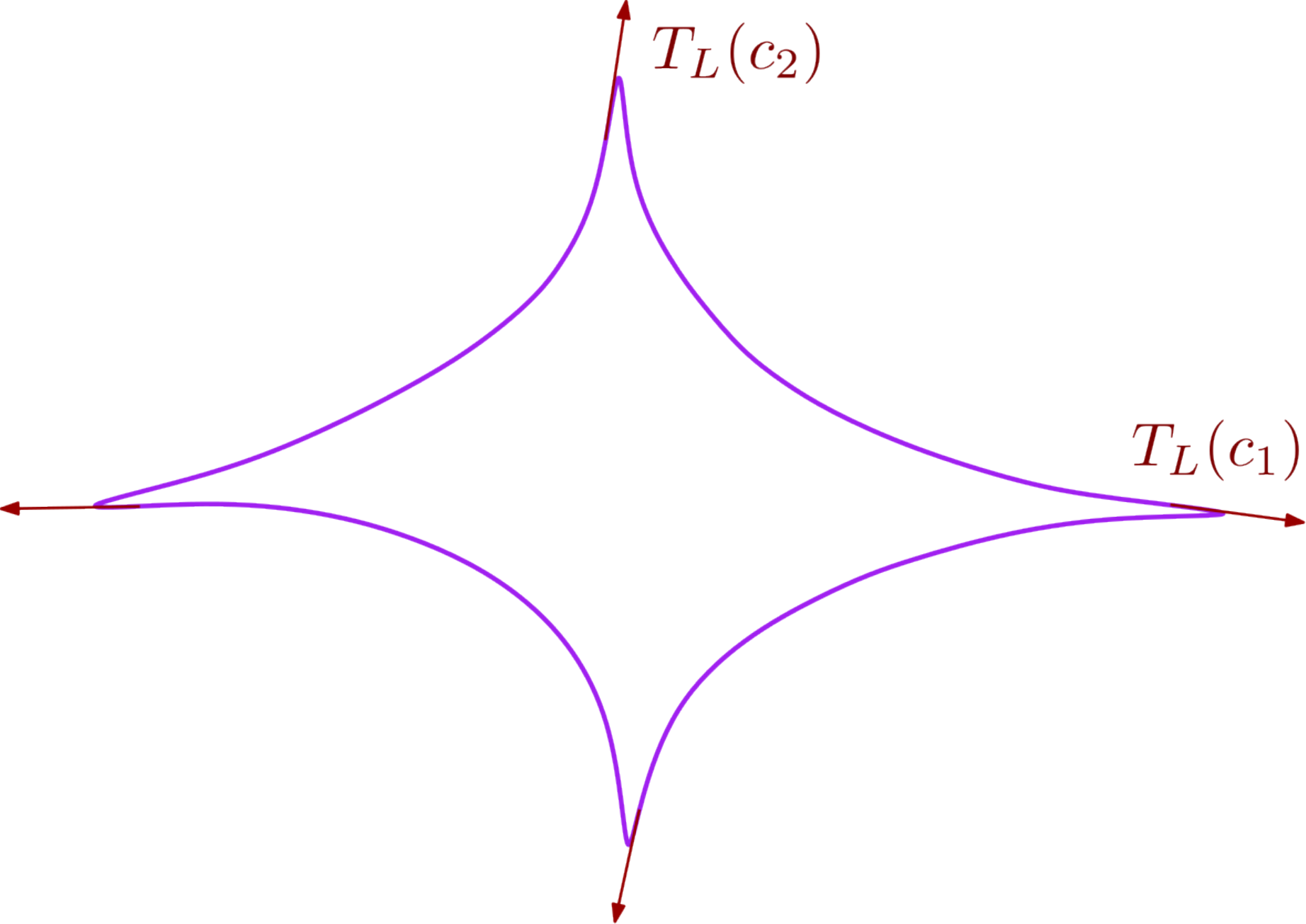} 
    \caption{ Top: The notation in Lemma~\ref{lem:focal-loops}. { Bottom: The smoothing of the focal set, where the corners at each cusp are now rounded off. }
    }
    \label{fig:focal-loop-setup}
\end{figure}

The sum of the angles for both $T_L(c_i)$ and $\dot{\mu} (x(c_i))$ have to sum up to $2\pi$, {because we have to go round. More precisely, there is a smoothing of the focal set, which removes the singularities of type $A_3$ where the tangent vector of the smoothed curve near the singularity coincides with $T_L(c_i)$, see Figure \ref{fig:focal-loop-setup}.    
This means that 
\eqref{eq:angleConsec} yields}
\begin{align}
2 \pi =\sum_{i=1}^{N} \angle T_L(c_i), T_L(c_{i+1}) = \sum_{i=1}^{N} \left(\pi - \angle \dot{\mu} (x(c_i)), \dot{\mu} (x(c_{i+1})) \right) = N \pi - 2 \pi ,
\nonumber
\end{align}
where $N$ is the number of singularities of type $A_3$. Hence we have that $N=4$. Moreover exactly two of these singularities are local minima and two are local maxima of curvature. 

Because $\M$ is convex and generic, the medial axis is bounded, but not a single point. Moreover, a result of Lieutier~\cite{lieutier2003any} means that the medial axis has the homotopy type of a point. All end points of the medial axis are local maxima of curvature, of which we have only two, and the medial axis is a(n open) curve, because by classification of singularities there are no end points except at $A_3$ singularities; self intersections are excluded as well by Lieutier's homotopy argument~\cite{lieutier2003any}.  Because $\M$ is convex we have that for points $x$ sufficiently far from $\M$ the persistence diagram for the function $d_\mathbb{E}^2 (\cdot , x)|_\M$ has only one point (using extended persistence or point at infinity for ordinary persistence), the medial axis and in fact the entire symmetry set has to lie in the interior of the focal set. {Additionally, this number can only increase upon crossing the focal set. } This means that the medial axis separates the interior of the focal set into two pieces, each of which has a $A_3$ singularity corresponding to a minimum of curvature in its closure. Using the classification of singularities again, we conclude that there are no other end points of the focal set except at the $A_3$ singularities, and hence we conclude that these singularities have to be connected by the symmetry set. Thanks to the Jordan curve theorem~\cite{jordan1893cours, Hales2007} this means that the medial axis has to be crossed, which thanks to the classification of singularities leads to singularity of type $A_1^2/ A_1^2$.   
\end{proof} 

{
\begin{remark}
We recall that the symmetry set of the union of multiple connected components $\M = \M_1 \cup \dots \cup \M_k$ is a superset of the symmetry sets of the individual connected components $\M_i$, while the focal set of the union of multiple connected components $\M$ is union of the focal sets. The third conclusion of Lemma \ref{lem:focal-loops} hence applies component-wise. 
\end{remark}
}

\begin{corollary}
\label{cor:TwoPointsNoTopology}
    Under the assumption of Lemma \ref{lem:focal-loops}, we have that for any $p$ the persistence diagram of $d_\mathbb{E}^2(\cdot ,p)|_\M$ contains at most two points. Moreover, the vineyard of $d_\mathbb{E}^2(\cdot ,\gamma(t))|_\M$, for any generic observation loop $\gamma(t)$, does not exhibit monodromy. 
\end{corollary}

\begin{proof} 
The first statement is immediate from the proof of Lemma~\ref{lem:focal-loops}. Outside the focal loop $\mathcal{F}$ there is only one point in the persistence diagram $d_\mathbb{E}^2(\cdot ,p)|_\M$. Upon crossing $\mathcal{F}$ that number increases to two, but there are, by assumption, no other parts of the focal set where this number can increase. 

Thanks to the elder rule, one of the points in the persistence diagram is always born first and dies last in extended persistence, in ordinary persistence it is at infinity, see Figure~\ref{fig:Cor22}. By genericity of $\gamma$ there are no self intersections of the vines. This means that there is always a point that is more persistent than the other, which separates the two and prevents monodromy.  
\end{proof} 

\begin{figure}[h!]
    \centering
    \includegraphics[width=0.3\linewidth]{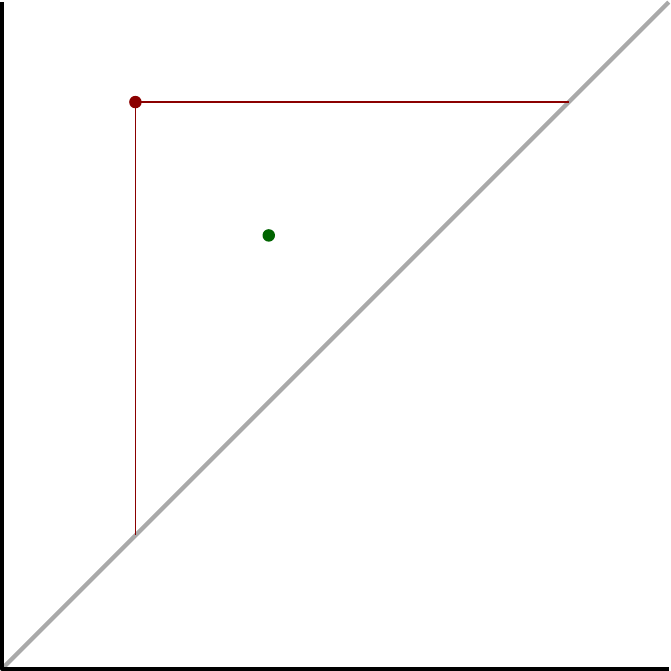}
    \caption{As noted in the proof of Corollary \ref{cor:TwoPointsNoTopology}, in extended persistence the low persistence point (green) is confined in the region induced by the high persistence point (red).  
    \label{fig:Cor22}  } 
\end{figure}

\section{Singularities in the symmetry set as generators of the braiding}

\subsection{Elementary moves for observation loops with respect to the generalized symmetry set} 
\label{sec:ElementaryMovesObs}

Let $\gamma_\mu (t)$ be a family of (smooth) loops where the loop is parametrized by $t \in [0,2\pi]/(0 \sim 2\pi)$, that is we identify $0$ and $2\pi$ yielding a circle and the family in parametrized by $\mu \in [0,1]$ and $\mathcal{M}$ a generic manifold (curve). Often $\gamma_0 (t)$ will be a point. We will also refer to this family as an isotopy. Moreover we assume that $\gamma_\mu$ is generic in the following sense: 
\begin{definition}
\label{def:genericGammaMu}
An isotopy $\gamma_\mu$ is called generic relative to a generic curve $\mathcal{M}$ if for any $\mu$ there is {at most} one $t$ such that {exactly one of the following holds}:
\begin{enumerate}
\item  $\gamma_\mu (t)$ lies on a singularity in the generalized symmetry set of codimension $2$, and the intersection is {locally generic, in the sense that it cannot be perturbed away by a small local change.  } 
\item The curve is tangent to a manifold piece of the generalized symmetry set at $\gamma_\mu (t)$.
\end{enumerate}
\end{definition}

This notion of generic ensures that each curve of the isotopy crosses the generalized symmetry set transversely except at isolated times, where it is briefly tangent to it before ``sliding over" a co-dimension 2 singularity or a curve of the generalized symmetry set.
To describe the behaviour of the vineyard under the isotopy, it is helpful to introduce a generalization of Reidemeister moves of a curve $\gamma$ with respect to the singularities of the generalized symmetry set. We will call these elementary moves.

\begin{figure}[h!]
    \centering
    \includegraphics[width=\linewidth]{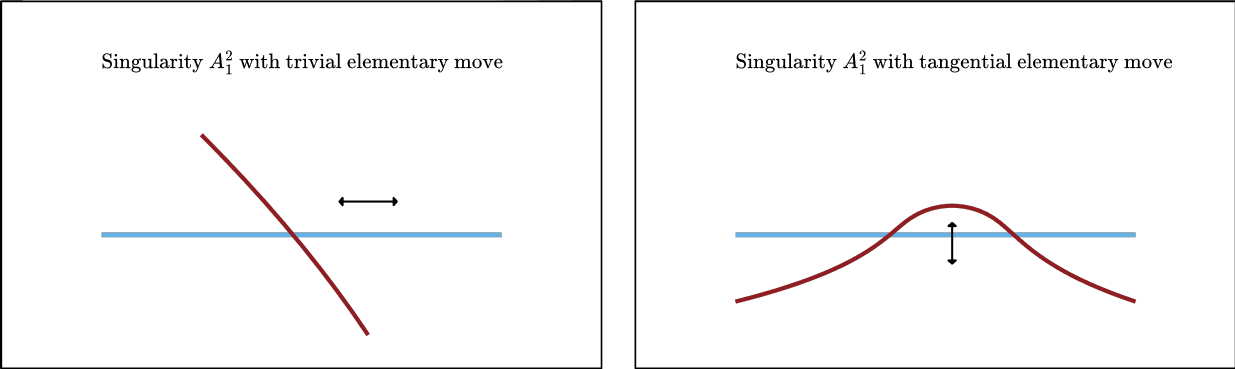}
    \caption{The elementary moves for a singularity of type $A_1^2$.     \label{fig:ReidemeisterA12}  } 
\end{figure}

\begin{figure}[h!]
    \centering
    \includegraphics[width=\linewidth]{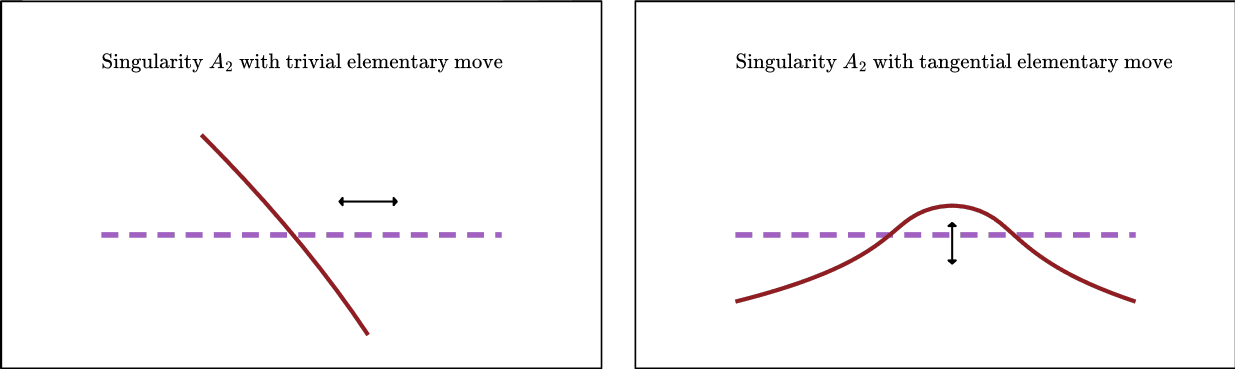}
    \caption{The elementary moves for a singularity of type $A_2$. Please note that we reduced the list using reflection symmetry, even if there is one side of the focal set, which corresponds to a birth and another which corresponds to a death.  } 
    \label{fig:ReidemeisterA2} 
\end{figure}

\begin{figure}[h!]
    \centering
    \includegraphics[width=\linewidth]{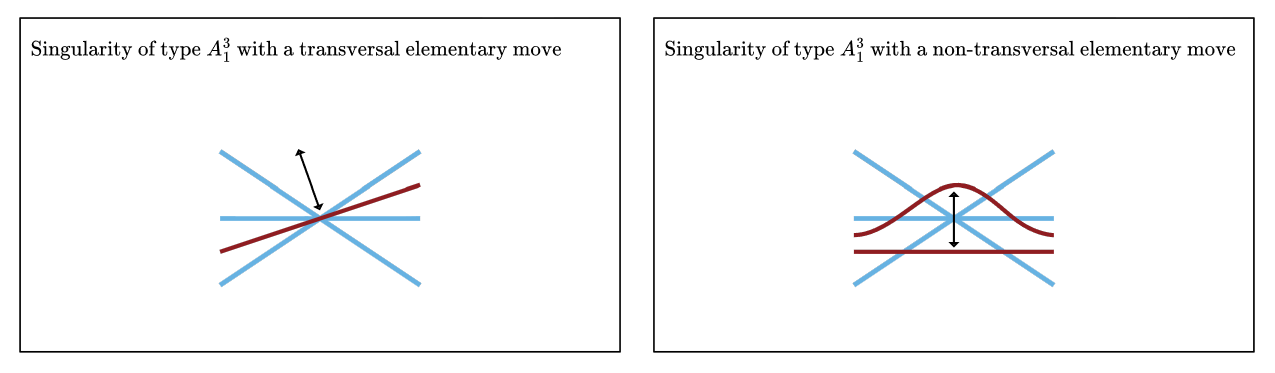}
    \caption{The elementary moves for a singularity of type $A_1^3$.  Please note that we reduced the list using symmetry. \label{fig:ReidemeisterA13}  } 
\end{figure}

\begin{figure}[h!]
    \centering
    \includegraphics[width=\linewidth]{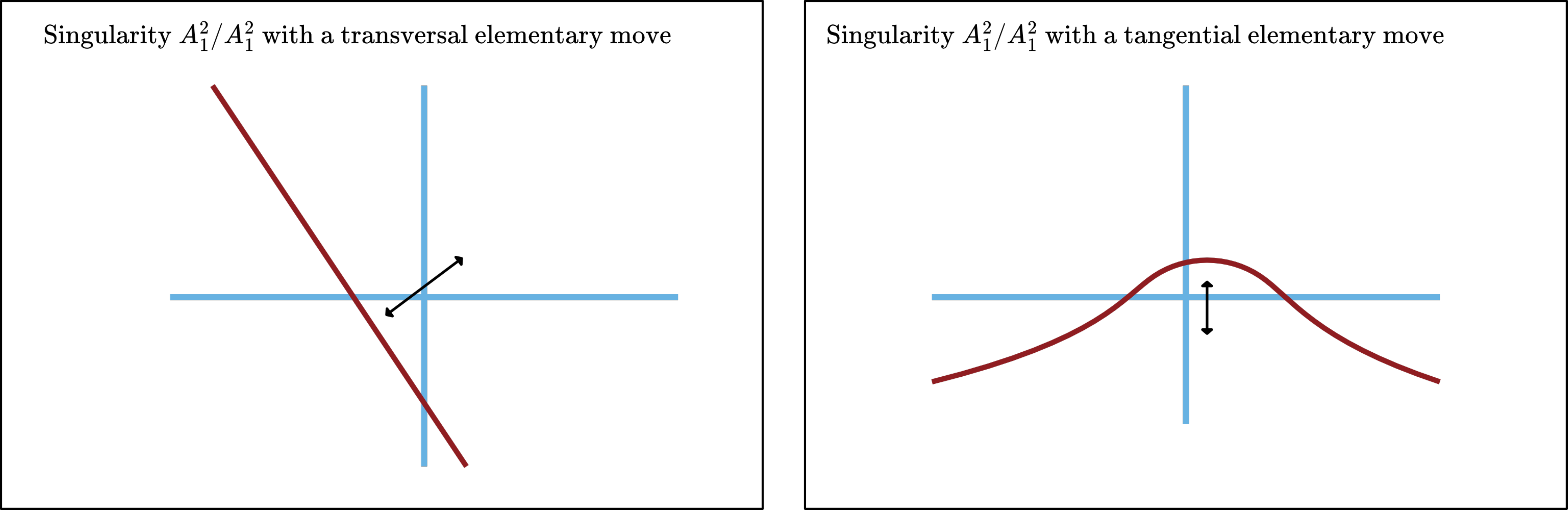}
    \caption{The elementary moves for a singularity of type $A_1^2/A_1^2$.  Please note that we reduced the list using symmetry. \label{fig:ReidemeisterA12A12}  } 
\end{figure}

\begin{figure}[h!]
    \centering
    \includegraphics[width=\linewidth]{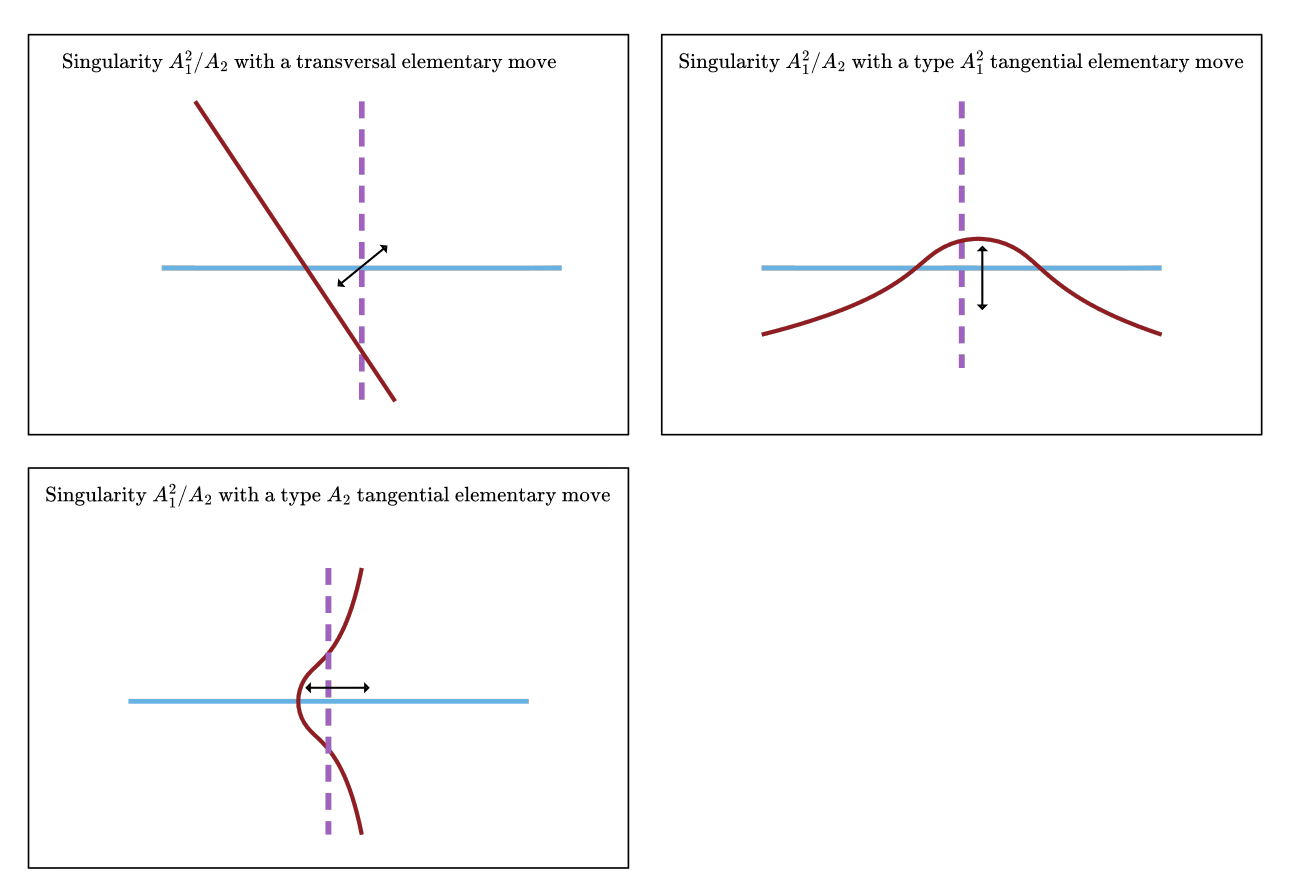}
    \caption{The elementary moves for a singularity of type $A_1^2/A_2$.  Please note that we reduced the list using reflection symmetry, even if there is one side of the focal set, which corresponds to a birth and another which corresponds to a death. \label{fig:ReidemeisterA12A2}  } 
\end{figure}

\begin{figure}[h!]
    \centering
    \includegraphics[width=\linewidth]{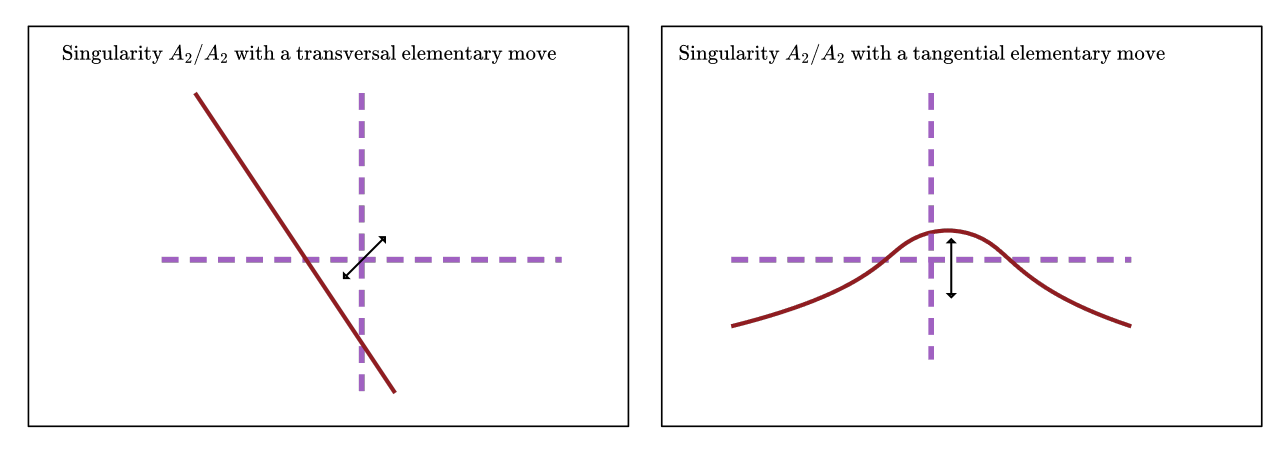}
    \caption{The elementary moves for a singularity of type $A_2/A_2$.  Please note that we reduced the list using reflection symmetry, even if there is one side of the focal set, which corresponds to a birth and another which corresponds to a death. \label{fig:ReidemeisterA2A2}  } 
\end{figure}

\begin{figure}[h!]
    \centering
    \includegraphics[width=\linewidth]{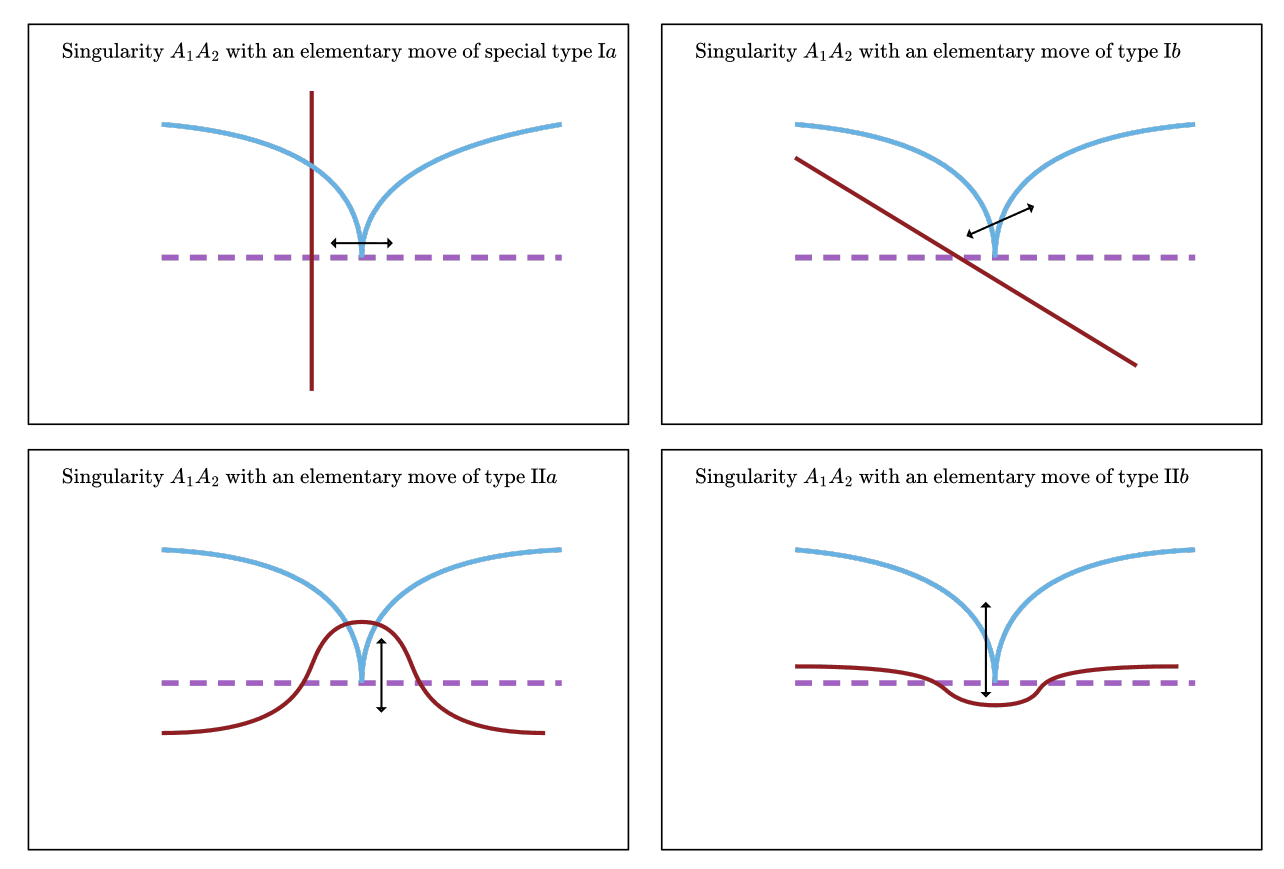}
    \caption{The elementary moves for a singularity of type $A_1A_2$. 
    \label{fig:ReidemeisterA1A2}  } 
\end{figure}

\begin{figure}[h!]
    \centering
    \includegraphics[width=\linewidth]{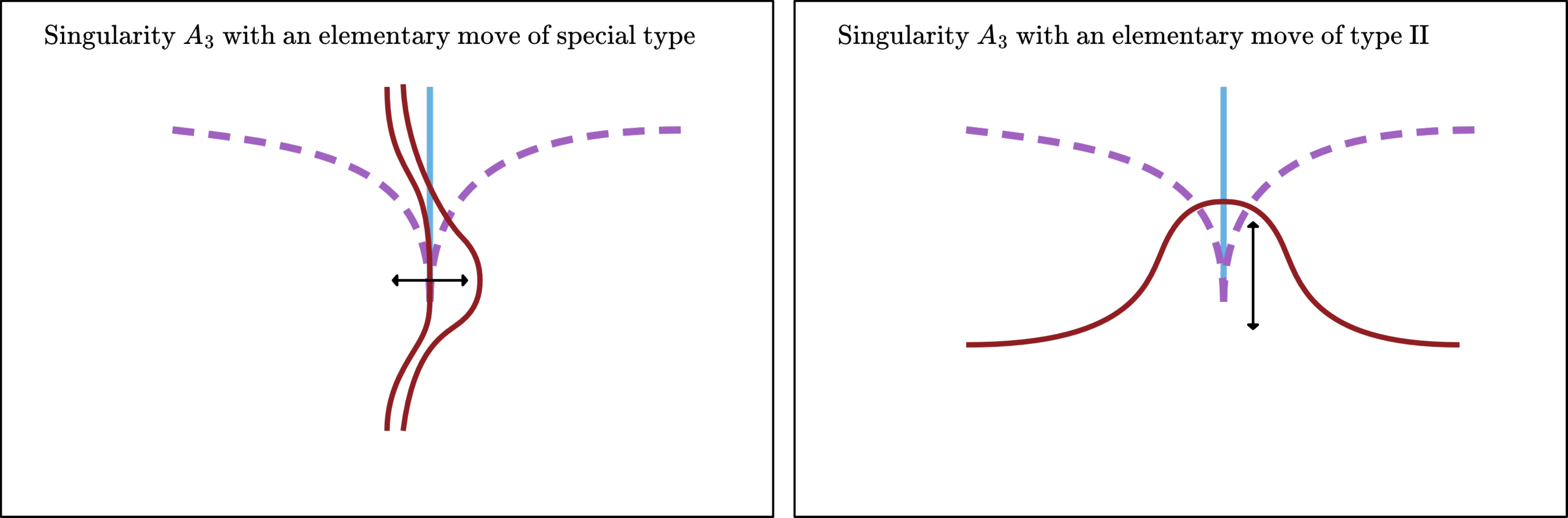}
    \caption{The elementary moves for a singularity of type $A_3$. Because the observation curve has to be tangent to the symmetry set at the point of singularity, the move of special type is not strictly speaking generic, but it is topologically possible. We stress that because of the cusp nature of the singularity the observation loop near the singularity should intersect the focal set either twice or not at all unless it is tangential to the symmetry set exactly at the singularity. \label{fig:ReidemeisterA3}  } 
\end{figure}

\begin{lemma}\label{lem:MoveObservationLoopClassification}
 The lists given in Figures \ref{fig:ReidemeisterA12}-\ref{fig:ReidemeisterA3} are exhaustive for a (locally) generic family $\gamma_\mu$. We have included two
 non-generic special elementary moves one for the singularity type $A_1A_2$ and one for the singularity of type $A_3$, that are referred to as moves of special type. 
\end{lemma}

\begin{proof} 
Because we assume that $\gamma_\mu$ is smooth and does not have any self intersections, it suffices to consider the second order Taylor approximation of the curve $\gamma_\mu(t)$ with respect to $t$ (or more formally its $2$-jet). This means that any stratum of the focal or symmetry set that are smooth can only be slided along (see Figure \ref{fig:ReidemeisterA12} on the left) or crossed in a quadratic manner like a Reidemeister $\textrm{II}$ move (see Figure \ref{fig:ReidemeisterA12} on the right).
\footnote{This argument can be made more formal using techniques from differential topology, in particular transversality~\cite{Hirsch1976}.}
This explains the elementary moves for the singularities $A_1^2$, $A_2$, $A_1^3$, $A_1^2/A_1^2$, $A_1^2/A_2$, and $A_2/A_2$. 

For the singularity of type $A_1A_2$ we need to additionally note that the symmetry set cusp plays a special role, because the left and right limit of the tangent space are the same. This in particular means that any smooth family of curves $\gamma_\mu(t)$, for which $\gamma_{\mu^*} (t^*)$ is tangent to the focal set at the tip of the cusp there exist $\varepsilon_1, \varepsilon_2>0$, such that $\gamma_{\mu}|_{[t^*-\varepsilon_1 , t^*+\varepsilon_1]}$, with $\mu\in [\mu^*-\varepsilon_2 , \mu^*+\varepsilon_2]$  either intersects the symmetry set not at all or twice. 

For the singularity of type $A_3$ we note that the symmetry set has an end point, which can be crossed transversally or tangentially, where  "tangentially" here means that observation loop is tangent to the affine hull of the limit of the tangent line of the symmetry set (or equivalently the focal set)  as one goes towards the tip of the singularity.  
The transversal types are equivalent up to ambient diffeomorphism to moves of type $\textrm{II}$ (as depicted in Figure~\ref{fig:ReidemeisterA3}), because the curve $\gamma_\mu$ has to intersect the focal set either twice or not at all, as for the $A_1A_2$ case.  
For the tangential type, the curve $\gamma_\mu$ intersects the focal set only once, as depicted in Figure \ref{fig:ReidemeisterA3}. 
\end{proof}

\begin{proposition}
The elementary moves on the singularities listed in Figures \ref{fig:ReidemeisterA12}-\ref{fig:ReidemeisterA3} can be reduced to the elementary moves of the following types:
\begin{itemize}
    \item The tangential elementary moves on the singularities of type $A_1^2$ and of type $A_2$.
    \item The transversal elementary moves on singularities of type $A_1^2/A_1^2$, $A_1^2/A_2$, and $A_2/A_2$, as well as $A_1^3$.   
    \item The elementary move of special type $\textrm{I}a$ on singularities of type $A_1A_2$.
    \item The elementary move of special type on a singularity of type $A_3$. 
\end{itemize}
\label{prop:reduction}
\end{proposition}

\begin{proof}
The proof is conducted per singularity type.
\subparagraph{Type $A_1^3$ singularity} The non-transversal move on a type singularity $A_1^3$ can be written as the composition of a non-transversal move on a $A_1^2$ singularity in combination with the transversal move on a $A_1^3$ singularity, see Figure \ref{fig:EquivalenceA13}.  

\begin{figure}[h!]
    \centering
    \includegraphics[width=\linewidth]{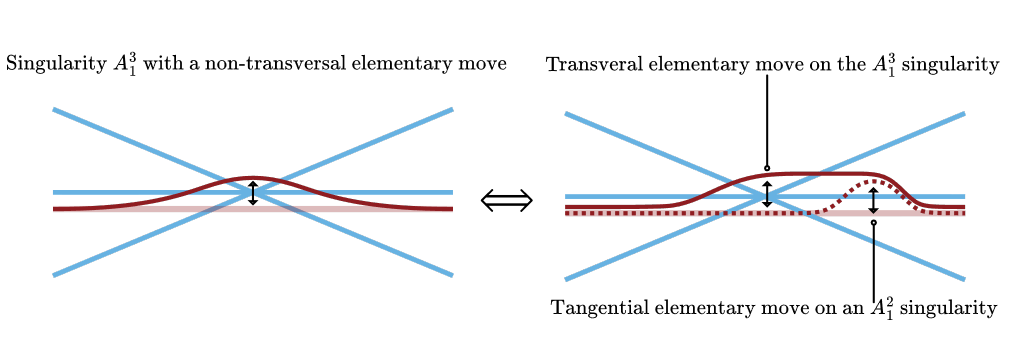}
    \caption{The non-transversal move on $A_1^3$ can be written as a combination of a non-transversal move on $A_1^2$ and a transversal move on $A_1^3$.  } 
    \label{fig:EquivalenceA13} 
\end{figure}

\subparagraph{Type $A_1^2/A_2$ singularity} The tangential moves of type $A_1^2$ on the singularity of type $A_1^2/A_2$ can be written as the combination of a tangential move on a singularity of type $A_1^2$ and a transversal move on a singularity of type $A_1^2/A_2$. This is shown in Figure \ref{fig:EquivalenceA12SA2}. 
By a symmetric construction it is clear that the tangential moves of type $A_2$ on the singularity of type $A_1^2/A_2$ can be written as the combination of a tangential move on a singularity of type $A_2$ and a transversal move on a singularity of type $A_1^2/A_2$. In conclusion only the elementary move of transversal type is required to capture all topological changes of the curve with respect to the observation loop.

\begin{figure}[h!]
    \centering
    \includegraphics[width=\linewidth]{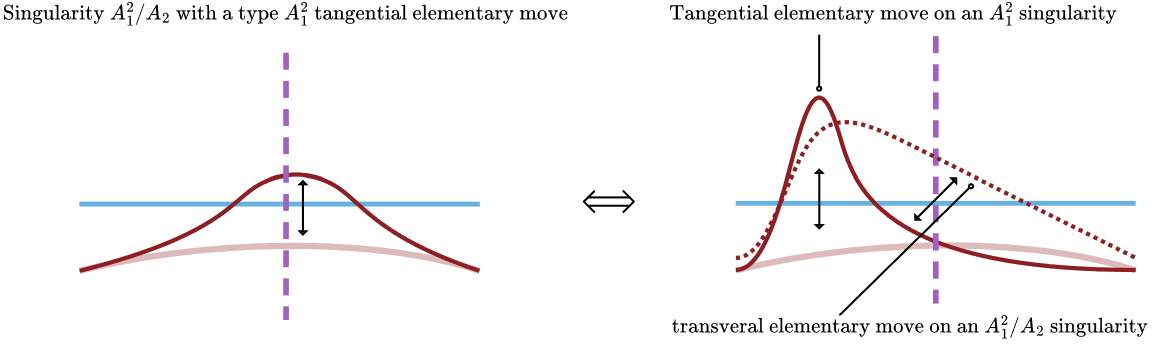}
    \caption{ The tangential moves of type $A_1^2$ on the singularity of type $A_1^2/A_2$ can be written as a combination of a non-transversal move on $A_1^2$ and a transversal move on $A_1^2/A_2$.  } 
    \label{fig:EquivalenceA12SA2} 
\end{figure}

\subparagraph{Type $A_1^2/A_1^2$ and type $A_2/A_2$ singularities} 
By the same construction as for the singularities of type $A_1^2/A_2$ only the transversal elementary move on the singularity of type $A_1^2/A_1^2$ and the transversal elementary move on the singularity of type $A_2/A_2$ are required to capture all topological changes. 

\subparagraph{Type $A_1A_2$ singularity} Every elementary move on a singularity of type $A_1A_2$ can be written as a combination of an elementary move of type $\textrm{i}a$ and some other elementary moves on singularities of type $A_1^2$ and/or type $A_2$, see Figures \ref{fig:EquivalenceA1A2First}, \ref{fig:EquivalenceA1A2Second}, and \ref{fig:EquivalenceA1A2Third}. 

\begin{figure}[h!]
    \centering
    \includegraphics[width=\linewidth]{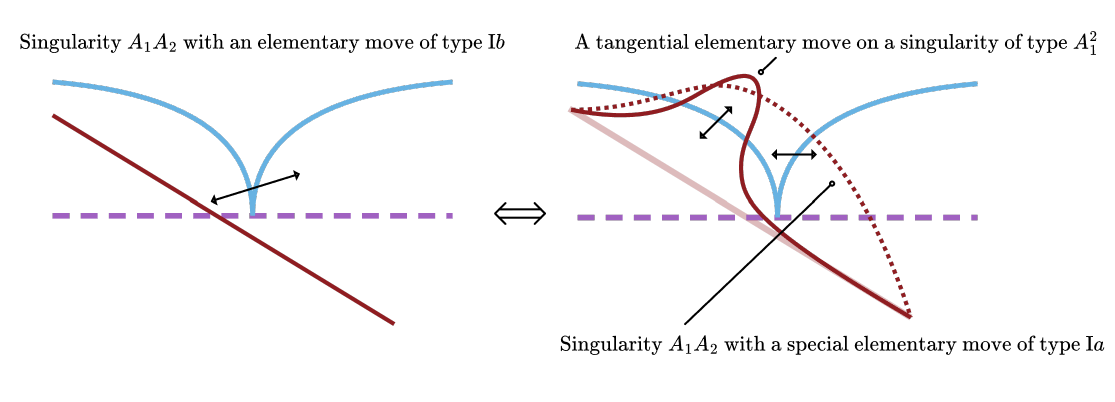}
    \caption{The elementary move of type $\textrm{I}b$ on a singularity of type $A_1A_2$ can be written as a combination of a tangential move on a singularity of type $A_1^2$ and a special elementary move of type $\textrm{I}a$ of a singularity of type $A_1A_2$.} 
    \label{fig:EquivalenceA1A2First} 
\end{figure}

\begin{figure}[h!]
    \centering
    \includegraphics[width=\linewidth]{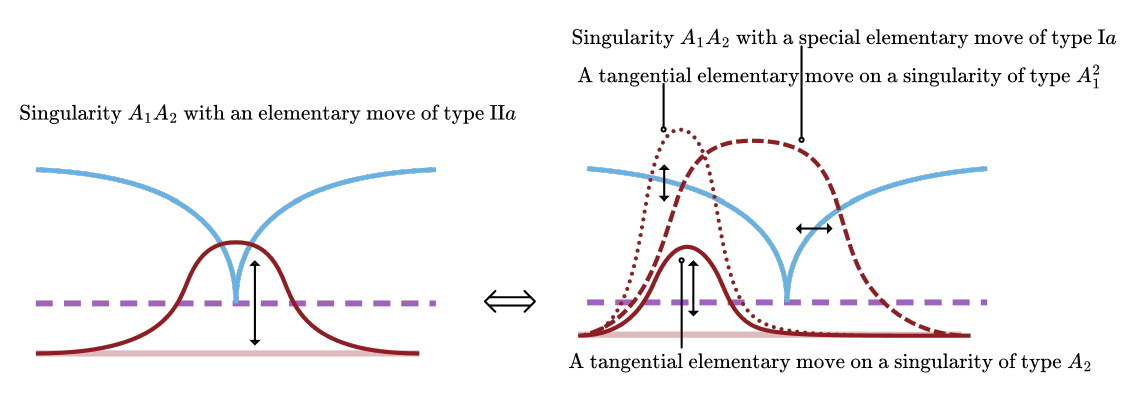}
    \caption{The elementary move of type $\textrm{II}a$ on a singularity of type $A_1A_2$ can be written as a combination of a tangential move on a singularity of type $A_2$, then a tangential move on a singularity of type $A_1^2$ and a special elementary move of type $\textrm{I}a$ of a singularity of type $A_1A_2$.  } 
    \label{fig:EquivalenceA1A2Second} 
\end{figure}

\begin{figure}[h!]
    \centering
    \includegraphics[width=\linewidth]{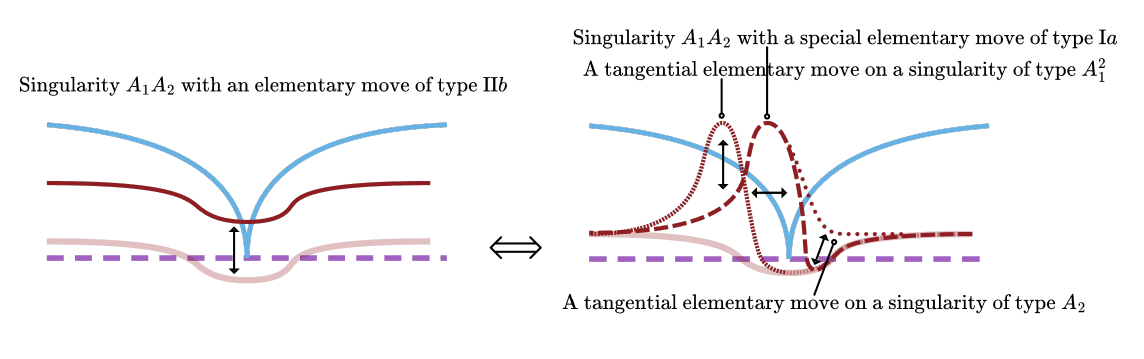}
    \caption{The elementary move of type $\textrm{II}b$ on a singularity of type $A_1A_2$ can be written as a combination of a tangential move on a singularity of type $A_1^2$, then a special elementary move of type $\textrm{I}a$ of a singularity of type $A_1A_2$ and finally a tangential move on a singularity of type $A_1^2$.   } 
    \label{fig:EquivalenceA1A2Third} 
\end{figure}

\subparagraph{Type $A_3$ singularity} 
The elementary move of type $\textrm{II}$  on an $A_3$ singularity can be written as a combination of a tangential elementary move on a singularity of type $A_2$ and an elementary move of special type, see Figure \ref{fig:EquivalenceA1A2Third}. We stress once more that the move of special type is non-generic, but is included to simplify a later proof. 

\begin{figure}[h!]
    \centering
    \includegraphics[width=\linewidth]{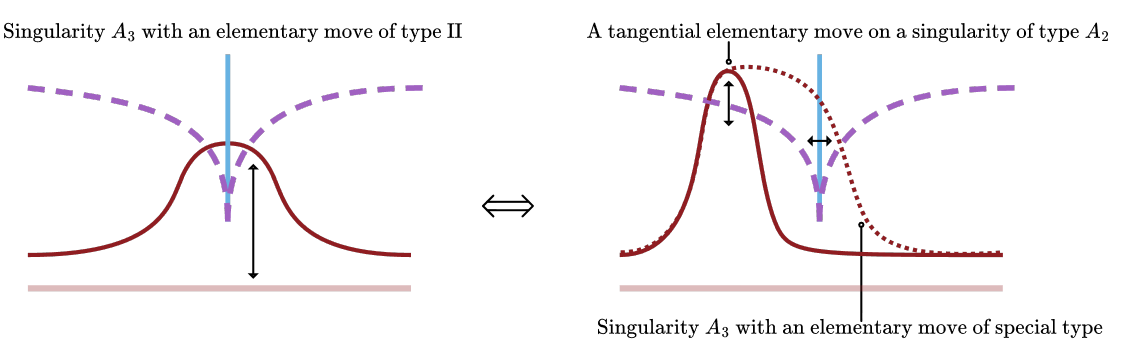}
    \caption{An elementary move of type $\textrm{II}$ on a singularity of type $A_3$ can be written as a combination of a tangential move on a singularity of type $A_2$ and an elementary move of special type on $A_3$. } 
    \label{fig:EquivalenceA3Typeii} 
\end{figure}
\end{proof}

\subsection{From elementary moves of the observation loop to changes in the vineyard} \label{sec:FromLoopToVineyard}

The entire section is dedicated to proving one of our main results, namely showing which elementary moves of the observation loop induce which changes in the vineyard. This in particular shows that the list of elementary moves on the vineyard, as given in Section \ref{sec:VineyardTransitions}, is complete in the context of the radial transformation.

\MovesOnVineyard*

{
The theorem has the following immediate corollary:
\begin{corollary}
The topology of a vineyard for an observation loop $\gamma$ can be computed by considering a generic isotopy $\gamma_\mu$ from a constant curve to $\gamma$ and keeping track of the elementary moves which $\gamma_\mu$ makes. We stress that for a number of singularities (like the $A_1^2/A_1^2$ singularity) a local computation is needed to check which of the options listed occurs.  
\end{corollary}
}

Before proving the theorem, we will make some observations which helps to streamline the proof c.f. \cite[Section 2]{local}.  
\VineIntersectionCriterion*

\begin{observation}\label{obs:dictionary}
We note the following:
\begin{itemize} 
\item {By the classification of singularities, recalled in Section \ref{sec:Singularities_Plane}, all the singularities are isolated. } 
\item Let $p \in \mathbb{R}^2$ and $f_p := d_{\mathbb{E}}^2(\cdot,p)|_{\mathcal{M}}$. Two distinct points of $\mathrm{Dgm}(f_p)$ share a coordinate if and only if two distinct critical points of $f_p$ have the same critical value, that is, if and only if a single circle centred at $p$ is tangent to $\mathcal{M}$ in at least two places and hence $p$ lies on the symmetry set. 
\item Two critical points $x_1,x_2$ with a common critical value $c$ always give rise to two \emph{distinct} points of the diagram: 
if both are minima or both are maxima this is immediate, since a point of the diagram has a single birth and a single death coordinate; and if one is a minimum $m$ and the other a maximum $M$, they cannot be paired with one another, because the component born at $m$ and destroyed at $M$ satisfy {the strict inequality $f_p(m) < f_p(M)$. } 
\item By contrast, $f_p$ fails to be Morse exactly when $p$ is a centre of curvature of $\mathcal{M}$, that is, when $p$ lies on the focal set. So along an isotopy the two mechanisms by which the combinatorics of the diagram can change are separated by the two parts of $\mathrm{gsym}(\mathcal{M})$: concurrences are created and destroyed on the symmetry set, while points enter and leave the diagonal on the focal set.
\end{itemize} 
\end{observation}

In all figures of this section the vineyard is computed with extended persistence, so that each vine  above the diagonal 
is accompanied by its reflection in the diagonal plane; any event in a vine occurs equally in its extended counterpart. Since the figures are two dimensional projections of a three dimensional object, not every event is legible from the viewing direction chosen, and vines that appear to cross need not meet. Where this matters we say which copy or which coordinates carry the event in question.
The examples shown here use the curve $\M$ constructed in \cite[Section 3.2]{local}.

\begin{proof}[Proof of Theorem \ref{prop:MovesOnVineyard}] 
The proof is conducted per singularity type.

\subparagraph{Type $A_1^2$ singularity} 

Let $p$ lie on a manifold piece of the symmetry set of type $A_1^2$, so that the circle centred at $p$ is tangent to $\M$ at 2 points $x_1,x_2$, and $f_p$ has two Morse critical points with a common critical value $c$. By Observation \ref{obs:dictionary} these contribute one coordinate equal to $c$ to each of two distinct points of $\mathrm{Dgm}(f_p)$. 
Because of the isolation of singularities, nothing else happens.

{
For the \emph{trivial} move of Figure~\ref{fig:ReidemeisterA12} (left), $\gamma_\mu$ meets the stratum transversally in a single point for all $\mu$ near $\mu^*$, and the crossing time $t(\mu)$ varies continuously with $\mu$. Since the 2 diagram points share only the coordinate $c$, each crossing carries exactly one concurrence, so a single concurrence of fixed type is present throughout and only the time at which it occurs changes. No vine reaches the diagonal, and by Observation~\ref{ob:A12intersection} no two vines intersect, so the vineyard is merely deformed isotopically and the move induces no {topological} change; see Figure~\ref{fig:A12trivialresult}.
}

\begin{figure}[h!]
    \centering
    \includegraphics[width=\linewidth]{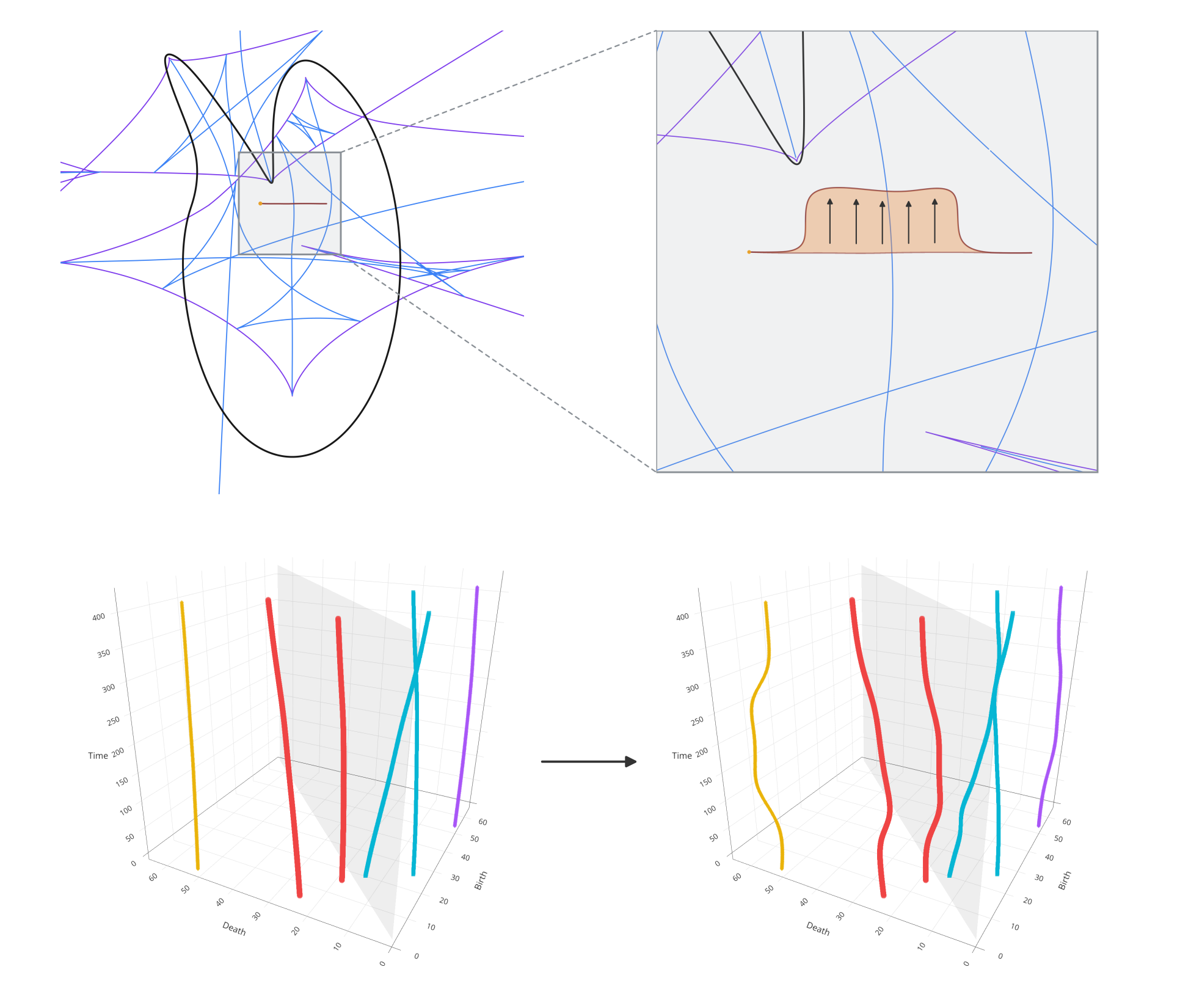}
    \caption{The trivial elementary move on a singularity of type $A_1^2$. Top: the curve $\gamma_\mu$ (dark red) is deformed in the direction of the arrows, so that its transversal crossing with a manifold piece of the symmetry set (blue) slides along that piece; the box on the left is enlarged on the right. Bottom: the vineyard before (left) and after (right) the move. The two cyan (and red) vines retain their single birth--death concurrence throughout and only the time at which it occurs changes, so the vineyard is deformed isotopically.} 
    \label{fig:A12trivialresult} 
\end{figure}

{
For the \emph{tangential} move of Figure~\ref{fig:ReidemeisterA12} (right), $\gamma_{\mu^*}$ is tangent to the stratum at $t^*$, and by genericity the tangency is quadratic (Definition~\ref{def:genericGammaMu}). On one side of $\mu^*$ the curve misses the stratum locally, while on the other it crosses it transversally at two times $t_1(\mu) < t^* < t_2(\mu)$ with $t_1(\mu),t_2(\mu) \to t^*$ as $\mu \to \mu^*$. The two vines determined by $x_1$ and $x_2$ therefore have no concurrence on the first side and exactly two, of the same type and emerging from the single time $t^*$, on the second. This is precisely a concurrence pair creation in the sense of Definition~\ref{def:ChangesVineyard}, respectively an annihilation if the direction of the isotopy is reversed. Since only one coordinate of the two vines is ever shared, the vines again do not intersect and the topology of the vineyard is unaffected; see Figure~\ref{fig:A12tangentialresult}.
}

\begin{figure}[h!]
    \centering
    \includegraphics[width=\linewidth]{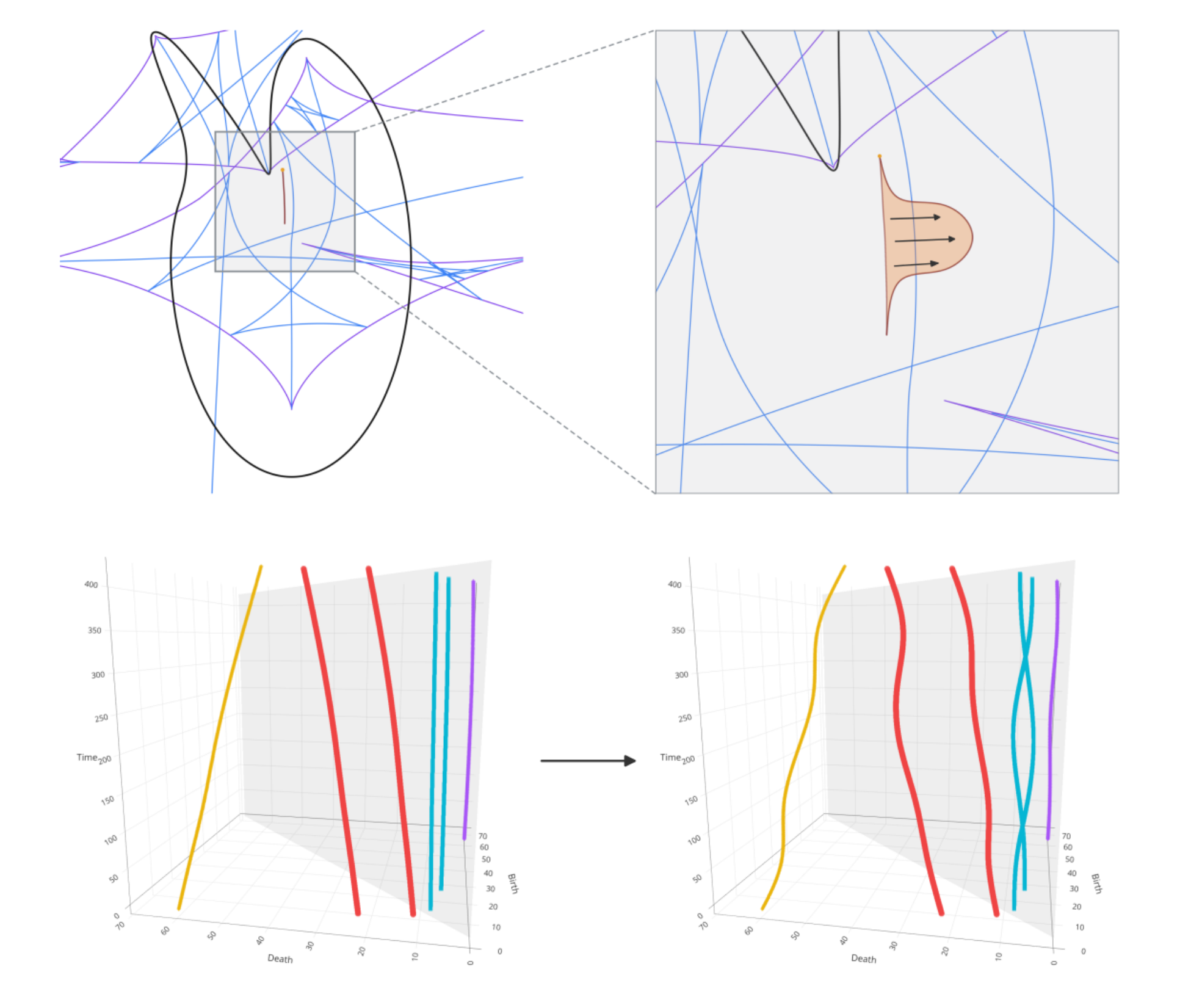}
    \caption{The tangential elementary move on a singularity of type $A_1^2$. Top: the curve $\gamma_\mu$ (dark red) is pushed in the direction of the arrows until it is tangent to a manifold piece of the symmetry set (blue), and then past it, so that two transversal crossings appear. Bottom: the vineyard before (left) and after (right) the move. The two cyan (and red) vines have no concurrence before the move and two afterwards, which is the concurrence pair creation of Definition~\ref{def:ChangesVineyard}. Note that they only appear to cross: the concurrence is between the birth coordinate of the one and the death coordinate of the other, while their remaining coordinates stay distinct, so the vines do not intersect.} 
    \label{fig:A12tangentialresult} 
\end{figure}

\subparagraph{Type $A_2$ singularity}

We start by introducing some notation. Let $p$ lie  on a singularity of type $A_2$. This means that, by the classification of singularities recalled in Section \ref{sec:Singularities_Plane}, $p$ lies on a manifold piece of the focal set, so that $p$ is the centre of curvature of a point $x \in \M$, and $f_p$ has a degenerate critical point at $x$ of type $A_2$. On one side of the focal set this degenerate critical point splits into maximum and minimum, while on the other side it disappears.
Consequently, a manifold piece of the focal set separates a neighbourhood $U$ of $p$ into $2$ parts $U_{+}$ and $U_{-}$, where for $q\in U_{+}$ the diagram $\mathrm{Dgm}(f_q)$ has one more point on one part than for $q \in U_{-}$. This extra point in the diagram pairs the additional maximum and minimum.  As $q$ approaches the focal set the 2 critical values converge, so the point tends to the diagonal. No concurrence is involved: by Observation~\ref{obs:dictionary} concurrences are created only on the symmetry set  which is locally not present, because singularities are (generically) isolated.

For the \emph{trivial} move of Figure~\ref{fig:ReidemeisterA2} (left), $\gamma_\mu$ meets the stratum transversally in a single point for all $\mu$ near $\mu^*$, and the crossing time $t(\mu)$ varies continuously with $\mu$. A single vine therefore enters or leaves the diagonal throughout, and only the time at which it does so changes. By Observation~\ref{ob:A12intersection} the vines move isotopically, so the move induces no change in the vineyard; see Figure~\ref{fig:A2trivialresult}.

\begin{figure}[h!]
    \centering
    \includegraphics[width=\linewidth]{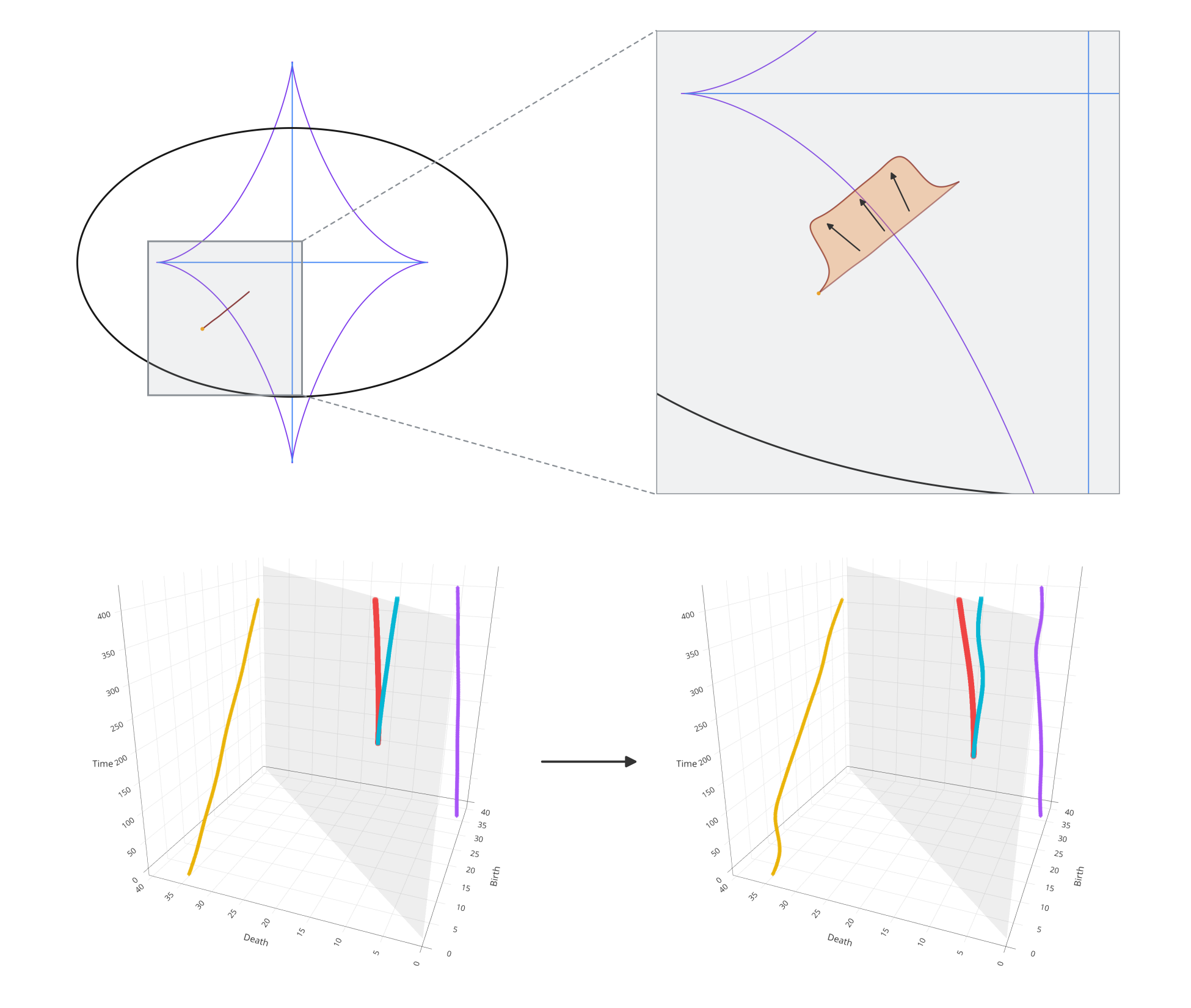}
    \caption{The trivial elementary move on a singularity of type $A_2$. Top: the curve $\gamma_\mu$ (dark red) is deformed in the direction of the arrows, so that its transversal crossing with a manifold piece of the focal set (purple) slides along that piece; the box on the left is enlarged on the right. Bottom: the vineyard before (left) and after (right) the move. The vines that meet the diagonal do so at slightly different times, but no vine is created, destroyed, merged or split, so the vineyard is deformed isotopically.} 
    \label{fig:A2trivialresult} 
\end{figure}

{
For the \emph{tangential} move of Figure~\ref{fig:ReidemeisterA2} (right), $\gamma_{\mu^*}$ is tangent to the stratum at $t^*$ with 2nd order tangency (Definition~\ref{def:genericGammaMu}). On one side of $\mu^*$ the curve misses the stratum locally, while on the other it crosses it transversally at two times $t_1(\mu) < t^* < t_2(\mu)$ with $t_1(\mu),t_2(\mu) \to t^*$. Which change this induces depends on the side of the focal set from which the curve approaches. If $\gamma_\mu$ { starts on the $U_{-}$ side, }
then for $t \in (t_1,t_2)$ an additional point is present; it is born from the diagonal at $t_1$ and returns to it at $t_2$, and by Convention~\ref{convention:Annihilation} its endpoints are connected forward in time along the diagonal. This is a vine creation, respectively an annihilation in the reverse direction. 

If instead $\gamma_\mu$ starts {on the side $U_{+}$,} 
then the vine that was present throughout the local time interval reaches the diagonal at $t_1$ and re-emerges at $t_2$, so that it is cut in two; by Convention~\ref{convention:MergeSplit} the resulting endpoints are reconnected forward in time and we obtain either {two vines or a vine that starts and ends on the diagonal.} This is a split, respectively a merge in the reverse direction, and the two diagonal endpoints that coincide at $\mu^*$ are the merger point of Definition~\ref{def:ChangesVineyard}. In either case the change is exactly one of those listed, which is what was claimed; see Figure~\ref{fig:A2tangentialresult}.
}

\begin{figure}[h!]
    \centering
    \includegraphics[width=\linewidth]{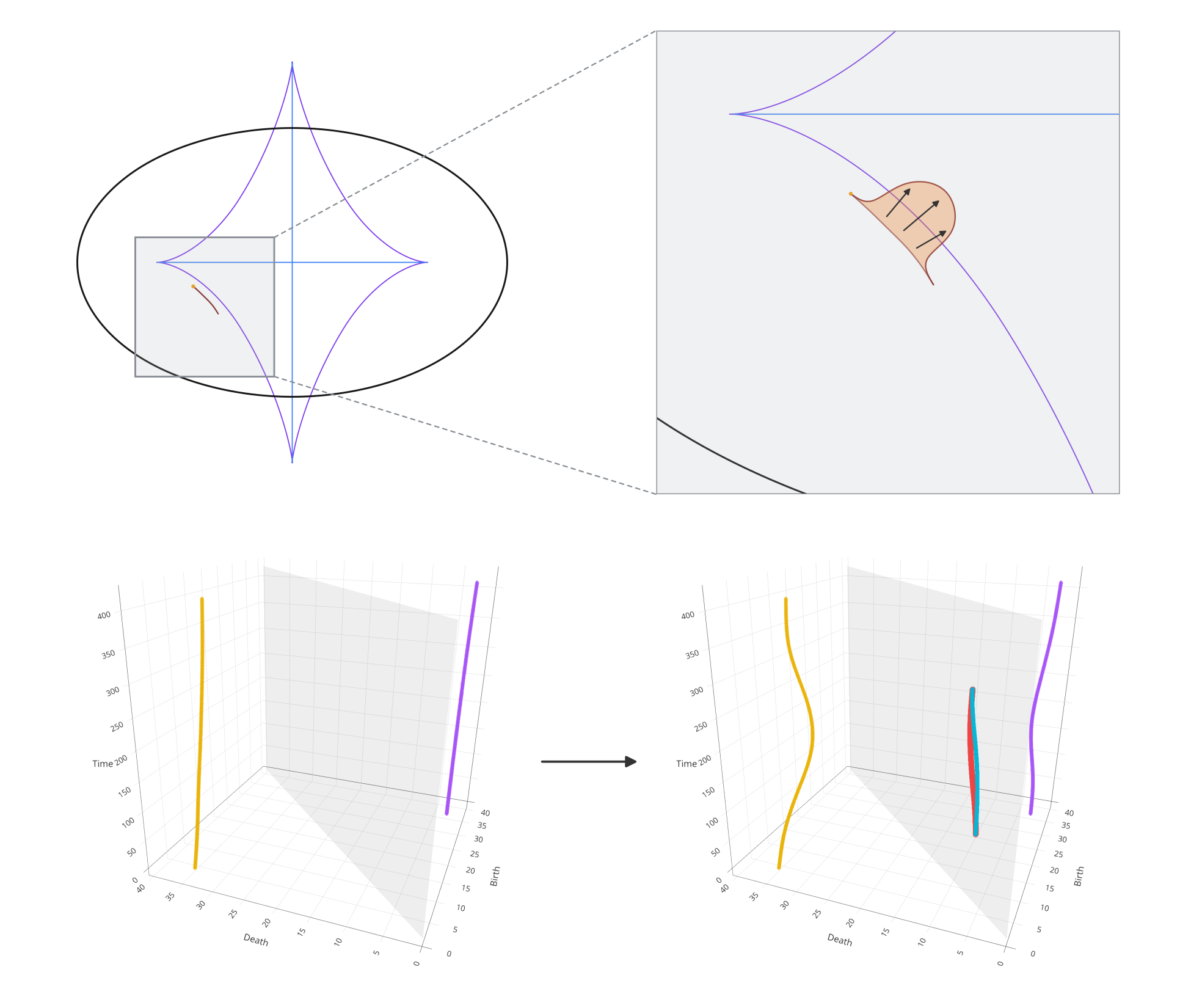}
    \caption{The tangential elementary move on a singularity of type $A_2$. Top: the curve $\gamma_\mu$ (dark red) is pushed in the direction of the arrows until it is tangent to a manifold piece of the focal set (purple), and then past it, so that 2 transversal crossings occur. Here $\gamma_\mu$ approaches from the side on which the persistence diagram has one point fewer. Bottom: the vineyard before (left) and after (right) the move. A new vine of low persistence appears close to the diagonal plane, born from the diagonal at the first crossing and returning to it in the second; this is the vine creation of Definition~\ref{def:ChangesVineyard}.} 
    \label{fig:A2tangentialresult} 
\end{figure}

\subparagraph{Type $A_1^2/A_1^2$ singularity}

Let $p$ be a singularity of type $A_1^2/A_1^2$. 
As we have seen in the recap in Section \ref{sec:Singularities_Plane}, two manifold pieces $S_1,S_2$ of the symmetry set intersect in the point $p$ and there are two concentric circles of radius $r_1,r_2$ centred at $p$ that are tangent to $\M$ in two places each. For $p'$ near $p$ this gives rise to $4$ critical values of $d_\mathbb{E}^2 (\cdot, p')$, namely  $r_{1,a}^2, r_{1,b}^2 \simeq r_1^2$ and $r_{2,a}^2, r_{2,b}^2 \simeq r_2^2$, which form the birth and/or death values of between two and four different persistence points. As we have seen in Observation~\ref{ob:A12intersection}, these points form only two persistence points, if and only if there is a vine intersection. If not, then there are only concurrence related local changes in the vineyard (rearrangements in fact, for the transversal move as we will discuss below).

{
For the \emph{transversal} move of Figure~\ref{fig:ReidemeisterA12A12} (left), $\gamma_\mu$ crosses both $S_1$ and $S_2$ transversally for $\mu$ near $\mu^*$ and passes over their intersection at $\mu = \mu^*$. Writing $t_1(\mu)$ and $t_2(\mu)$ for the crossing times, we have $t_1(\mu^*) = t_2(\mu^*) = t^*$, and the order of $t_1$ and $t_2$ is interchanged as $\mu$ passes $\mu^*$. Each crossing carries one concurrence, so the effect on the vineyard is governed by the dichotomy above, {that is if the critical points form two (case 1) or more (namely $3$ or $4$, case 2).} In the second case the two concurrences concern different pairs of vines, no two vines ever share more than one coordinate, and the move merely interchanges the order in time of the two concurrences: this is a concurrence--concurrence rearrangement in the sense of Definition~\ref{def:ChangesVineyard}, and by Observation~\ref{ob:A12intersection} the vines move isotopically throughout. In the first case the two concurrences concern the same pair of vines, one in the birth and one in the death coordinate. At $\mu = \mu^*$ these two concurrences occur at the same time $t^*$, so both coordinates of the two vines agree and the vines meet; for $\mu \neq \mu^*$ they are separated in time and the vines are disjoint. The move is therefore a vine intersection type change. { As we have seen in the construction of \cite[Section 3.2]{local} a single singularity of type $A_1^2/A_1^2$ can give rise to monodromy and hence,} the connections between the pieces of the two vines before and after $t^*$ may differ,  {as we see in Figure \ref{fig:A12A12transversalresult}}. 
}

\begin{figure}[h!]
    \centering
    \includegraphics[width=\linewidth]{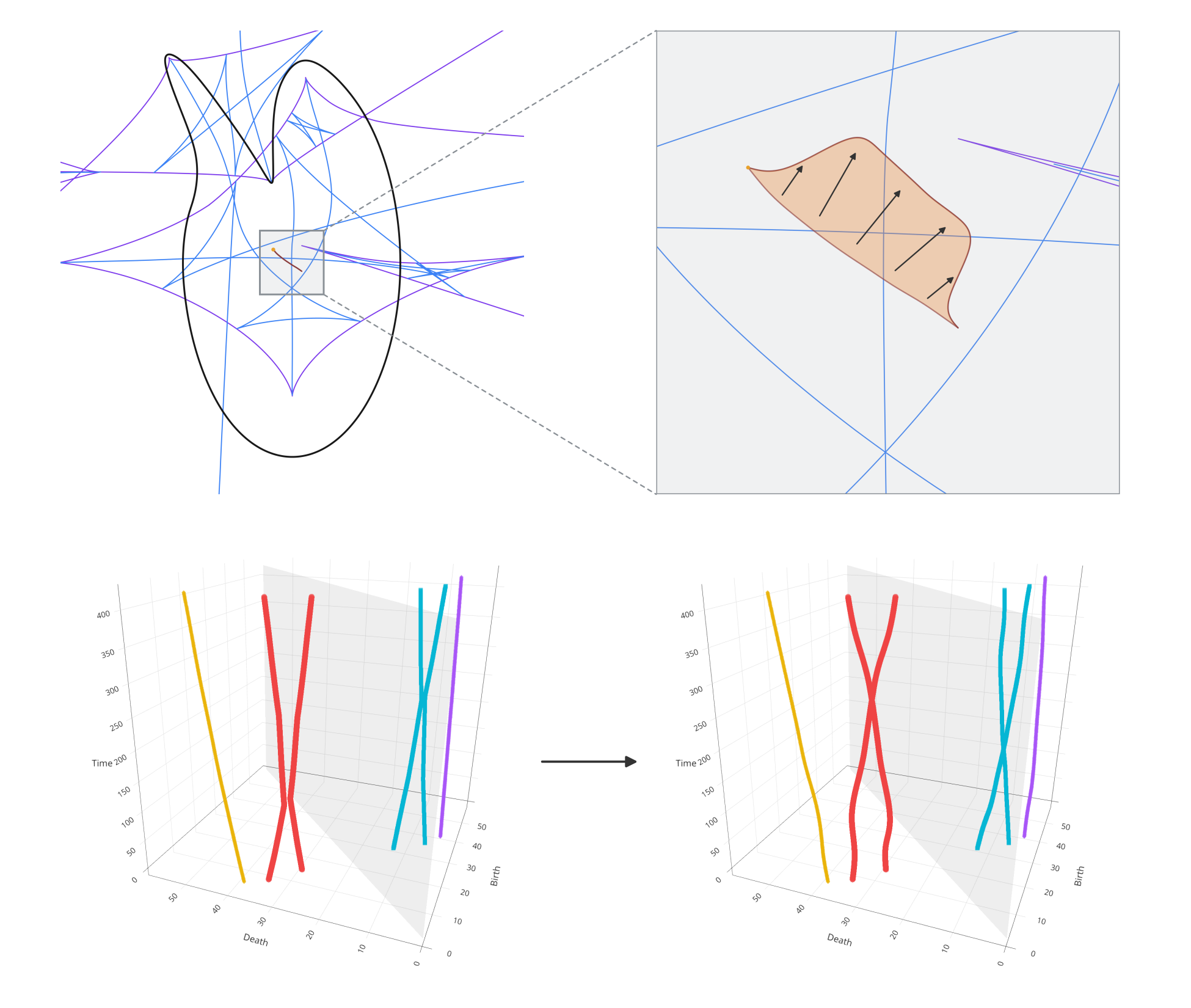}
    \caption{The transversal elementary move on a singularity of type $A_1^2/A_1^2$. Top: the curve $\gamma_\mu$ (dark red) is pushed in the direction of the arrows across the transverse intersection of two manifold pieces of the symmetry set (light blue); the box on the left is enlarged on the right. Bottom: the vineyard before (left) and after (right) the move. The order in time of the two concurrences is interchanged, which is the concurrence--concurrence rearrangement of Definition~\ref{def:ChangesVineyard}. The vines appear to cross in this projection, but at each concurrence only one of their two coordinates agrees, so they do not meet.} 
    \label{fig:A12A12transversalresult} 
\end{figure}

For the \emph{tangential} move of Figure~\ref{fig:ReidemeisterA12A12} (right), no separate argument is needed: by Proposition~\ref{prop:reduction} it is the composition of a tangential move on a singularity of type $A_1^2$ and a transversal move on a singularity of type $A_1^2/A_1^2$, so the induced change in the vineyard is the composition of a concurrence pair creation with one of the two changes above. Figure~\ref{fig:A12A12tangentialresult} shows this: the pair of concurrences created by the tangency straddles, in time, the concurrence already present from the transversal crossing of the other branch.

\begin{figure}[h!]
    \centering
    \includegraphics[width=\linewidth]{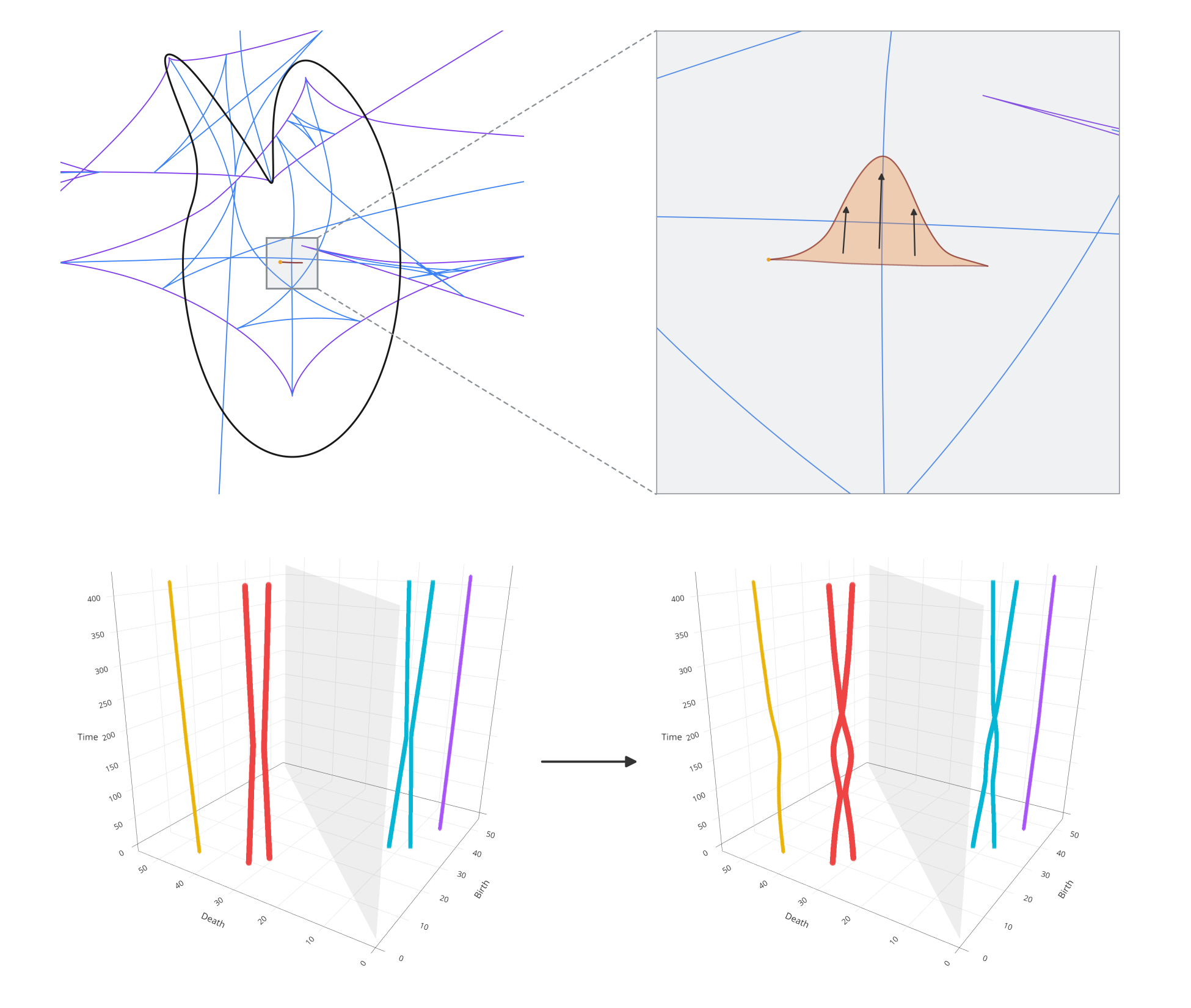}
    \caption{The tangential elementary move on a singularity of type $A_1^2/A_1^2$. Top: the curve $\gamma_\mu$ (dark red) is pushed in the direction of the arrows until it is tangent to one manifold piece of the symmetry set (light blue) at the point where the second piece crosses it, and then past it. Bottom: the vineyard before (left) and after (right) the move. A pair of concurrences is created, straddling the concurrence carried by the transversal crossing of the other branch; by Proposition~\ref{prop:reduction} this move is a composition of the tangential move on $A_1^2$ and the transversal move on $A_1^2/A_1^2$.} 
    \label{fig:A12A12tangentialresult} 
\end{figure}

\subparagraph{Type $A_2/A_2$ singularity}

Let $p$ be a singularity of type $A_2/A_2$. As we have seen in the recap in Section \ref{sec:Singularities_Plane}, two concentric circles centred at $p$, of radii $r_1\neq r_2$, are both osculating circles of $\M$. Locally two manifold pieces $F_1,F_2$ of the focal set cross transversally at $p$. As we have seen before in the discussion of the $A_2$ singularity, the point where an (observation) curve intersects the focal set corresponds to the point where a vine leaves the diagonal in the vineyard. Moreover, because $r_1\neq r_2$ the two vines concerned leave the diagonal at different points. No other elementary change related to concurrences can occur due to the isolation of singularities, as noted in Observation \ref{obs:dictionary}.

For the \emph{transversal} move of Figure~\ref{fig:ReidemeisterA2A2} (left), $\gamma_\mu$ crosses both $F_1$ and $F_2$ transversally for $\mu$ near $\mu^*$. Writing $t_1(\mu)$ and $t_2(\mu)$ for the 2 crossing times, we have $t_1(\mu^*)=t_2(\mu^*)=t^*$, and their order is interchanged as $\mu$ passes $\mu^*$. Each crossing is a time at which a vine enters or exits the diagonal, so the move interchanges the order in time of 2 such events: this is a vine diagonal--diagonal rearrangement in the sense of Definition~\ref{def:ChangesVineyard}. It {can induce monodromy}, and this is what distinguishes the present case from the concurrence--concurrence rearrangement of the $A_1^2/A_1^2$ case and the vine diagonal--concurrence rearrangement of the $A_1^2/A_2$ case: in those the reordering involves a concurrence, which concerns coordinate values only, while here both events are times at which a vine meets the diagonal. Indeed, the times at which the two vines are off-diagonal are intervals whose relative position changes with the passage: on one side of $\mu^*$ they are disjoint, or one contains the other, while on the other side they overlap without containment. 
By Convention~\ref{convention:Annihilation}, vines are connected forward in time  to their own end points, hence the connection can change {depending on whether the vine needs to be connected forward in time along the diagonal with a short piece or one of approximately length $2\pi$. }
Lemma~\ref{lem:A2SA2Monodromy} shows how this reordering gives rise to monodromy when the curve follows a loop of the focal set through such a singularity. See Figure~\ref{fig:A2A2transversalresult}.

\begin{figure}[h!]
    \centering
    \includegraphics[width=\linewidth]{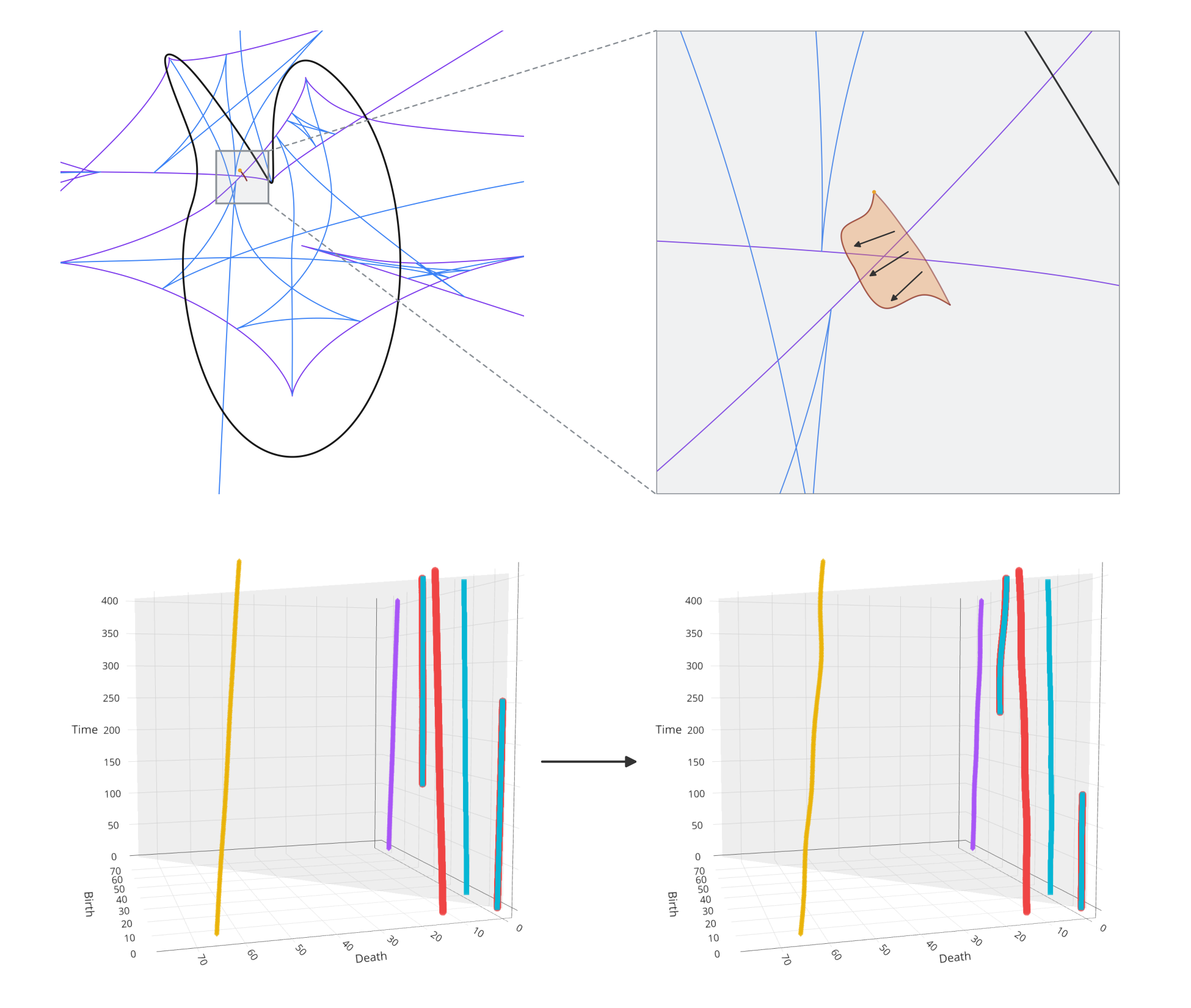}
    \caption{The transversal elementary move on a singularity of type $A_2/A_2$. Top: the curve $\gamma_\mu$ (dark red) is pushed in the direction of the arrows across the transverse intersection of two manifold pieces of the focal set (purple); the box on the left is enlarged on the right. Bottom: the vineyard before (left) and after (right) the move. The order in time in which two vines enter and exit the diagonal plane is interchanged, which is the essential vine diagonal--diagonal rearrangement of Definition~\ref{def:ChangesVineyard}.} 
    \label{fig:A2A2transversalresult} 
\end{figure}

{
For the \emph{tangential} move of Figure~\ref{fig:ReidemeisterA2A2} (right), no separate argument is needed: by Proposition~\ref{prop:reduction} it is the composition of a tangential move on a singularity of type $A_2$ with a transversal move on $A_2/A_2$, so the induced change is a vine creation or split, followed by a vine diagonal--diagonal rearrangement; see Figure~\ref{fig:A2A2tangentialresult}.
}

\begin{figure}[h!]
    \centering
    \includegraphics[width=\linewidth]{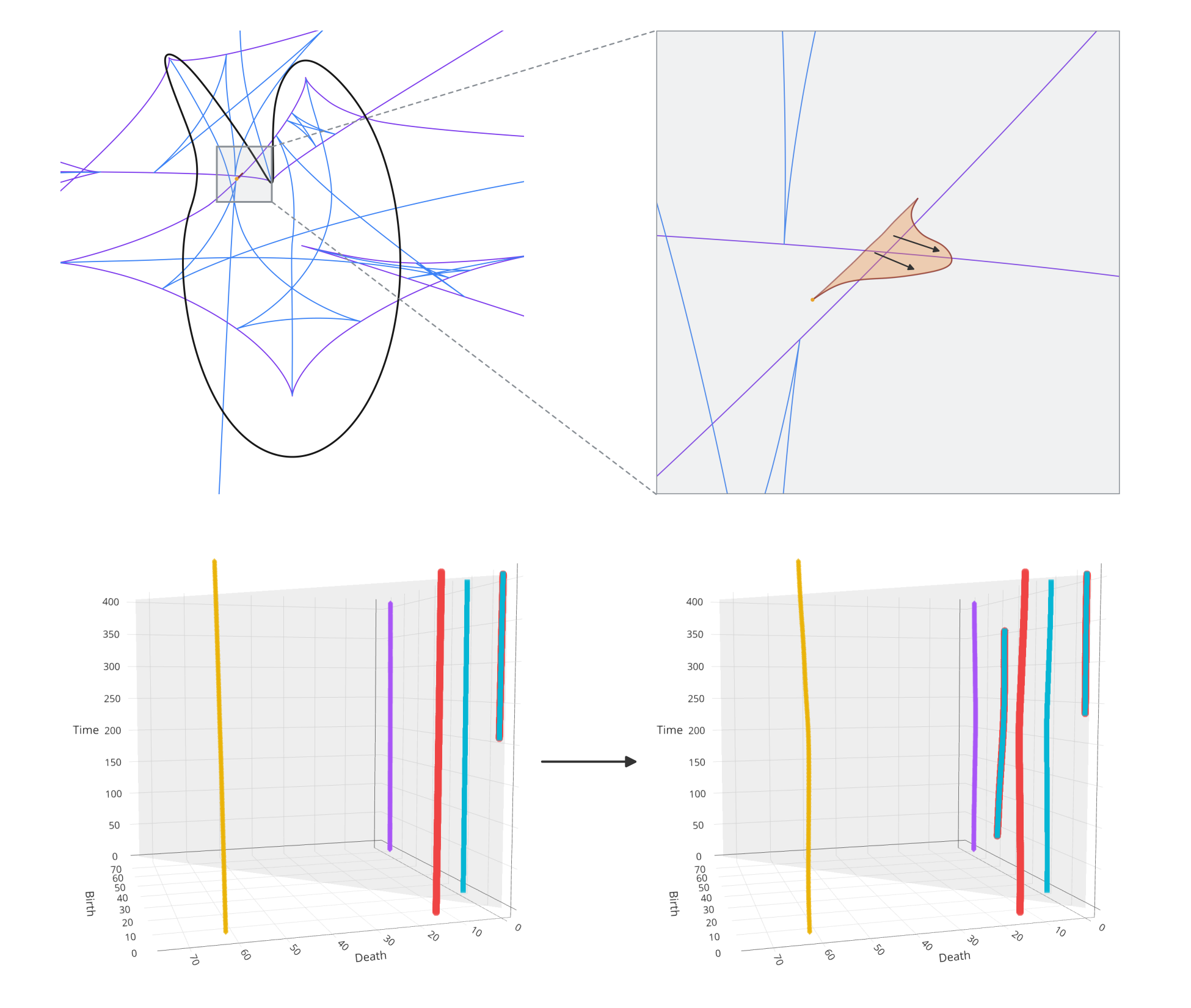}
    \caption{The tangential elementary move on a singularity of type $A_2/A_2$. Top: the curve $\gamma_\mu$ (dark red) is pushed in the direction of the arrows until it is tangent to one manifold piece of the focal set (purple) at the point where the second piece crosses it, and then past it. Bottom: the vineyard before (left) and after (right) the move. By Proposition~\ref{prop:reduction} this move is a composition of the tangential move on $A_2$ with the transversal move on $A_2/A_2$, and the induced change is the corresponding composition.} 
    \label{fig:A2A2tangentialresult} 
\end{figure}

\subparagraph{Type $A_1^2/A_2$ singularity}

The classification of singularities recalled in Section \ref{sec:Singularities_Plane}, tells us that if 
$p$ is a singularity of type $A_1^2/A_2$ then there are two concentric circles centred at $p$, of radii $r_1 \neq r_2$ one of which is tangent to $\mathcal{M}$ at two points, the other osculates $\mathcal{M}$. Locally a manifold piece $S$ of the symmetry set and a manifold piece $F$ of the focal set cross transversally at $p$. As we have seen in the discussion of the $A_1^2$ singularity, crossing $S$ produces a concurrence. As we have seen in the discussion of the $A_2$ singularity, crossing $F$ makes a vine enter or leave the diagonal. Since the radii differ, the concurrence and entry into the diagonal occur at different (birth and death) values.

For the \emph{transversal} move of Figure~\ref{fig:ReidemeisterA12A2} (left), $\gamma_\mu$ crosses both $S$ and $F$ transversally for $\mu$ near $\mu^*$ and passes over their intersection at $\mu = \mu^*$. Writing $t_S(\mu)$ and $t_F(\mu)$ for the two crossing times, we have $t_S(\mu^*) = t_F(\mu^*) = t^*$, and their order is interchanged as $\mu$ passes $\mu^*$. By the above, the vine that enters or leaves the diagonal at $t_F$ is not one of the two vines that share a coordinate at $t_S$, so no vine is created, destroyed, merged or split by the passage, and no two vines meet. The move therefore interchanges the time at which a vine reaches the diagonal with the time at which a concurrence takes place, which is a vine diagonal--concurrence rearrangement in the sense of Definition~\ref{def:ChangesVineyard}; see Figure~\ref{fig:A12A2transversalresult}.

\begin{figure}[h!]
    \centering
    \includegraphics[width=\linewidth]{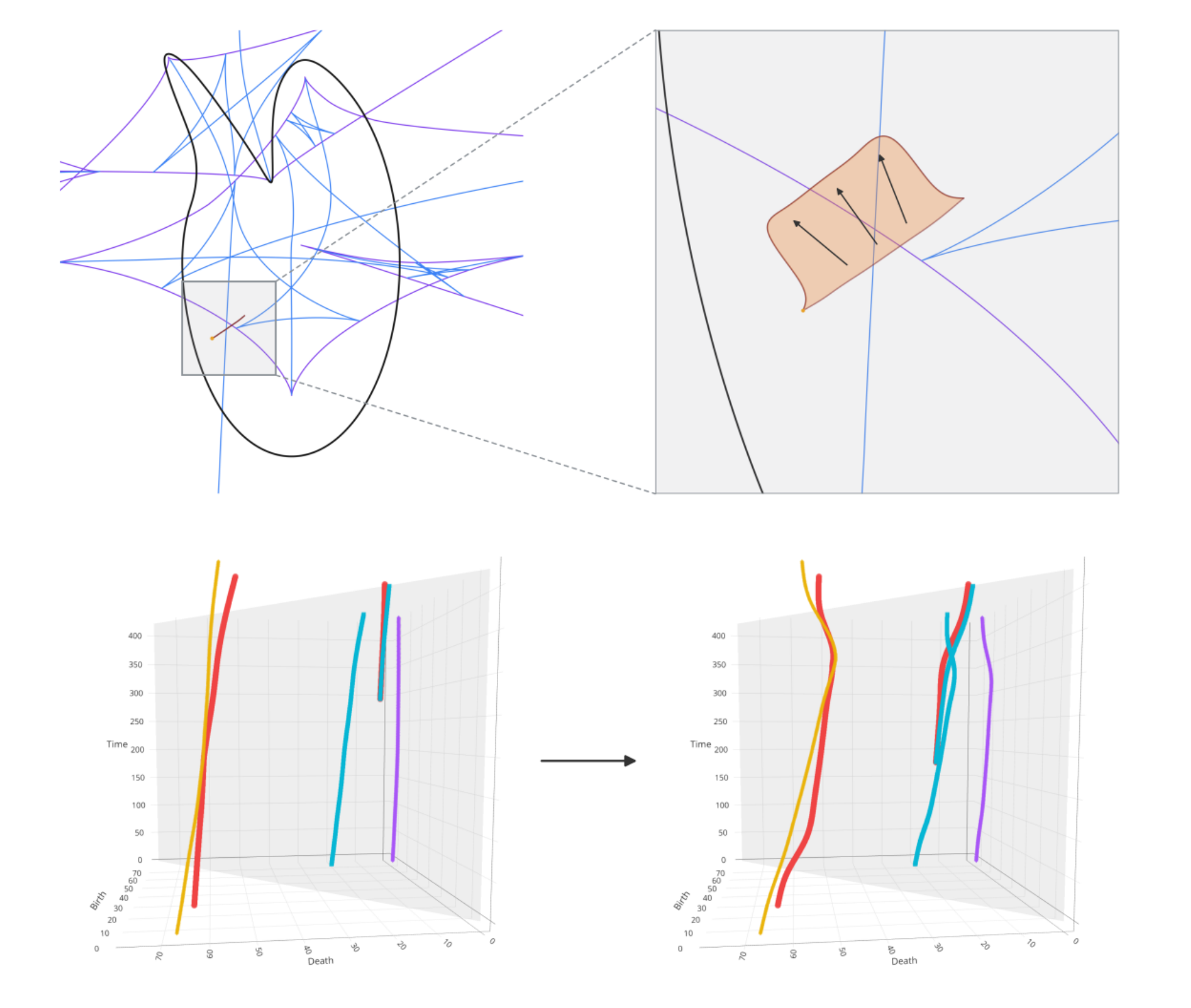}
    \caption{The transversal elementary move on a singularity of type $A_1^2/A_2$. Top: the curve $\gamma_\mu$ (dark red) is pushed in the direction of the arrows across the transverse intersection of a manifold piece of the symmetry set (light blue) with a manifold piece of the focal set (purple); the box on the left is enlarged on the right. Bottom: the vineyard before (left) and after(right) the move. The time at which a vine meets the diagonal plane and the time at which a concurrence occurs are interchanged, which is the vine diagonal--concurrence rearrangement of Definition~\ref{def:ChangesVineyard}. The two events involve different vines, so the topology of the vineyard is unchanged.} 
    \label{fig:A12A2transversalresult} 
\end{figure}

{
For the two \emph{tangential} moves of Figure~\ref{fig:ReidemeisterA12A2} (right), no separate argument is needed. By Proposition~\ref{prop:reduction} the tangential move of type $A_1^2$ is the composition of a tangential move on a singularity of type $A_1^2$ with a transversal move on $A_1^2/A_2$, and symmetrically the tangential move of type $A_2$ is the composition of a tangential move on a singularity of type $A_2$ with a transversal move on $A_1^2/A_2$. The induced changes are therefore the corresponding compositions: a concurrence pair creation, respectively a vine creation or split, followed by a vine diagonal--concurrence rearrangement; see Figures~\ref{fig:A12A2tangentialA12result} and~\ref{fig:A12A2tangentialA2result}.
}

\begin{figure}[h!]
    \centering
    \includegraphics[width=\linewidth]{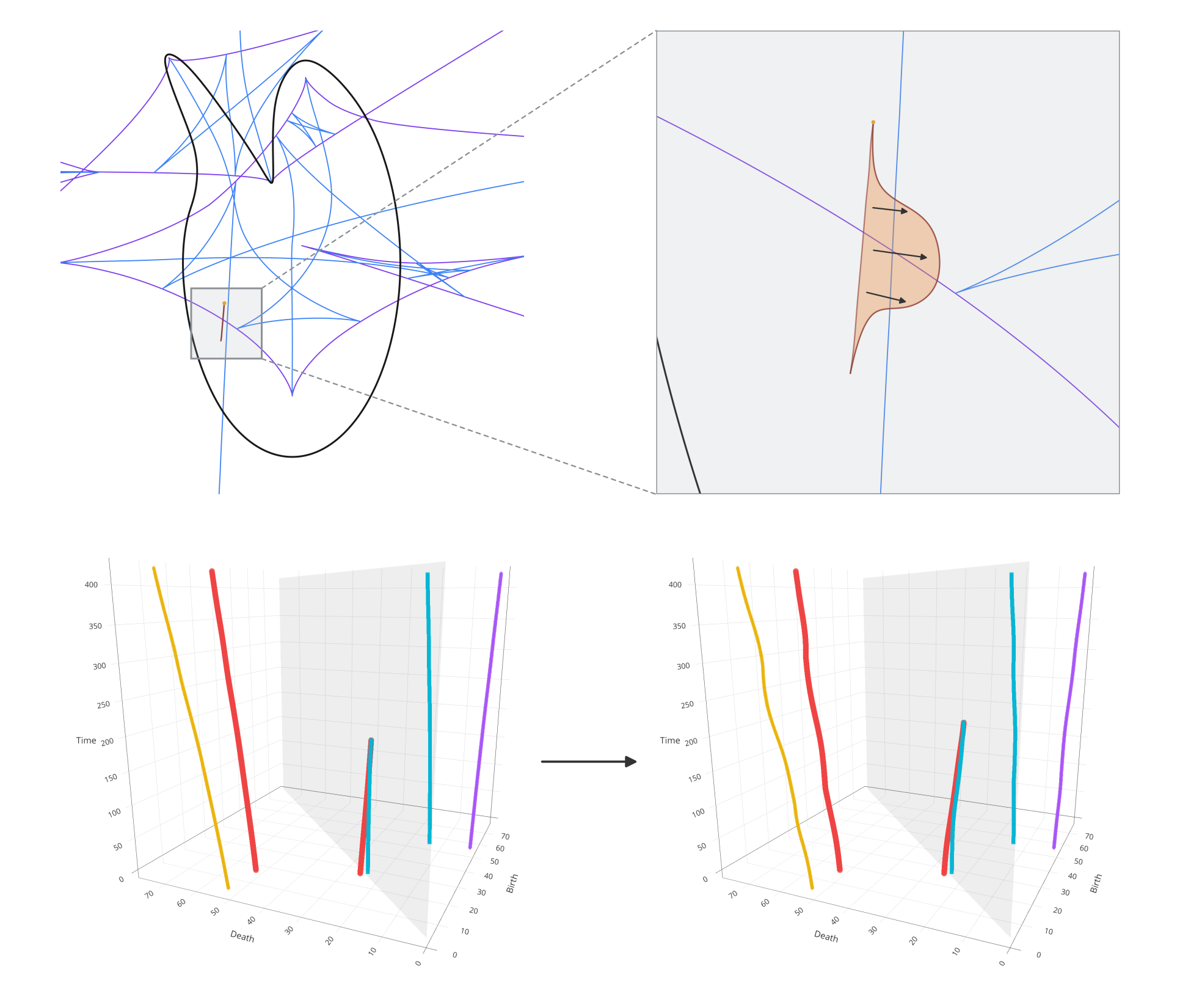}
    \caption{The tangential elementary move of type $A_1^2$ on a singularity of type $A_1^2/A_2$. Top: the curve $\gamma_\mu$ (dark red) is pushed in the direction of  the arrows until it is tangent to the manifold piece of the symmetry set (light blue) at the point where the focal set (purple) crosses it, and then past it. Bottom: the vineyard before (left) and after (right) the move. By Proposition~\ref{prop:reduction} this move is a composition of the tangential move on $A_1^2$ with the transversal move on $A_1^2/A_2$, and the induced change is the corresponding composition of a concurrence pair creation with a vine diagonal--concurrence rearrangement. The concurrence pair is not legible from this viewing direction; it involves the birth coordinates alone, which this projection compresses.} 
    \label{fig:A12A2tangentialA12result} 
\end{figure}

\begin{figure}[h!]
    \centering
    \includegraphics[width=\linewidth]{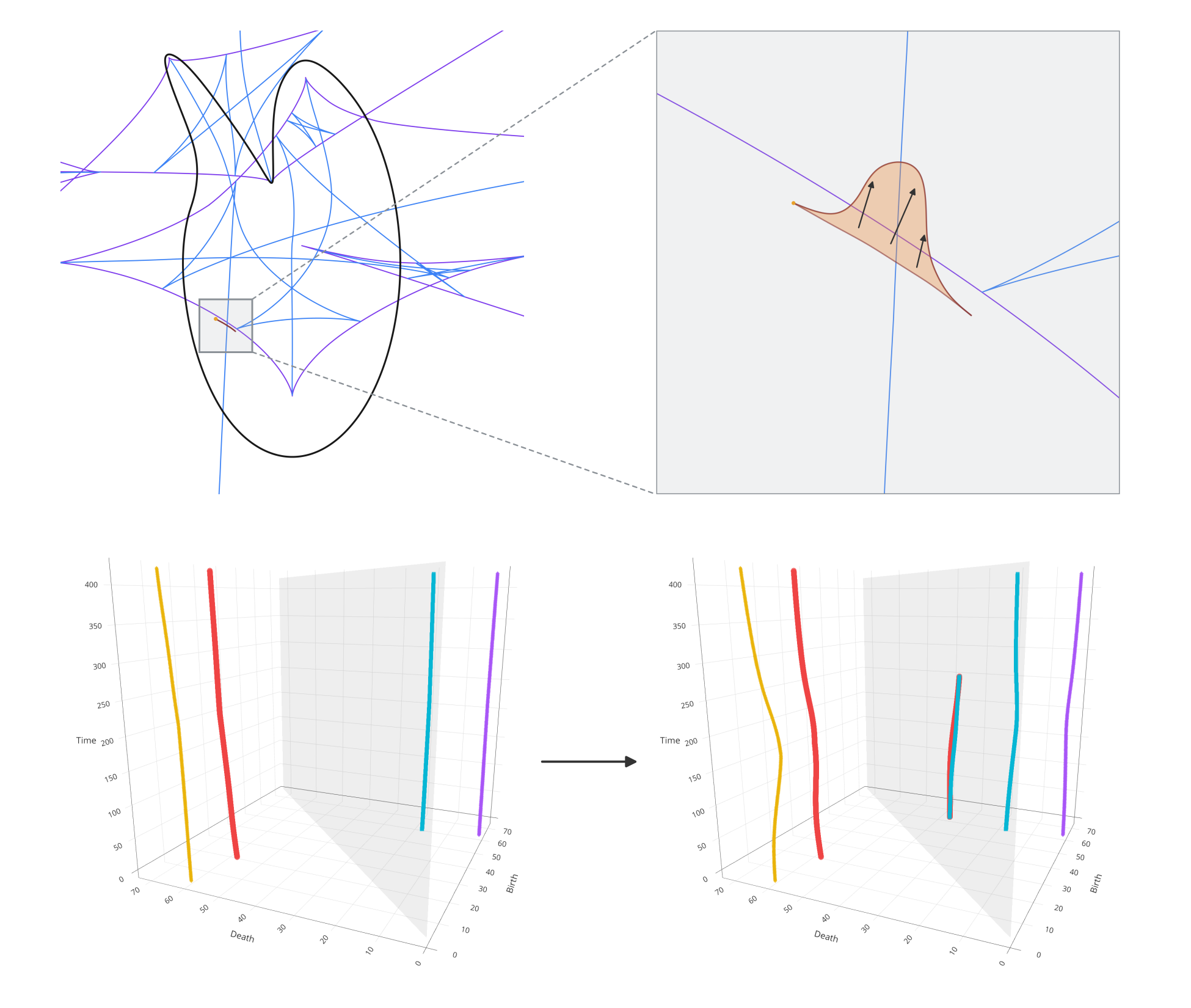}
    \caption{The tangential elementary move of type $A_2$ on a singularity of type $A_1^2/A_2$. Top: the curve $\gamma_\mu$ (dark red) is pushed in the direction of the arrows until it is tangent to the manifold piece of the focal set (purple) at the point where the symmetry set (light blue) crosses it, and then past it. Bottom: the vineyard before (left) and after (right) the move. A short vine of low persistence appears close to the diagonal plane, so the change is a vine creation followed by a vine diagonal--concurrence rearrangement, in accordance with Proposition~\ref{prop:reduction}.} 
    \label{fig:A12A2tangentialA2result} 
\end{figure}

\subparagraph{Type $A_1^3$ singularity}

Let $p$ be a singularity of type $A_1^3$, which as we have seen in Section \ref{sec:Singularities_Plane} refers to a single circle centred at $p$ which is tangent to $\mathcal{M}$ at three points. This means that three coordinates of three points in the persistence diagram of $d_\mathbb{E}^2 (\cdot , p ')_\M$ are almost equal for $p'$ near $p$. The fact that there are three distinct points follows from the third point of Observation \ref{obs:dictionary}.  
In particular, by Observation \ref{ob:A12intersection} the vines do not intersect. 
Locally three manifold pieces $S_{12}, S_{13}, S_{23}$ of the symmetry set intersect at $p$, each individual one dimensional stratum corresponding to a concurrence, as we have discussed for the $A_1^2$ singularities.

{
For the \emph{transversal} move of Figure~\ref{fig:ReidemeisterA13} (left), $\gamma_\mu$ crosses all three pieces transversally for $\mu$ near $\mu^*$ and passes over their common point at $\mu = \mu^*$. Writing $t_{12}, t_{13}, t_{23}$ for the three crossing times, these coincide at $\mu = \mu^*$ and their order is reversed as $\mu$ passes $\mu^*$. The three concurrences involved concern the three pairs among the same three vines, and the move interchanges the order in time in which they occur: this is a triple concurrence rearrangement in the sense of Definition~\ref{def:ChangesVineyard}. Since only single coordinates are ever shared, the vines do not meet and, by Observation~\ref{ob:A12intersection}, they move isotopically throughout, so neither the topology of the vineyard nor its monodromy is affected; see Figure~\ref{fig:A13transversalresult}.
}

\begin{figure}[h!]
    \centering
    \includegraphics[width=\linewidth]{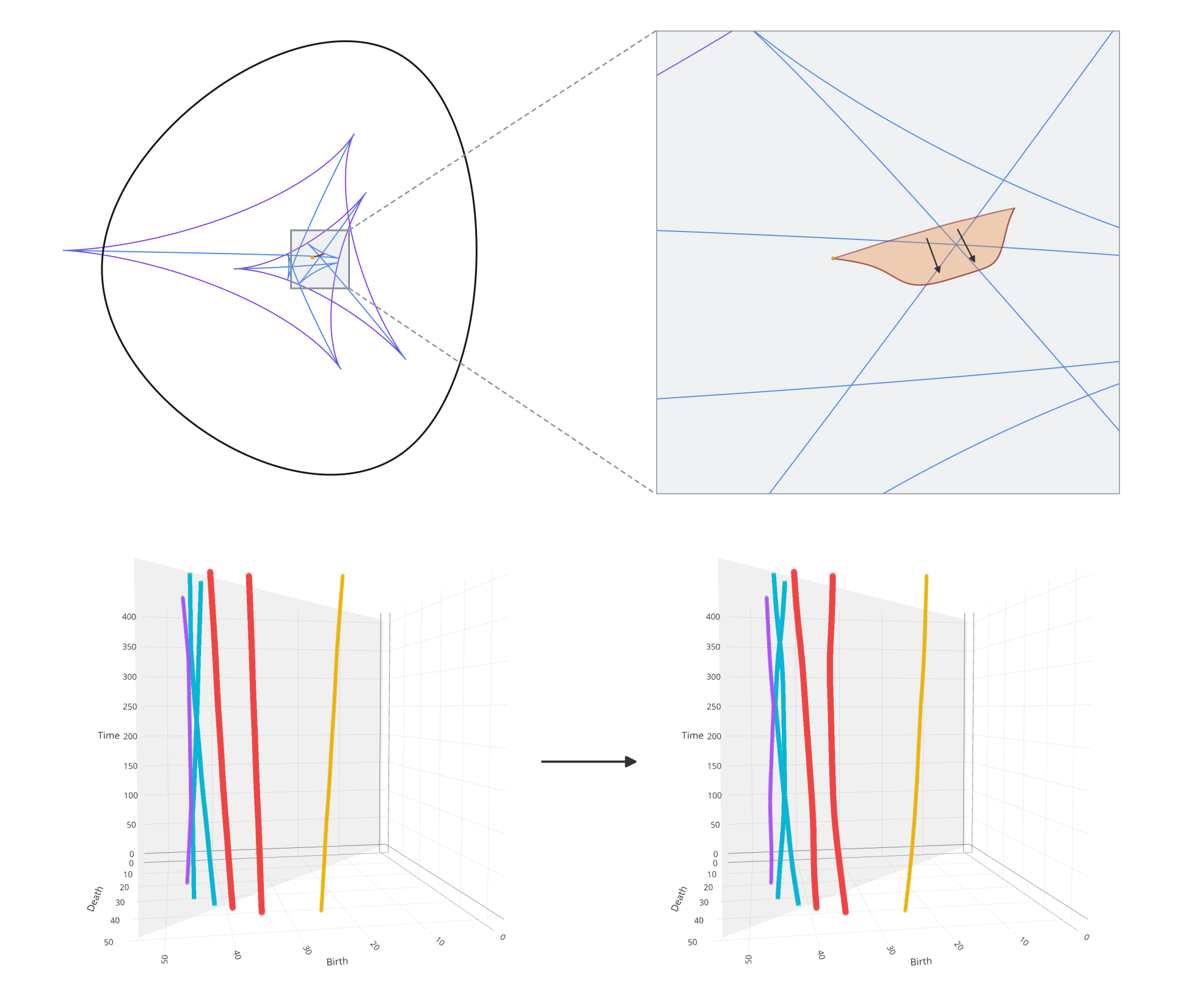}
    \caption{The transversal elementary move on a singularity of type $A_1^3$. Top: the curve $\gamma_\mu$ (dark red) is pushed in the direction of the arrows across the point where three manifold pieces of the symmetry set (light blue) meet; the box on the left is enlarged on the right. Bottom: the vineyard before (left) and after (right) the move. The order in time of the three concurrences among the same three vines is reversed, which is the triple concurrence rearrangement of Definition~\ref{def:ChangesVineyard}. The vines appear to cross in this projection, but at each concurrence only one of their two coordinates agrees, so they do not meet.} 
    \label{fig:A13transversalresult} 
\end{figure}

For the \emph{non-transversal} move of Figure~\ref{fig:ReidemeisterA13} (right), no separate argument is needed: by Proposition~\ref{prop:reduction} it is the composition of a non-transversal move on a singularity of type $A_1^2$ with a transversal move on $A_1^3$, as shown in Figure~\ref{fig:EquivalenceA13}, so the induced change is a concurrence pair creation followed by a triple concurrence rearrangement; see Figure~\ref{fig:A13nontransversalresult}.

\begin{figure}[h!]
    \centering
    \includegraphics[width=\linewidth]{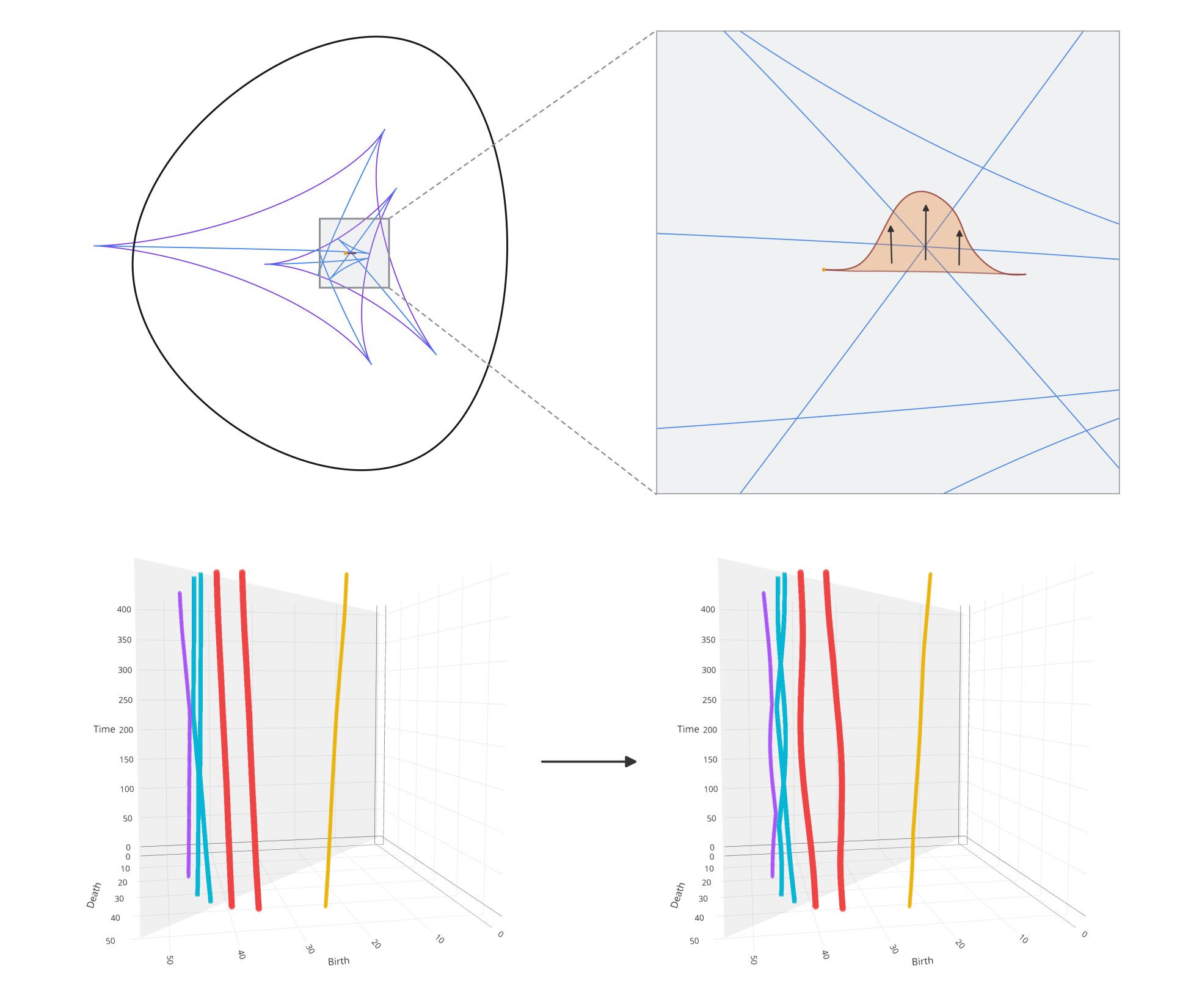}
    \caption{The non-transversal elementary move on a singularity of type $A_1^3$. Top: the curve $\gamma_\mu$ (dark red) is pushed in the direction of the arrows until it is tangent to one of the three manifold pieces of the symmetry set (light blue) at their common point, and then past it. Bottom: the vineyard before (left) and after (right) the move. By Proposition~\ref{prop:reduction} this move is a composition of the non-transversal move on $A_1^2$ with the transversal move on $A_1^3$, and the induced change is the corresponding composition of a concurrence pair creation with a triple concurrence rearrangement.} 
    \label{fig:A13nontransversalresult} 
\end{figure}

\subparagraph{Type $A_1A_2$ singularity}

Let $p$ be a singularity of type $A_1A_2$. As we've recalled in Section \ref{sec:Singularities_Plane}, this means the focal set $F$ is locally a manifold (curve) that goes though $p$ and symmetry set $S$ has a cusp singularity with $p$ at its tip. The symmetry set lies on the side of $F$, where there is one point more in the persistence diagram. As we have seen above the focal set gives rise to vine leaving or entering the diagonal, while the symmetry set induces a concurrence.

{
For the elementary move of type $\textrm{I}a$ of Figure~\ref{fig:ReidemeisterA1A2}, $\gamma_\mu$ is slid across the cusp point, so that the branch of the cusp it crosses is exchanged for the other one while its crossing with $F$ persists. On both sides of $\mu^*$ the curve crosses $F$ once and one branch of the cusp once and in the same order (in $t$), so the vineyard has, throughout, one vine entering  or 
leaving the diagonal and one concurrence between the vine that is close to the diagonal and some other vine.}
The move therefore gives an instance of a vine diagonal--concurrence configuration in which no reordering takes place, and the induced change in the vineyard is trivial; see Figure~\ref{fig:A1A2iaresult}.

\begin{figure}[h!]
    \centering
    \includegraphics[width=\linewidth]{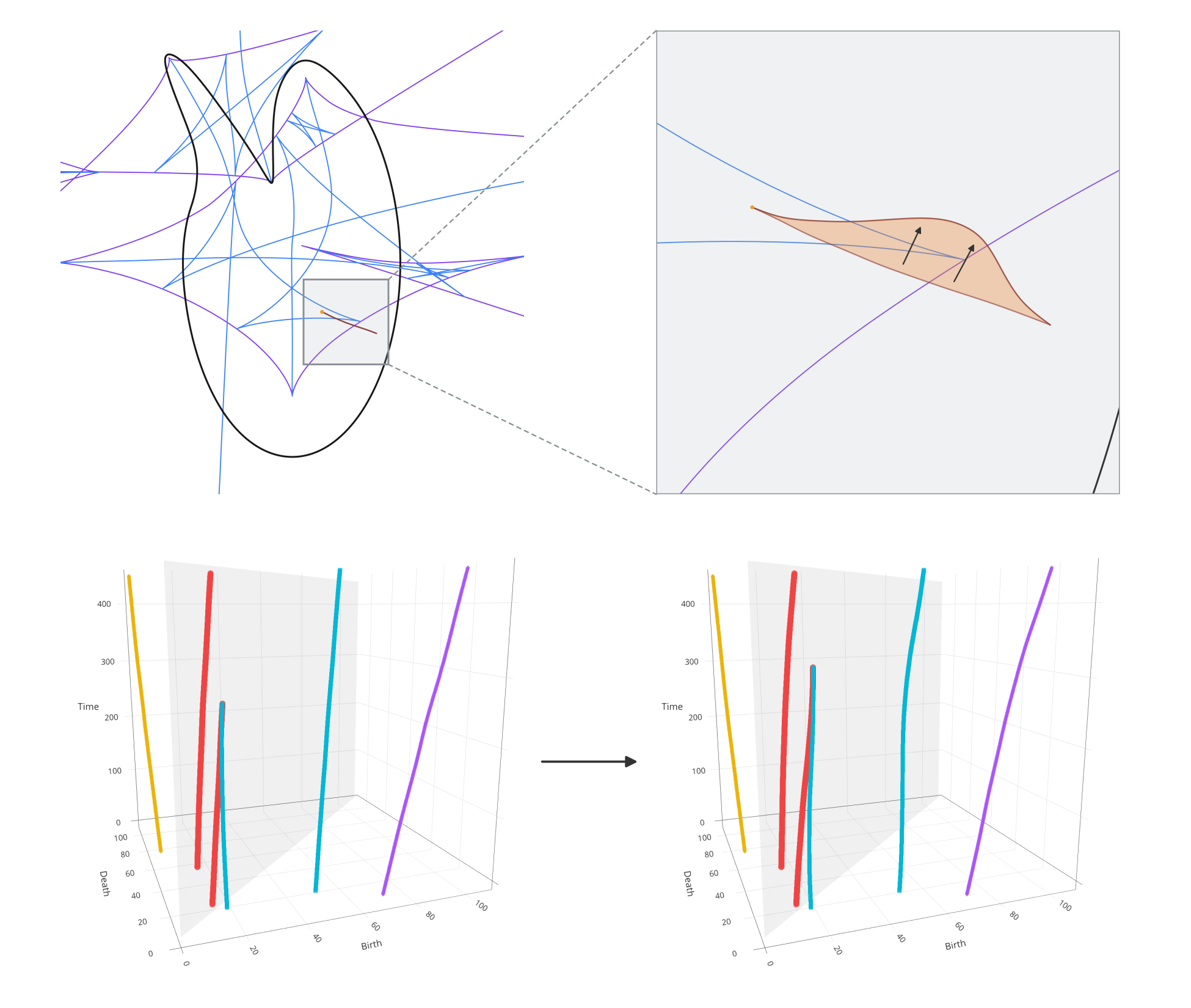}
    \caption{The elementary move of type $\textrm{I}a$ on a singularity of type $A_1A_2$. Top: the curve $\gamma_\mu$ (dark red) is slid in the direction of the arrows across the point where a cusp of the symmetry set (light blue) touches a manifold piece of the focal set (purple); the box on the left is enlarged on the right. Bottom: the vineyard before (left) and after (right) the move. The short vine and its concurrence with the neighbouring vine are present on both sides, and their order in time is unchanged, so the induced change is trivial.} 
    \label{fig:A1A2iaresult} 
\end{figure}

For the elementary moves of types $\textrm{I}b$, $\textrm{II}a$ and $\textrm{II}b$, no separate argument is needed: by Proposition~\ref{prop:reduction} each is a combination of a move of type $\textrm{I}a$ with tangential moves on singularities of type $A_1^2$ and of type $A_2$, as shown in Figures~\ref{fig:EquivalenceA1A2First}--\ref{fig:EquivalenceA1A2Second}. The induced changes are the corresponding combinations of a concurrence pair creation, respectively a vine creation or split, with the trivial change above; see Figures~\ref{fig:A1A2ibresult}, \ref{fig:A1A2iiaresult} and~\ref{fig:A1A2iibresult}.

\begin{figure}[h!]
    \centering
    \includegraphics[width=\linewidth]{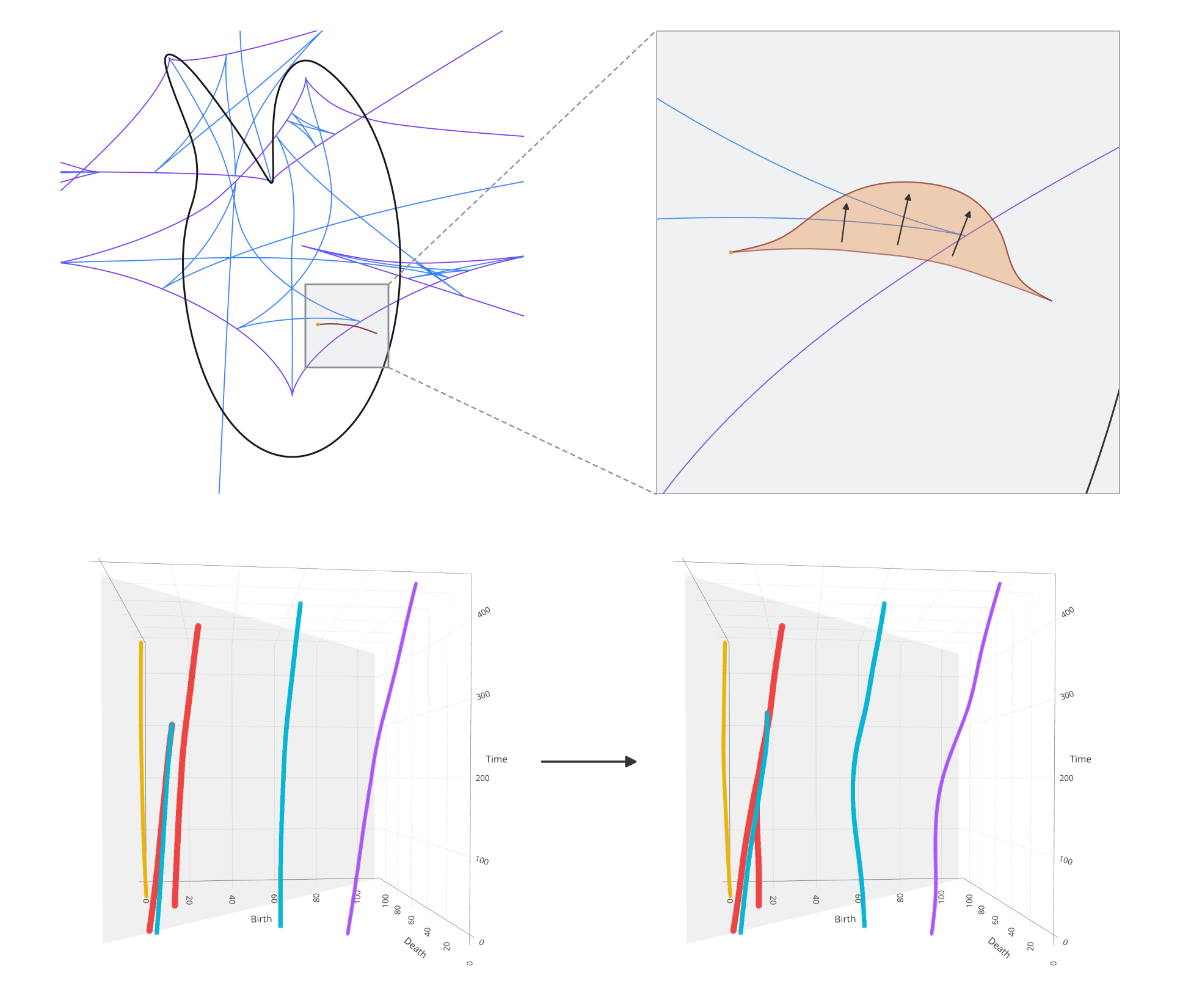}
    \caption{The elementary move of type $\textrm{I}b$ on a singularity of type $A_1A_2$. Top: the curve $\gamma_\mu$ (dark red) is pushed in the direction of the arrows transversally across the cusp point. Bottom: the vineyard before (left) and after (right) the move. By Proposition~\ref{prop:reduction} this move is a combination of a tangential move on $A_1^2$ with a move of type $\textrm{I}a$, and the induced change is the corresponding combination.} 
    \label{fig:A1A2ibresult} 
\end{figure}

\begin{figure}[h!]
    \centering
    \includegraphics[width=\linewidth]{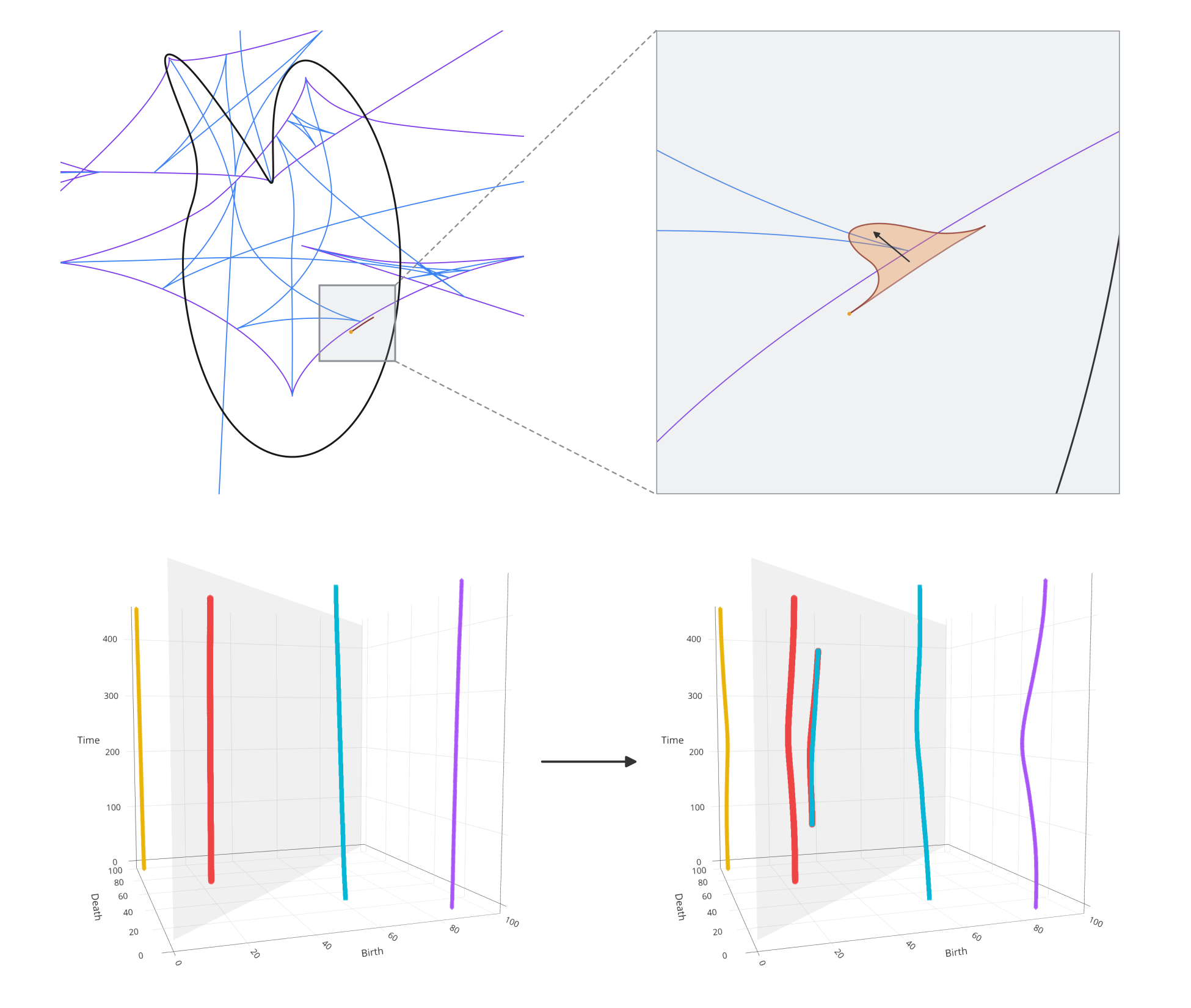}
    \caption{The elementary move of type $\textrm{II}a$ on a singularity of type $A_1A_2$. Top: the curve $\gamma_\mu$ (dark red) is pushed in the direction of the arrow over the cusp point. Bottom: the vineyard before (left) and after (right) the move. A short vine of low persistence appears, so the change is a vine creation together with the trivial change of the move of type $\textrm{I}a$, in accordance with Proposition~\ref{prop:reduction}.} 
    \label{fig:A1A2iiaresult} 
\end{figure}

\begin{figure}[h!]
    \centering
    \includegraphics[width=\linewidth]{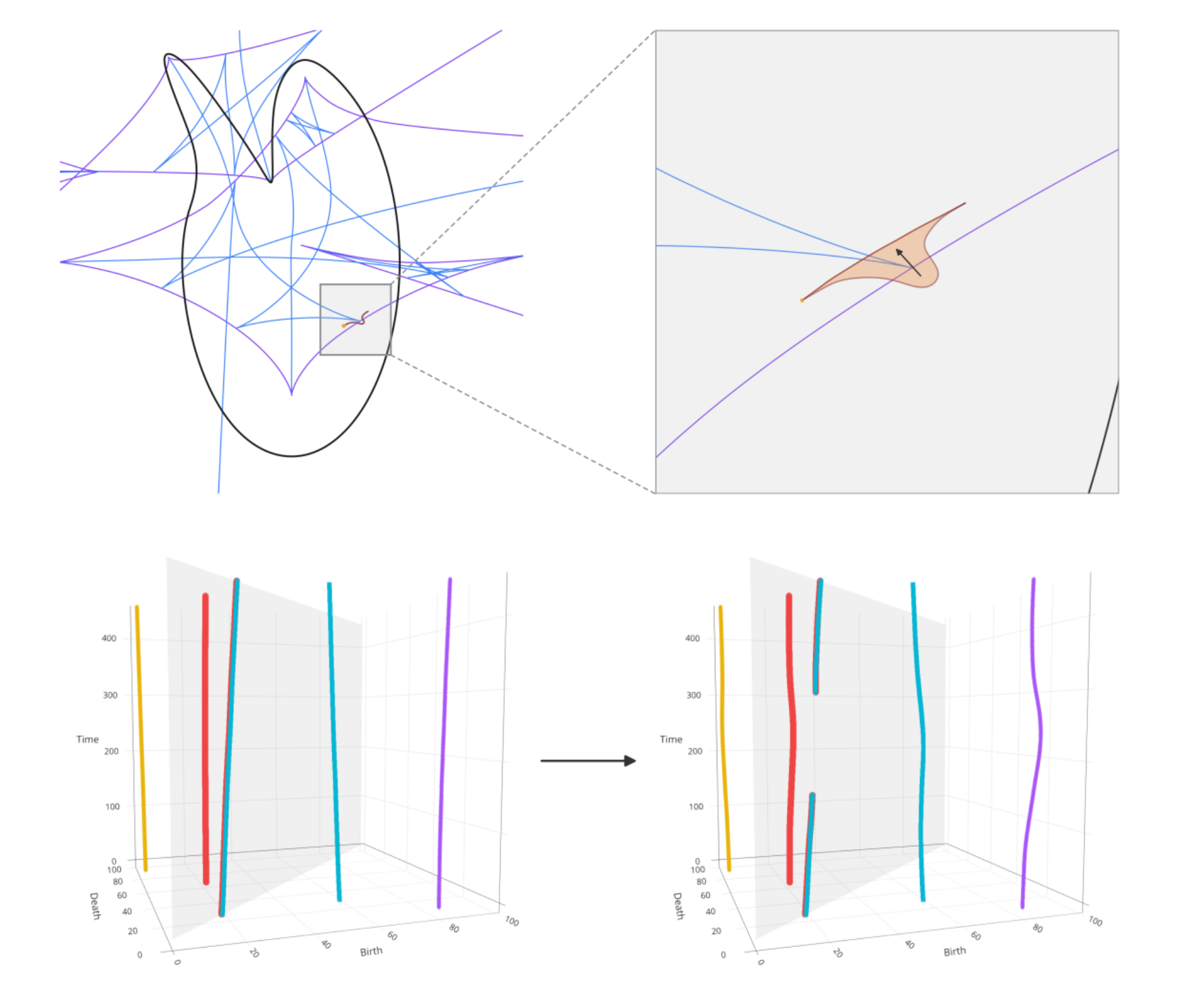}
    \caption{The elementary move of type $\textrm{II}b$ on a singularity of type $A_1A_2$. Top: the curve $\gamma_\mu$ (dark red) is pushed in the direction of the arrow so that it dips past the cusp point. Bottom: the vineyard before (left) and after (right) the move. The vine present throughout the local time interval is cut in two, so the change is a split together with the trivial change of the move of type $\textrm{I}a$, in accordance with Proposition~\ref{prop:reduction}.} 
    \label{fig:A1A2iibresult} 
\end{figure}

\subparagraph{Type $A_3$ singularity}

Let $p$ be a singularity of type $A_3$. As we've recalled in Section \ref{sec:Singularities_Plane}, this means the focal set $F$ is locally a cusp that has $p$ at its tip and symmetry set $S$ has ends at $p$, between the two branches of the focal set. The symmetry set lies on the side of $F$, where there is one point more in the persistence diagram. As we have seen in the discussion of the $A_1^2$ and $A_2$ singularities, the focal set gives rise to vine leaving or entering the diagonal, while the symmetry set induces a concurrence.

For the elementary move of \emph{special type} of Figure~\ref{fig:ReidemeisterA3} (left), $\gamma_\mu$ is tangent to the symmetry set exactly at the singularity, and meets the focal set only once, as noted in Figure~\ref{fig:ReidemeisterA3}. A vine is therefore present on one side of that crossing throughout the move. For $\mu$ before $\mu^*$ the curve misses the branch of the symmetry set emanating from $p$, so this vine carries no concurrence. At $\mu=\mu^*$, the curve reaches the end point of the symmetry set, and there the point at which the vine leaves the diagonal comes to share a coordinate with the vine carried by the remaining contact point. For $\mu$ beyond $\mu^*$, the two vines share that coordinate over an interval of times and then separate again. Thus a single coordinate of each of two vines begins to coincide, and the limiting position of this concurrence at $\mu = \mu^*$ is the point at which the short vine ends on the diagonal. This is precisely a single concurrence creation in the sense of Definition~\ref{def:ChangesVineyard}, and the concurrence is a knee, as that definition requires. As noted in Lemma~\ref{lem:MoveObservationLoopClassification} and Figure~\ref{fig:ReidemeisterA3}, the move is not strictly generic, since the tangency must occur at the singularity itself; it is included because it simplifies the reduction below. See Figures~\ref{fig:A3special1result} and~\ref{fig:A3special2result}.

\begin{figure}[h!]
    \centering
    \includegraphics[width=\linewidth]{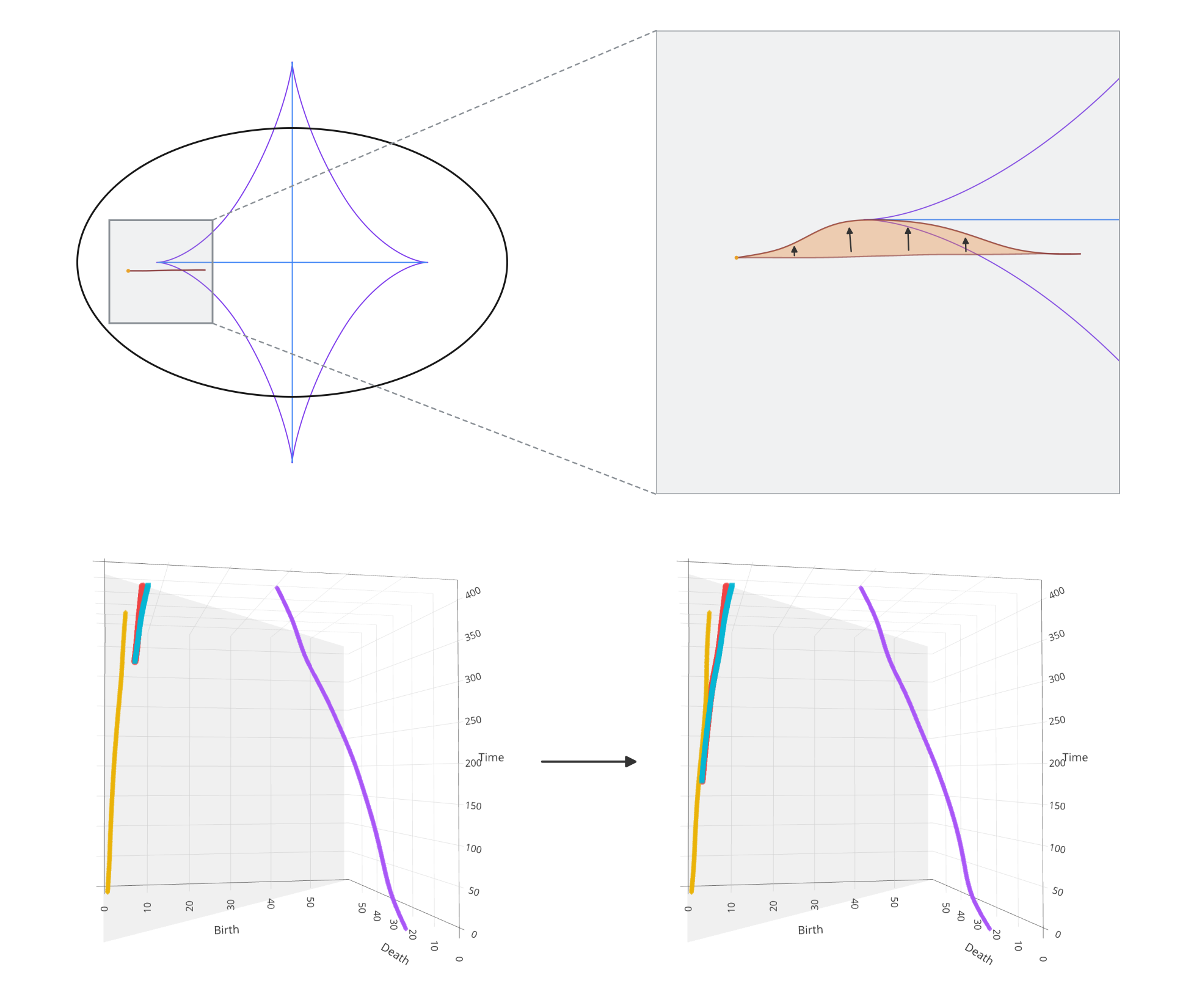}
    \caption{The elementary move of special type on a singularity of type $A_3$, before the transition. Top: the curve $\gamma_\mu$ (dark red) is pushed in the direction of the arrows towards the tip of the cusp of the focal set (purple), at which an end point of the symmetry set (light blue) lies; the box on the left is enlarged on the right. Bottom: the vineyard before (left) and after (right) this part of the deformation. The short vine carries no concurrence, but the point at which it leaves the diagonal plane approaches the neighbouring vine.} 
    \label{fig:A3special1result} 
\end{figure}

\begin{figure}[h!]
    \centering
    \includegraphics[width=\linewidth]{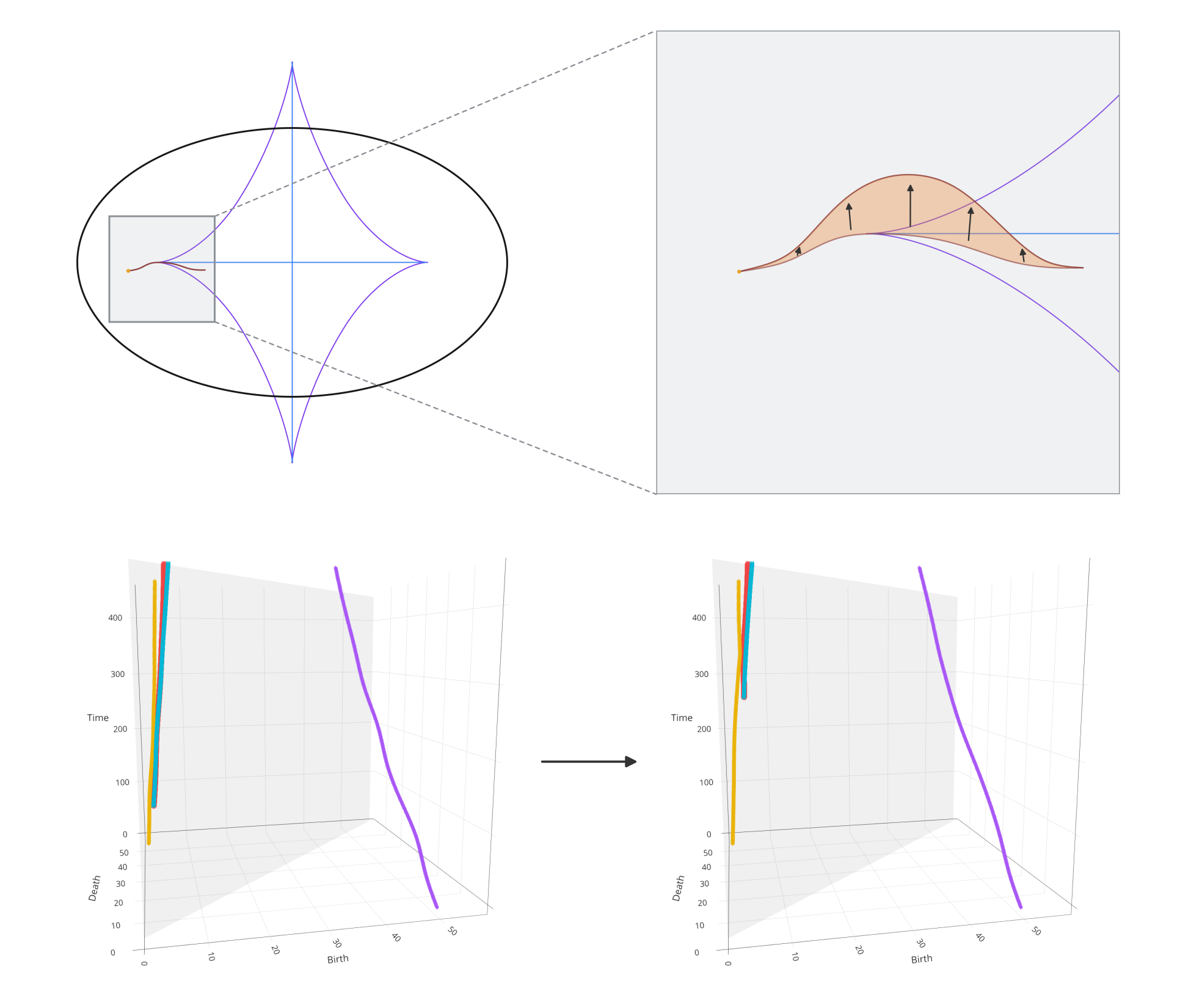}
    \caption{The elementary move of special type on a singularity of type $A_3$, at the transition. Top: the curve $\gamma_\mu$ (dark red) becomes tangent to the symmetry set (light blue) exactly at its end point, which lies at the tip of the cusp of the focal set (purple). Bottom: the vineyard before (left) and after (right) the move. The end point at which the short vine leaves the diagonal plane comes to share a coordinate with the neighbouring vine, after which the two separate again; this is the single concurrence creation of Definition~\ref{def:ChangesVineyard}, and the concurrence is a knee.} 
    \label{fig:A3special2result} 
\end{figure}

For the elementary move of type $\textrm{II}$ of Figure~\ref{fig:ReidemeisterA3} (right), no separate argument is needed: by Proposition~\ref{prop:reduction} it is the composition of a tangential elementary move on a singularity of type $A_2$ with a move of special type as shown in Figure~\ref{fig:EquivalenceA3Typeii}, so the induced change is a vine creation followed by a single concurrence creation; see Figure~\ref{fig:A3typeiiresult}.
\end{proof}

\begin{figure}[h!]
    \centering
    \includegraphics[width=\linewidth]{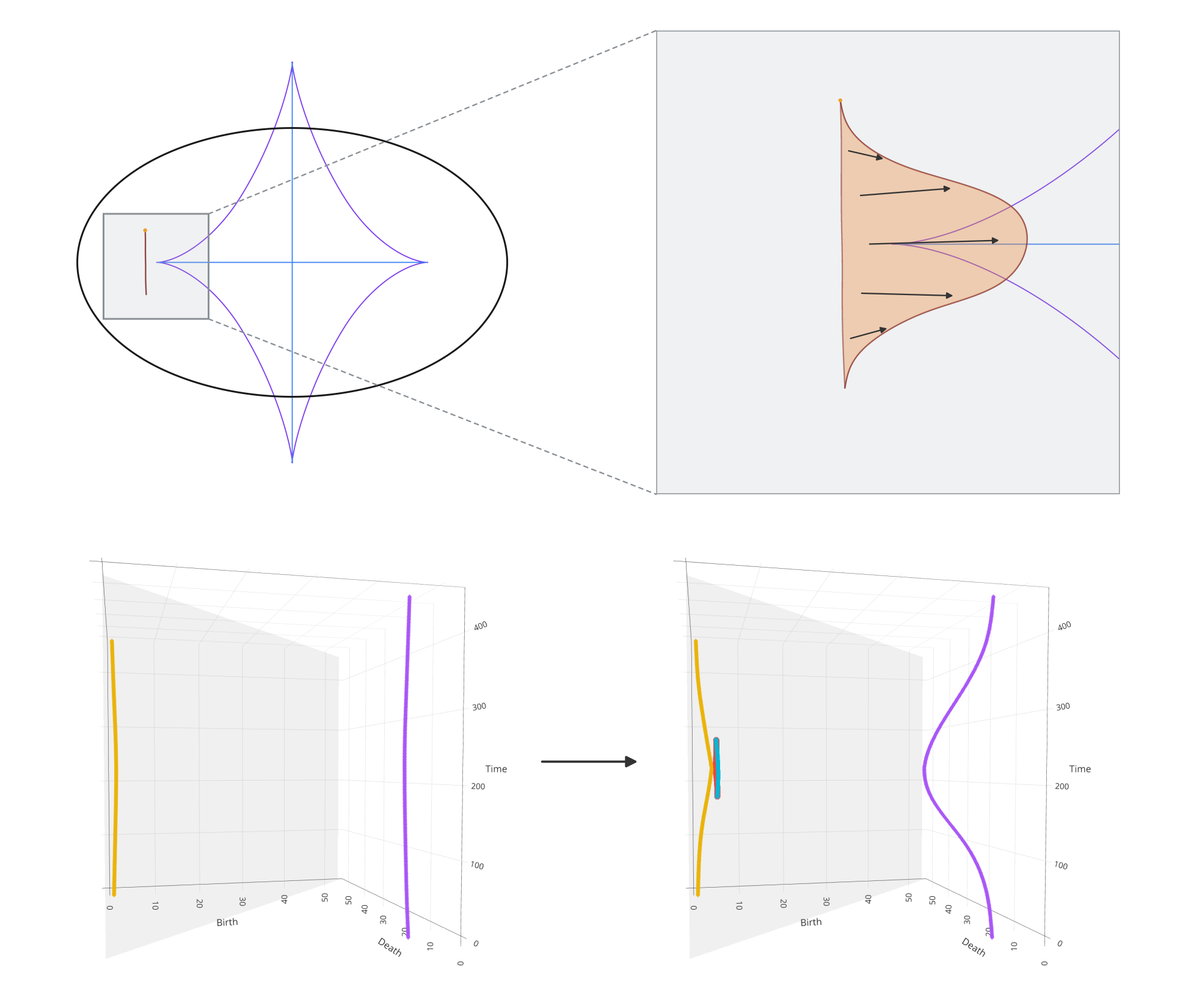}
    \caption{The elementary move of type $\textrm{II}$ on a singularity of type $A_3$. Top: the curve $\gamma_\mu$ (dark red) is pushed in the direction of the arrows across the cusp of the focal set (purple). Bottom: the vineyard before (left) and after (right) the move. A short vine appears, carrying a concurrence; by Proposition~\ref{prop:reduction} this move is the composition of a tangential elementary move on $A_2$ with the move of special type, and the induced change is the corresponding composition.} 
    \label{fig:A3typeiiresult} 
\end{figure}

\begin{corollary}
    Note that Theorem \ref{prop:MovesOnVineyard} implies any (generic) change in the vineyard can be written as a combination of the elementary moves described in Definition \ref{def:ChangesVineyard}. 
\end{corollary}

\section{From elementary moves to global topology}
\label{sec:MainResult} 

In this section we zoom out, that is to say we consider the global behaviour of a loop. 
However, our local analysis suffices to understand the transitions in a vine:  
\begin{observation}
Using partition of unity functions (with respect to $t$) on $\gamma_\mu(t)$ one can reduce to the setting where $\gamma_\mu(t)$ only changes in a small neighbourhood of a point for some subinterval for the parameter $\mu$. This implies that any change in the topology of the vines is a combination of the local changes we have described in the previous sections. 
\end{observation}
\begin{definition}
\label{def:GeneralizedGenericIsotopy}
We say that $\gamma_\mu(t)$ is a generalized generic isotopy relative to a generic curve $\mathcal{M}$, if locally its behaviour is described by the extended list of elementary moves of Lemma \ref{lem:MoveObservationLoopClassification}. 
\end{definition}

In particular, Theorem \ref{prop:MovesOnVineyard} has the following consequence: 
\TopologyMonodromyFromMoves*

Before we discuss the proof of the proposition, we return to the example given in Figure~\ref{fig:bishop-staff2} for the proof of Lemma \ref{lem:A2SA2Monodromy} which illustrate the different roles of $A_2$ and $A_2/A_2$ singularities.

\begin{remark} \label{rem:BishopRevisited} 
In the case of the example of Figure \ref{fig:bishop-staff2} there are two interesting isotopies to the curve $\gamma$ indicated in dark red: 
\begin{itemize}
    \item The first isotopy starts (of course) from a point, then it crosses the focal set twice with two $A_2$ tangential moves. This creates two vines. Then there is a transversal move on the $A_2/A_2$ singularity. This means that the two vines that we created before start to overlap in time (that is the time interval for which they are off-diagonal for both overlaps). Then finally a $A_2$ move merges these two vines. Because the time intervals overlap before the merge, the time for which the resulting vine is off-diagonal is more than $2\pi$ and hence there is monodromy. In conclusion, the monodromy appears at an $A_2$ vine merge.  
    \item The second isotopy again starts with a point. It then crosses the focal set once creating a single vine. Finally there is a transversal move on the $A_2/A_2$ singularity. Here the vine, as discussed in Lemma \ref{lem:A2SA2Monodromy}, starts to exhibit monodromy because the length of the vine exceeds $2\pi$. In conclusion, the monodromy appears at a transversal move on the $A_2/A_2$ singularity.
\end{itemize}
\end{remark}

We are now ready to give the proof. 
\begin{proof}[Proof of Proposition \ref{prop:TopologyMonodromFromMoves}]
Because concurrences refer to specific coordinates, but not to the topology of the vine, it follows that all changes to the vineyard that involve concurrences, as given in Theorem~\ref{prop:MovesOnVineyard}, affect neither the topology nor the monodromy. 

Clearly, the vine creation/annihilation changes the topology of the vineyard, because the number of vines either increases or decreases, but thanks to Convention \ref{convention:Annihilation} the vine thus created or annihilated is unlinked. We further note that since the (off-diagonal part of the) vine is so short, there cannot be monodromy associated to the vine. 

Although splits can clearly change the topology of the vines (as the number of connected components changes), it cannot introduce monodromy where there is none before the split as a direct consequence of Convention \ref{convention:Annihilation}. 
We stress that this may introduce improper intersections on the diagonal, but these are ignored by convention. 

As the inverse operation of the split, the merge changes the topology of the vineyard, because the Euler characteristic of the vineyard changes. The example in Figure \ref{fig:bishop-staff2} and in particular the first isotopy of Remark \ref{rem:BishopRevisited}, explain how a merge can create monodromy without the creation of topology. 

The fact that even small loop enclosing singularity of type $A_1^2/A_1^2$ can induce monodromy is the main result of \cite{local}. We refer to that paper for the full analysis. 

As we have seen in the example of Figure \ref{fig:bishop-staff2} and the second isotopy of Remark \ref{rem:BishopRevisited} exhibit how a transversal move on a singularity of type $A_2/A_2$ can create monodromy. 
\end{proof} 


We can now state our second main result:
\MainTHM*

\begin{proof} 
Because the vines as $\mu=0$ are straight lines, we have no linking or monodromy at $\mu=0$. 
We now use Lemma \ref{lem:focal-loops} to note that $\gamma_\mu$ did not contain any closed cycle of focal set for any $\mu$. To prove this claim we note that if $\gamma_\mu$ did contain, then $\gamma_\mu$ should have passed over: 
\begin{itemize}
    \item Either a singularity of type $A_2/A_2$ at some earlier point in the isotopy (if the cycle is not a smooth curve), which is excluded by assumption.
    \item Or the singularity of type $A_1^2/A_1^2$ which lies inside a smooth loop of $\mathcal{F}$, according to Lemma \ref{lem:focal-loops}, which is also excluded by assumption. 
\end{itemize}
This means that focal set in the interior of $\gamma_{\mu}$ consists of a number of curves {(potentially with cusps, which can be smoothed, cf. Figure \ref{fig:focal-loop-setup})} for which the two end points lie for each lie on $\gamma_\mu$. We can now use the Jordan curve theorem~\cite{jordan1893cours, Hales2007} to conclude that the focal set splits the interior of $\gamma_\mu$ into a number of topological disks. We now consider $\mu$ to be fixed and consider a new isotopy ${\gamma}_\nu$, which isotopes a point in the interior of $\gamma_\mu$ to the curve $\gamma_\mu$, such that it incorporates the topological disks induced by the focal set inductively, see Figure \ref{fig:gammanu}. We do so using one tangential elementary move on $A_2$. By again using 
Lemma \ref{lem:focal-loops} and Proposition \ref{prop:TopologyMonodromFromMoves}, we see that inside each topological disk there are no singularities that can change the topology of the vineyard. Moreover, because only one tangential elementary move on $A_2$ is used, only one creation or split event occurs per topological disk. More precisely, a manifold piece of the focal set divides the ambient space into two parts, on one side there is one more point in the persistence diagram than on the other. If $\gamma_{\nu}$ before the tangential elementary move on $A_2$ lies on the side 
with one point less, then the elementary move yields a creation, if it lies on the side with one point more in the persistence diagram, then it yields a split. 
Although splits and creations can introduce new connected components, they cannot introduce monodromy or linking, as noted in Proposition \ref{prop:TopologyMonodromFromMoves}. The result now follows. 
\end{proof} 

\begin{figure}[h!]
    \centering
    \includegraphics[width=0.6\linewidth]{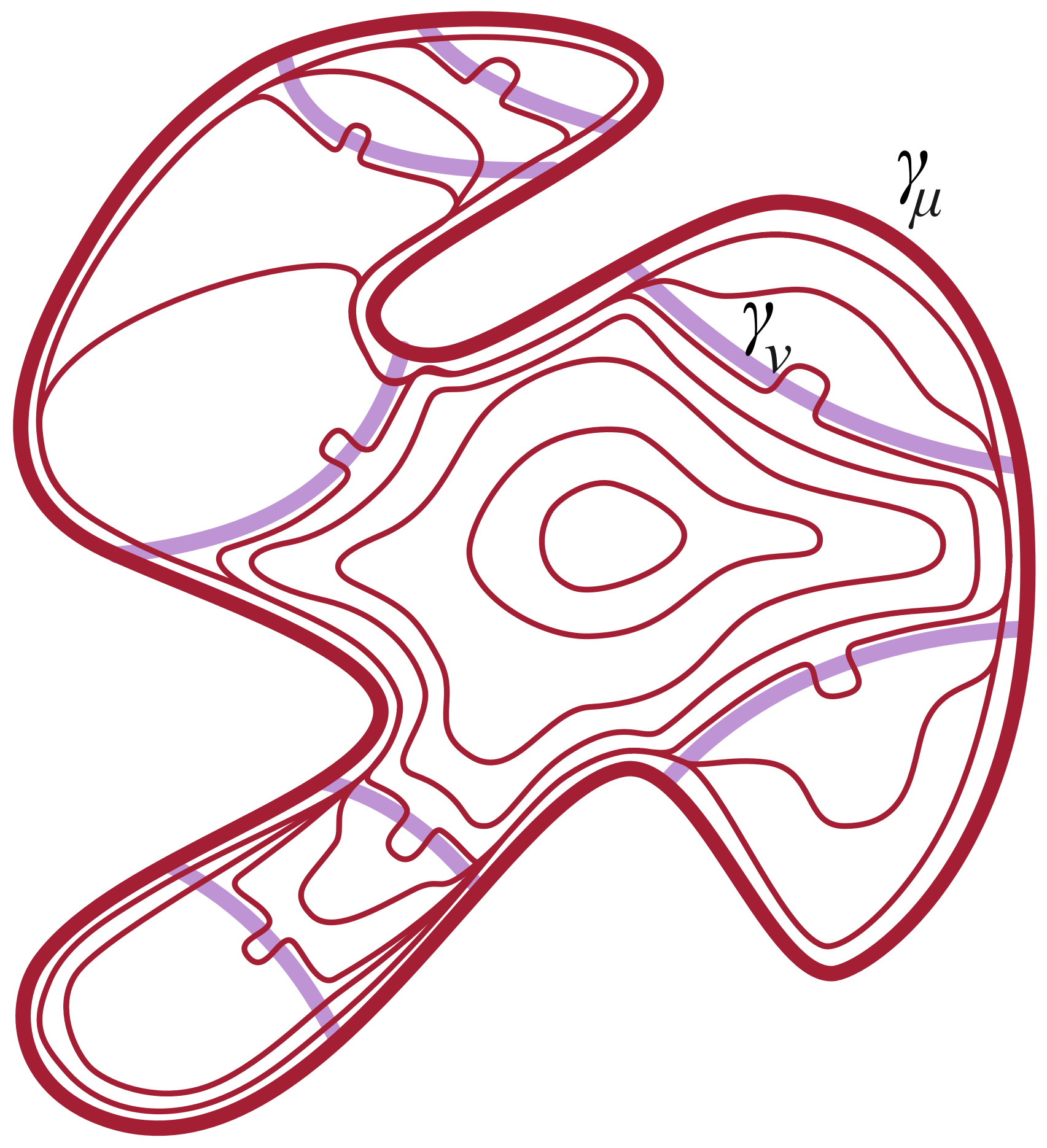}
    \caption{ The sequence of elementary moves with regard to the focal set induced by the ``isotopy''~$\gamma_\nu$ to the curve $\gamma_\mu$.  } 
    \label{fig:gammanu} 
\end{figure}

\begin{remark} \label{rem:Generators} 
As we have seen in Section 3 of \cite{local} and Theorem \ref{prop:MovesOnVineyard} of this paper, the singularities of type $A_1^2/A_1^2$ can (if the topology of $\M$ is right) generate a single interchange between two points in the vineyard. In a braid theoretical sense, this gives generators of the braid that forms the vineyards (or closed vineyard after identifying the ends of the braid). Moreover, if the interior of an observation loop $\gamma$ encloses no focal set, then each singularity of type $A_1^2/A_1^2$ can be combined relatively straightforward, by starting with a constant loop (point) inside the observation loop and  isotoping it to the observation loop and looking which order of generators it induces using Theorem \ref{prop:MovesOnVineyard}. However, given the fact that the focal set is ubiquitous in the plane, this setting is not very likely to occur (in practical settings).
\end{remark}

\end{document}